\documentclass[11pt]{article}
\usepackage[utf8]{inputenc}
\usepackage{cite}
\usepackage{amsmath,amssymb,amsbsy,amstext, amsthm, simplewick, amsfonts}
\usepackage{hyperref}
\usepackage{graphicx}
\usepackage{upgreek}
\usepackage{bm}
\usepackage{framed}
\usepackage{bbm}
\usepackage{textcomp}
\usepackage{tikz}
\usepackage{pifont}
\usetikzlibrary{matrix,shapes,fit,tikzmark,calc}
\usepackage{adjustbox}
\usepackage{makecell}
\usepackage{bbold}
\usepackage{tcolorbox}
\usepackage{physics}
\usepackage{empheq}
\usepackage[normalem]{ulem}

\numberwithin{equation}{section}
\def\L{\textrm{L}}
\def\S{\textrm{S}}
\def\a{\alpha}

\def\be {\begin{equation}}
\def\ee {\end{equation}}

\def\e{\eta}

\def\fI{f}

\def\P{\pi}
\def\Pc{\pi_c}
\def\ba{\begin{eqnarray}}
\def\ea{\end{eqnarray}}

\def\m{\mu}

\def\s{\sigma}

\def \bfx {\textbf{x}}

\def\vpi {\varphi}
\def \bfk {\textbf{k}}
\def \bfq {\textbf{q}}
\def \del {\partial}

\usepackage[framemethod=default]{mdframed}
\newmdenv[skipabove=7pt,
skipbelow=7pt,
rightline=false,
leftline=false,
topline=false,
bottomline=false,
backgroundcolor=gray!10,
linecolor=gray,
innerleftmargin=5pt,
innerrightmargin=5pt,
innertopmargin=5pt,
innerbottommargin=5pt,
leftmargin=0cm,
rightmargin=0cm,
linewidth=4pt]{eBox}
\newmdenv[skipabove=7pt,
skipbelow=7pt,
rightline=false,
leftline=false,
topline=false,
bottomline=false,
backgroundcolor=gray!10,
linecolor=gray,
innerleftmargin=5pt,
innerrightmargin=5pt,
innertopmargin=-5pt,
innerbottommargin=5pt,
leftmargin=0cm,
rightmargin=0cm,
linewidth=4pt]{eBox2}

\usepackage{array}

\definecolor{blue3}{RGB}{31, 119, 180}
\definecolor{red3}{RGB}{	214, 39, 40}
\definecolor{orange3}{RGB}{255, 127, 14}
\definecolor{green3}{RGB}{44, 160, 44}

\usepackage{colortbl}
\definecolor{lightgreen}{cmyk}{0.2, 0, 0.2, 0.2}
\definecolor{lightgray}{cmyk}{0.1,0.2,0,0.1}
\definecolor{lightgray2}{cmyk}{0.1,0.1,0,0.1}

\makeatletter
\newlength{\apb@width}
\newcommand{\autoparbox}[2][c]{\settowidth{\apb@width}{#2}\parbox[#1]{\apb@width}{#2}}

\makeatother


\def\Mpl{M_{\text{Pl}}}

\def\Mp{M_{\rm pl}}

\def \bfp {\textbf{p}}

\def\fNL{f_\textrm{NL}}
\def\F{{\cal F}}

\def\beq{\begin{equation}}
\def\eeq{\end{equation}}

\allowdisplaybreaks[1]
\begin{document}



\begin{titlepage}
\setcounter{page}{1} \baselineskip=15.5pt 
\thispagestyle{empty}

\begin{center}
{\fontsize{18}{18} \centering \bf Cosmological Bootstrap in Slow Motion \vspace{0.1cm}
\;}\\
\end{center}

\vskip 18pt
\begin{center}
\noindent
{\fontsize{12}{18}\selectfont Sadra Jazayeri\footnote{\tt
			jazayeri@iap.fr} and S\'{e}bastien Renaux-Petel\footnote{\tt
			renaux@iap.fr}}
\end{center}

\begin{center}
  \vskip 8pt
  \textit{ Institut d'Astrophysique de Paris, GReCO, UMR 7095 du CNRS et de Sorbonne Universit\'{e},\\ 98 bis
boulevard Arago, 75014 Paris, France}
\end{center}

\vspace{0.4cm}

Speed matters. How the masses and spins of new particles active during inflation can be read off from the statistical properties of primordial density fluctuations is well understood. However, not when the propagation speeds of the new degrees of freedom and of the curvature perturbation differ, which is the generic situation in the effective field theory of inflationary fluctuations. Here we use bootstrap techniques to find exact analytical solutions for primordial 2-,3- and 4-point correlators in this context. We focus on the imprints of a heavy relativistic scalar coupled to the curvature perturbation that propagates with a reduced speed of sound $c_s$, hence strongly breaking de Sitter boosts. We show that akin to the de Sitter invariant setup, primordial correlation functions can be deduced by acting with suitable weight-shifting operators on the four-point function of a conformally coupled field induced by the exchange of the massive scalar. However, this procedure requires the analytical continuation of this seed correlator beyond the physical domain implied by momentum conservation.
We bootstrap this seed correlator in the extended domain from first principles, starting from the boundary equation that it satisfies due to locality. We further impose unitarity, reflected in cosmological cutting rules, and analyticity, by demanding regularity in the collinear limit of the four-point configuration, in order to find the unique solution.
Equipped with this, we unveil that heavy particles that are lighter than $H/c_s$ leave smoking gun imprints in the bispectrum in the form of resonances in the squeezed limit, a phenomenon that we call the low speed collider. We characterise the overall shape of the signal as well as its unusual logarithmic mass dependence, both vividly distinct from previously identified signatures of heavy fields. Eventually, we demonstrate that these features can be understood in a simplified picture in which the heavy field is integrated out, albeit in a non-standard manner resulting in a single-field effective theory that is non-local in space. Nonetheless, the latter description misses the non-perturbative effects of spontaneous particle production, well visible in the ultra-squeezed limit in the form of the cosmological collider oscillations, and it breaks down for masses of order the Hubble scale, for which only our exact bootstrap results hold.


\end{titlepage}


\newpage
\setcounter{tocdepth}{3}
\tableofcontents

\newpage




\section{Introduction}
\begin{figure}
    \centering
    \includegraphics[scale=0.6]{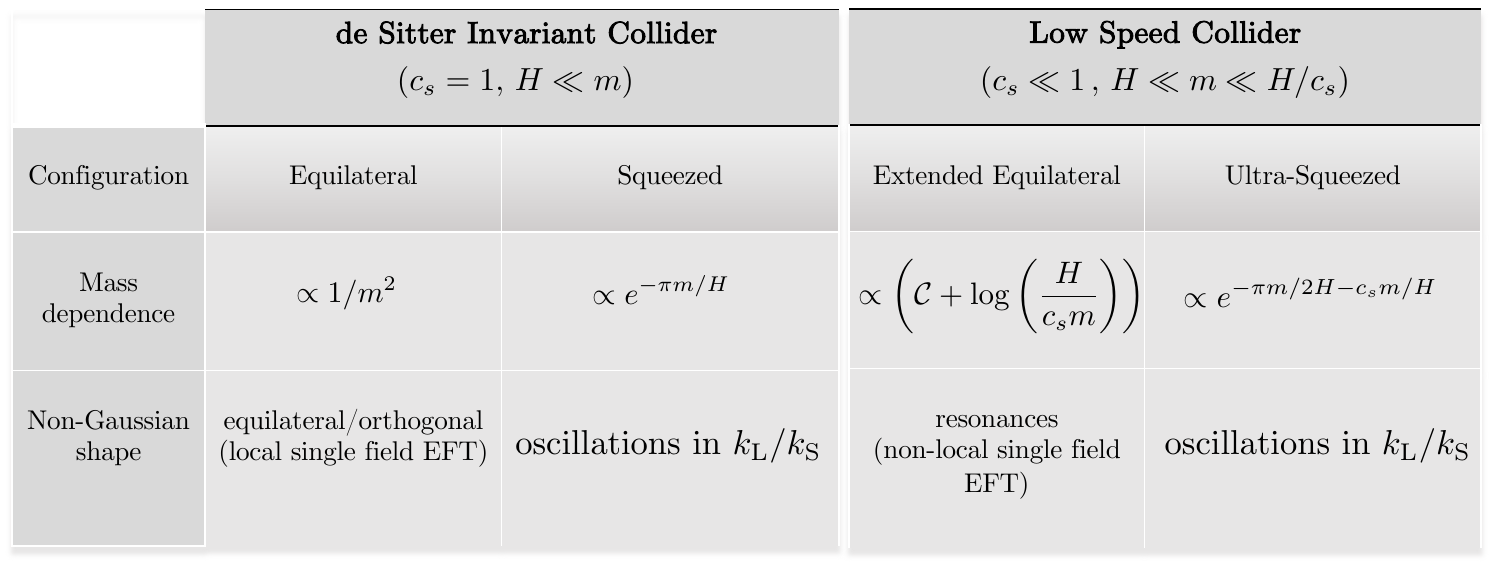}
    \caption{In this table, we compare the qualitative properties of the imprints of heavy fields
    on the bispectrum between the two setups of the de Sitter invariant (more generally $c_s=1$) and low speed colliders.
    In the low speed collider signal, the massive field propagates faster than the curvature perturbation and its generic sound speed $c_\sigma$ is set to unity by rescaling the spatial coordinates. We highlight that the mass dependence of the low speed collider signal depends on the precise kinematical configuration in the extended equilateral region
    ${\cal O}(1)c_s \frac{m}{H}\lesssim k_\L/k_\S \leqslant 1$, with the quoted one corresponding to the equilateral limit.}
    \label{fig:tableintro}
\end{figure}

The exponential expansion of the universe in its earliest epoch not only generates the seeds of primordial perturbations, but is also a generous particle factory that produces species of all types which can be as heavy as the Hubble scale during inflation. Such massive states can leave observable imprints on soft limits of cosmological correlators if they are coupled to the curvature perturbation. From this perspective, inflation is a natural ``cosmological particle collider" that can probe energy scales as high as $H\simeq 10^{14}\, \text{Gev}$, beyond the reach of any conceivable terrestrial accelerator \cite{Chen:2009we, Chen:2009zp,Baumann:2011nk, Assassi:2012zq,Chen:2012ge, Pi:2012gf, Noumi:2012vr, Arkani-Hamed:2015bza, Lee:2016vti, Chen:2016uwp, Kehagias:2017cym,An:2017hlx, An:2017rwo,Kumar:2017ecc,Baumann:2017jvh, Bordin:2018pca, Kumar:2018jxz, Goon:2018fyu, Anninos:2019nib, Kim:2019wjo, Alexander:2019vtb, Hook:2019zxa, Kumar:2019ebj, Liu:2019fag, Wang:2019gbi,Wang:2019gok, Bodas:2020yho,Lu:2021gso,Sou:2021juh,Lu:2021wxu, Wang:2021qez,Tong:2021wai,Pinol:2021aun,Cui:2021iie,Reece:2022soh,Chen:2022vzh,Qin:2022lva}.
Much of the explorations in ``cosmological collider physics" have been restricted to situations in which the scalar fluctuations and additional matter fields propagate at equal speeds, namely the speed of light. This also includes the recent works on the ``cosmological bootstrap" allowing for exact analytical results and where de Sitter invariance plays a key role \cite{Maldacena:2011nz,Bzowski:2011ab,Bzowski:2012ih,Raju:2012zr, Raju:2012zs, Mata:2012bx, Bzowski:2013sza,Kundu:2014gxa,Kundu:2015xta,Ghosh:2014kba, Shukla:2016bnu,  Arkani-Hamed:2018kmz,Baumann:2019oyu,Sleight:2019mgd,Sleight:2019hfp, Baumann:2020dch,Sleight:2020obc,Sleight:2021iix, Sleight:2021plv,Fichet:2021xfn,Gomez:2021ujt,Heckelbacher:2022hbq, Armstrong:2022csc}.  In particular, de Sitter boost symmetry implies that all fields must be propagating at equal speeds. In such situations, the heavy degrees of freedom ($m\geq 3/2H$) affect the correlators in two qualitatively different ways: one is through a set of irrelevant EFT operators that emerge after integrating out the massive fields. The resulting momentum space correlators are characterised by their simple analytical structure, namely at tree-level they are rational functions of momenta with simple poles. The EFT signal is generically suppressed by inverse powers of the heavy state, namely with factors of $(H/m)^n$. However, the EFT expansion misses the intrinsically cosmological phenomenon of particle pair creation which induces effects that are non-perturbative in inverse powers of mass, for example through the famous Boltzmann factor $\exp(-\pi m/H)$. The resulting correlators exhibit non-analytic behaviours in momenta in the form of branch cuts. For the special case of the bispectrum, the EFT signal dominates the three-point function around the equilateral configuration, whereas the signal attributed to particle production, which manifests itself as oscillations in the ratio between the long and short mode momenta, dominates in the squeezed-limit (see Figure \ref{fig:tableintro}).\\ 

\begin{figure}
    \centering
    \includegraphics[scale=0.65]{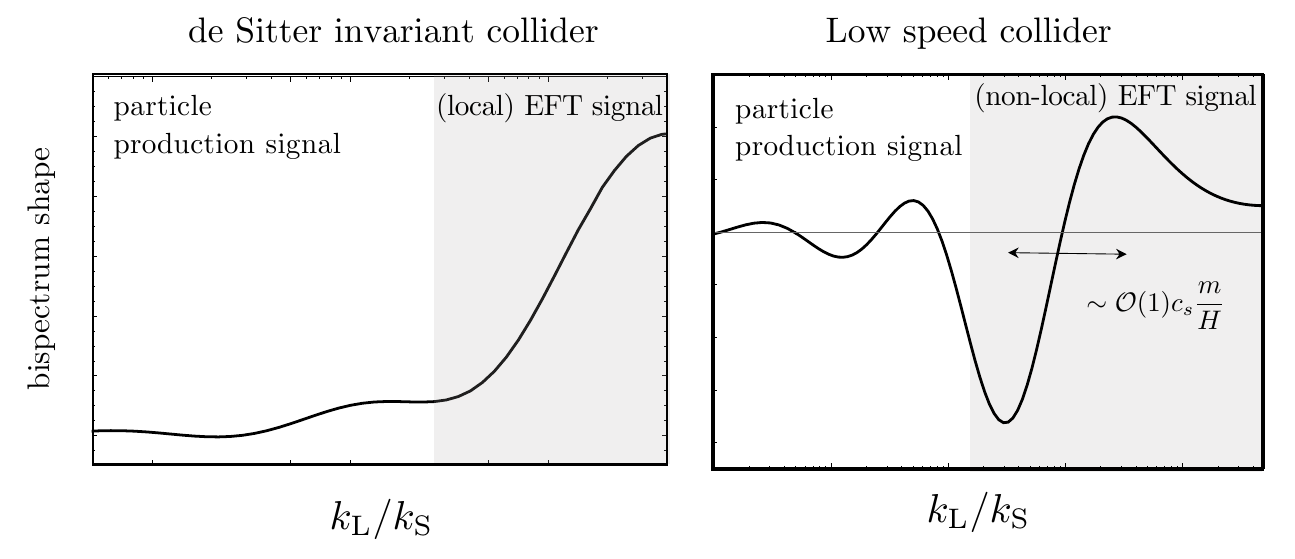}
    \caption{In de Sitter invariant and more generally equal speed setups (left schematic plot), heavy particles induce a non-Gaussian signal that around equilateral configurations (i.e. $k_\L/k_\S\sim 1$) can be captured by a local EFT description, while they leave characteristic oscillatory imprints in the squeezed limit of the bispectrum (i.e. $k_\L/k_\S\ll 1$). In the low speed collider signal (right plot), a supersonic particle lighter than $H/c_s$ manifests itself as a resonance in the extended equilateral configurations (defined by ${\cal O}(1)c_s \frac{m}{H}\lesssim k_\L/k_\S \leqslant 1$, while the associated particle production effect dominates the signal in the ultra-squeezed limit (i.e. $k_\L/k_\S\ll 2c_s$). Unless the mass is close to the Hubble scale, the resonance signal can be reproduced by adding an enough number of non-local EFT operators of the type discussed in Section \ref{non-localEFT}.}
    \label{fig:dSlowspeed}
\end{figure}

\noindent In this work, we are interested in scenarios where de Sitter boosts are strongly broken due to the sizeable coupling of the cosmological perturbations to the preferred time foliation during inflation. In most of such scenarios, the scalar fluctuations acquire a subluminal speed of propagation $c_s$, aka the speed of sound, hence explicitly breaking de Sitter boosts. We find that a reduced sound speed qualitatively changes the above picture, hence allowing for novel signatures of heavy states. In particular, we unveil that supersonic massive particles that are lighter than $H/c_s$ leave imprints as resonances in the squeezed limit of the bispectrum. We refer to this phenomenon as the \textit{low speed collider}, and we show that its signatures are vividly distinct from the usual EFT and particle production signals, both in their mass and kinematical dependencies. In more details, we point out that the overall size of the non-Gaussian signal attributed to the exchange of supersonic particles depends on $m$ only logarithmically as opposed to the case of $c_s=1$ where the bispectrum is dwarfed either by the power-law suppression in the equilateral configuration or by the Boltzmann exponential factor in the squeezed limit (see Figure \ref{fig:tableintro}). In this sense, the subluminal speed of propagation of the curvature perturbation makes its correlators more sensitive to the UV-physics, as particles that are much heavier than the Hubble scale (yet lighter than $H/c_s$) do not decouple. As for the shape of non-Gaussianity, we find that the bispectrum due to the exchange of such particles manifests peaky patterns in the squeezed limit, around $k_\L/k_\S\sim c_s\frac{m}{H}$, where $k_\L$ and $k_\S$ are the long and short modes.
The overall characteristics of the imprints left by massive fields in the bispectrum of the curvature perturbation, both for $c_s=1$ and a low speed of sound, are summarised in Figure \ref{fig:dSlowspeed}.\\

\noindent In fact, the main features of the squeezed limit bispectrum in the regime of $m< H/c_s$ can be understood in simple physical terms due to the existence of two characteristic times in the dynamics of the system. One is the time at which the short mode of $\zeta$ exits the sound horizon ($c_s k_\S/a(t_1)=H$) and freezes, prior to which it was oscillating like a massless field in flat space. The second instance is when the long mode's physical momentum of the exchanged field drops below its mass (i.e.~$k_\L/a(t_2)=m$), called mass-crossing in the following, after which the heavy field decays as $1/a^{3/2}(t)$ and before which it evolves like a massless field. For particles with $m\ll H/c_s$, $t_1$ can occur before $t_2$ such that between the two events the short mode of $\zeta$ interacts with the long mode of the massive field as if the latter was massless. This leads to an ``infrared''
growth of the three-point function in the interval $t_1<t<t_2$, which finally terminates as the massive field mode function starts to decay. By and large, when $t_1<t_2$, which is equivalent to $k_\L/k_\S\gtrsim c_s \frac{m}{H}$, non-Gaussianity takes a form very similar to the local shape. The opposite case with $t_2<t_1$ gives rise to the standard cosmological collider oscillations, manifesting themselves in what we call the ultra-squeezed limit such that $k_\L/k_\S\ll {\cal O}(1)c_s\frac{m}{H}$, and which encode the oscillations in $e^{i mt}$ of the massive field mode function with time after mass crossing. Furthermore, non-Gaussianity in this ultra-squeezed limit is suppressed by $(k_\L/k_\S)^{3/2}$ owing to the corresponding decay of the mode function. Eventually, one expects the signal to be maximal when the two characteristic times coincide (i.e. $t_1=t_2$), giving rise to resonances for $k_\L/k_\S\sim c_s \frac{m}{H}$, although the precise shape dependence of the signal in that region can only be found by the proper computation we do in this paper.\\

\noindent The standard approach to the computation of cosmological correlators is the \textit{in-in} formalism 
in which unpacking the unitary evolution of the Heisenberg operators of interest in perturbation theory results in a set of bulk time integral expressions for the correlators \cite{Weinberg:2005vy}. The lack of time translation symmetry in an expanding background often complicates the evaluation of such time integrals, even more so in the presence of massive fields. 
In fact, even equipped with de Sitter isometries, it was not until a few years ago that analytical expressions were provided for the simplest possible tree-level correlators involving massive fields (such as the four-point function of external massless fields mediated by heavy fields) \cite{Arkani-Hamed:2018kmz,Baumann:2019oyu}. More generally, the study of the cosmological collider physics has been largely limited to the squeezed limit tail of the bispectrum where the time integrals that describe the exchange of heavy fields factorises and can be easily computed. 
However, having analytical expressions for correlators has high theoretical and observational merits. Theoretically, it is interesting to study how a consistent time evolution that respects the celebrated physical principles of unitarity, locality and causality is encoded in the late time correlators, which are the fundamental observables in cosmology. This is not possible without having enough theoretical data on the structure of cosmological correlators, at least in perturbation theory. Furthermore, from an observational point of view, confronting the predictions of early universe models on non-Gaussianity with future data from CMB and LSS experiments requires templates that cover all kinematical configurations and as much theoretically motivated situations as possible (see e.g.~\cite{Achucarro:2022qrl}).\\

\noindent In recent years, inspired by the tremendous successes of on-shell methods in scattering amplitudes (for a pedagogical review see\cite{cheung2018tasi}),  a significant number of works have been devoted to a boundary viewpoint  on correlators. In this approach,  instead of following the dynamics of the system in time, the \textit{boundary correlators} are directly ``bootstrapped" by requesting consistency with  \textit{unitarity}, \textit{locality} and \textit{symmetries} (see the recent reviews \cite{Baumann:2022jpr,Benincasa:2022gtd} and also \cite{Hogervorst:2021uvp,DiPietro:2021sjt} for some efforts in the direction of non-perturbative bootstrap in cosmology).  The focus of the recent bootstrap literature has been the de Sitter isometric correlators, although similar methods have been invented and applied to boostless setups with massless fields and more general backgrounds  \cite{Arkani-Hamed:2017fdk,Arkani-Hamed:2018bjr,Benincasa:2018ssx,Benincasa:2019vqr,Benincasa:2020aoj,Pajer:2020wnj,Pajer:2020wxk,Benincasa:2020uph, Jazayeri:2021fvk,Baumann:2021fxj, Meltzer:2021zin,Bonifacio:2021azc,Cabass:2021fnw,Hillman:2021bnk,Bittermann:2022nfh}.  
In this study we extend the reach of the cosmological bootstrap program to boostless setups involving massive particles.  One of our core results is that, in a manner similar to \cite{Arkani-Hamed:2018kmz,Baumann:2019oyu} in de Sitter invariant setups, the boostless bispectra and trispectra of $\zeta$ due to the tree-level exchange of scalars with arbitrary masses and interactions can still be mapped onto the four-point function of a conformally coupled field induced by the same intermediate heavy field, through a set of bespoke ``weight-shifting" operators (see \cite{Pimentel:2022fsc} for an alternative approach).
In our case though, a key difference arises which is that in order to make a link between the dS invariant  four-point function of the conformally coupled field (which propagates at the speed of light) and the correlators
of $\zeta$ (which has a reduced speed of sound $c_s$) the former must be analytically continued beyond the physical domain allowed by momentum conservation. More specifically,  the ($s-$channel) four-point function of the conformally coupled field characterised by four spatial momenta $\bfk_a\, (a=1,\dots 4)$ can be expressed in terms of a function of the ratios $u=|\bfk_1+\bfk_2|/(k_1+k_2)$ and  $v=|\bfk_1+\bfk_2|/(k_3+k_4)$, both of which are smaller than unity due to the triangle inequality. Transforming to the bispectrum of external $\zeta$ fields with momenta $k_a\,(a=1,\dots,3)$ forces us to re-scale the external size of the four-point momenta by $c_s$ while leaving the intermediate momentum $|\bfk_1+\bfk_2|$ intact. This procedure is meaningful only if we think of the seed four-point correlator as a function of the ratios above, analytically continued beyond the respective unit disks (i.e. beyond $u\leq 1,v\leq 1$). Even with a known convergent series for the seed four-point inside the aforementioned unit disks, finding the analytical continuation outside is very challenging. In this paper, we bootstrap this seed four-point function directly in the region of interest by leveraging locality, unitarity and analyticity. In more detail, locality will be manifested as a set of \textit{boundary partial differential equations} that the seed four-point function must satisfy.\footnote{
Here by locality we mean the properties that the boundary correlators inherit from the local equations of motion of the bulk fields, in particular the exchanged massive field. See also \cite{Jazayeri:2021fvk} for locality constraints on the wavefunction coefficients of massless fields.}
The unitarity of the time evolution, encoded in an infinite set of algebraic equations for the wavefunction coefficients which are called \textit{cosmological cutting rules} \cite{COT,Meltzer:2020qbr,Cespedes:2020xqq, Goodhew:2021oqg,Melville:2021lst}, will be employed in this work in order to partially fix the homogeneous solution that can be added to the boundary PDE's we alluded to above. The remaining freedom in adding further homogeneous solutions will be removed by asking regularity of the four-point function in the collinear limit.\\  
\\
\noindent The non-perturbative effect of spontaneous particle production
cannot be described without taking into account the genuine dynamics of the heavy field. However, all other key features of the correlators, including the hallmark resonances and the logarithmic dependence on $c_s m/H$, can be reproduced and understood in a simplified single-field picture in which the heavy field is integrated out, albeit in a non-standard manner owing to the fact that it is relativistic at sound horizon crossing, resulting in a non-local (in space) effective theory for $\pi$. Such kind of descriptions have already been argued to provide (partial) UV-completions of the simplest setup of the EFT of inflation \cite{Gwyn:2012mw}, albeit without working out its predictions as we do here. The origin of this non-local EFT can be simply understood: because of the relative slow motion of $\pi$ with respect to the one of $\sigma$, one can approximately consider that the latter instantaneously responds to the dynamics of the former. 
By contrast to the standard integrating out procedure leading to a local action, valid only for $m \gg H/c_s$ and in which the kinetic and gradient terms of $\sigma$ are neglected compared to its mass term, the non-local EFT stems from solely neglecting the kinetic term of $\sigma$ compared to its gradient and mass term, keeping the latter two on equal footing.
The resulting non-local EFT can be used at arbitrary order in this expansion, with corrections to the leading-order result organized in positive powers of temporal derivatives and giving rise to an infinite set of non-local operators. These would-be corrections can be used to check the sanity of the EFT: while they lead to convergent results in the domain of validity of the EFT, they diverge around the resonances for $m$ close to the threshold value $3/2H$, a situation for which the EFT breaks down and only our exact bootstrap results are applicable. Remarkably, the (leading-order) non-local EFT provides one with simple templates for the bispectra: one-parameter families of shapes that depend on $\a =c_s m/H$, that generalise well known ones from the EFT of inflation recovered in the large $\alpha$ limit, while describing the physics of the 
low speed collider and the associated resonances for small $\a$, see Equations \eqref{nonlocalEFTs}.\\

\noindent \textbf{Roadmap and summary of the results}: \\
\begin{itemize}

\item In Section \ref{sec:setup-overview}, we introduce our setup of interest: the Goldstone boson of broken time translation during inflation $\pi$, propagating at the speed of sound $c_s$, coupled to a heavy scalar field $\sigma$ propagating at the speed of light. We concentrate on interactions between the two governed by the ubiquitous lowest-order quadratic coupling $\dot{\pi}\sigma$ as well as the cubic ones $\dot{\pi}^2\sigma$ and $(\partial_i \pi)^2\sigma$, with our results that can be straightforwardly generalised to other interactions with more derivatives.
We explain our motivations for focusing on the qualitatively new regime $m < H/c_s$ with  $c_s \ll 1$,\footnote{Up to a rescaling of coordinates, our results are more generally valid for any two sound speeds, with $c_s$ then corresponding to their ratio.} we give a qualitative overview of the most salient features expected in that regime on simple physical grounds, and we comment on possible UV-completions of our effective field theory.
 
\item In section \ref{WS-bootstrap}, we use diagrammatic rules to show how the building blocks of the correlators of interest can be related to the building block entering the seed four-point single-exchange diagram $\hat{F}$ of a conformally coupled field $\vpi$ interacting with $\sigma$ through the coupling $g \vpi^2 \sigma$. 
Gluing these building blocks together, it is then easy to deduce the (correction to the) power spectrum, the bispectrum and the trispectrum of $\pi$ (equivalently of the curvature perturbation $\zeta$), mediated by the exchange of the massive scalar field, from the seed correlator $\hat{F}$, upon considering a suitable soft limit and acting via a set of \textit{weight-shifting} operators. This schematically reads, for the bispectrum:
\begin{align}
    B_\zeta(k_1,k_2,k_3)={\cal W}\left(k_i,\dfrac{\partial}{\partial k_i}\right)\, \hat{F}(c_s k_1,c_s k_2, c_s k_3,k_4\to 0; k_3)\,,  
\end{align}
where, on the RHS, ${\cal W}$ depends on the operators that act at each vertex, and the rescaling of the external ``energies'' by the speed of sound $c_s < 1$  necessitates to deal with the analytical continuation of the four-point function $\hat{F}$ outside its domain of physical configurations. In this section, we also describe the bootstrap tools used in the following to explicitly compute that seed correlator.

\item Section \ref{seedfourpoint} is dedicated to the determination of $\hat{F}$. The latter (s-channel contribution) depends on the two variables $u=s/(k_1+k_2), v=s/(k_3+k_4)$ with $s=|\bfk_1+\bfk_2|$ the momentum of the exchanged $\sigma$ field. In the allowed kinematical domain, $u$ and $v$ are less than one, and it was sufficient in Ref.~\cite{Arkani-Hamed:2018kmz} to work out $\hat{F}$ inside the corresponding unit circle(s), which was done in terms of a double series expansion in $u$ and $u/v$ (assuming $u<v$). However, that expansion is not convergent outside the unit disk, whereas our weight-shifting operators require the evaluation of $\hat{F}$ at
 \begin{align}
        u\to \dfrac{k_3}{c_s(k_1+k_2)}\,,\qquad v\to \dfrac{1}{c_s}\,.
    \end{align}    
We hence set out to determine $\hat{F}$ from first principles in terms of a new, convenient and rapidly convergent series representation around $u,v=\infty$. We use two bootstrap tools to achieve this goal. We leverage locality through a \textit{boundary partial differential equation} that $\hat{F}(u,v)$ satisfies. Then we solve this equation as a series expansion within the strip of $1<|u|<|v|$, i.e. 
     \begin{align}
         \hat{F}\supset \sum_{m,n=0}^{\infty}\left( a_{m n}+b_{m n}\,\log(u) \right)u^{-m}\,\left(\frac{u}{v}\right)^n\,,\qquad 1<|u|<|v|\,,
     \end{align}
where the unusual logarithmic term is forced upon us by the structure of the boundary equation. We further exploit unitarity in the form of a \textit{cosmological cutting rule} for the four-point function in order to fix the remaining freedom in adding a homogeneous solution of the boundary equation. The final result $\hat{F}(u,v)$, which is the seed to all the correlators of interest, has several characteristic features, mother of all the specificities of the bispectrum: (i) a bump around $u\sim m/H$ as long as $m/H \ll v$, (ii) oscillations for $u<1$ that encode the standard pair creation effect, (iii) Eventually, for $(u,v)\gg m/H$, $\hat{F}$ takes the following simple form
\begin{align}
         \hat{F}&=g^2\left(\dfrac{1}{u}+\dfrac{1}{v}\right)\left(\log(\frac{1}{C(\mu)}\dfrac{uv}{u+v})+1-\gamma_E\right)\,,\quad (u,v)\gg \frac{m}{H}\,,
\end{align}
exhibiting the logarithmic running with mass that we mentioned before, with $C(\mu) \approx m/H$.

     \item In section \ref{sec:correlators}, we extensively study the resulting correlators of $\pi$ upon acting with the weight-shifting operators. We first discuss the power spectrum and then move on to the bispectra generated by the two cubic interactions, whose shapes we characterise as a function of $c_s$ and $m/H$, focusing in turn, for the new regime of interest, on the ``generic'' triangular configurations $k_\L/k_\S\gg c_s m/H$, on the oscillations in the ultra-squeezed limit, and on the resonances occurring for $k_\L/k_\S \sim c_s m/H$. We also discuss the amplitude of the signal and the constraints set by perturbativity, finding that the resonances can be observably large $\fNL \sim (\rho/H)^2  (c_s m/H)^{-1}  \gg 1$, where $\rho/H \lesssim {\cal O}(m)$ is the amplitude of the quadratic coupling.

      \item The section \ref{non-localEFT} deals with the non-local EFT. We first discuss its regime of validity and work it out at the level of the seed theory of $\sigma$ coupled to the conformally coupled field. We then compute the corresponding four-point correlator $\hat{F}$ in this EFT, which offers a particularly transparent and clear picture of the physics of the low-speed collider, and we also point out its intrinsic limitations. Eventually we show how the non-local EFT enables one to derive simple one-parameter families of shape templates (presented in Equations \eqref{nonlocalEFTs}) that encapsulate the rich physics described in this paper when varying the speed of sound and the mass of the exchanged field.
   
\end{itemize}

\noindent

\subsection*{Notations}
We adopt the following definition for the Fourier transformed fields:
\begin{align}
    f(\bfx)=\int\,\dfrac{d^3 \bfk}{(2\pi)^3}\,f(\bfk)\,\exp(i\bfk.\bfx)\,,\quad f(\bfk)=\int d^3\bfx\,f(\bfx)\,\exp(-i\bfk.\bfx)\,.
\end{align}
The dS space will be charted by the following coordinates: 
\begin{align}
    ds^2=a^2(\e) \left(-d\e^2+d\bfx^2 \right)\,,\quad a(\e)=-\dfrac{1}{\e\,H}\,,
\end{align}
where $\e$ is the conformal time. We will denote the comic time by $t$.  
We denote the Goldstone boson of broken time translation during inflation by $\P$ and its canonically normalized field by $\pi_c$. 
The speed of propagation of $\pi$ will be indicated by $c_s$. The field $\vpi$ will refer to the conformally coupled field in dS (with $m_\vpi^2=2H^2$). 
Derivation with respect to the conformal time $\eta$ will be indicated by a prime. Four-point exchange diagram of a field with external momenta $\bfk_1,\dots\bfk_4$ will be characterised by the \textit{energy} variables
\footnote{Energy is not a conserved quantity in a time dependent background. Nevertheless, the terminology is useful in cosmology because, in the $\e\to -\infty$ limit, time translation is restored and the dispersion relation have the form $E\propto c_s|\bfk|/a(\e)$. Moreover, the total energy pole of the correlators contains the flat space amplitude in which $c_s|\bfk|$ plays the role of the energies of the particles that participate in the scattering process.}
\begin{align}
    s\equiv |\bfk_1+\bfk_2|\,,\quad t\equiv |\bfk_1+\bfk_3|\,.
\end{align}
and the external \textit{energies} $k_a\equiv |\bfk_a|\,,a=1,\dots 4$ (notice that $|\bfk_1+\bfk_4|$ is not an independent variable because of the conservation of momentum). We refer to the bispectrum of $\zeta$ by $B_\zeta(k_1,k_2,k_3)$ and to the four-point exchange diagram of $\vpi$ by $F(k_1,\dots, k_4,s,t,u)$, i.e. 
\begin{align}
    \left\langle \zeta(\bfk_1)\zeta(\bfk_2)\zeta(\bfk_3)\right\rangle &=B_\zeta(k_1,k_2,k_3)(2\pi)^3\,\delta^3(\sum_{a=1}^4\,\bfk_a)\,,\\ \nonumber
    \left\langle \vpi(\bfk_1)\dots \vpi(\bfk_4)\right\rangle &=F(k_a,s,t,u)\,(2\pi)^3\,\delta^3(\sum_{a=1}^4\,\bfk_a)\,.
\end{align}
The symbol $\langle \dots \rangle'$ will indicate a correlator with the factor $(2\pi)^3\delta^3(\bfk_1+\dots)$ stripped off. 
We use natural units throughout and define the Planck mass as $\Mpl^2=1/(8\pi G_N)$.

\section{Physical setup, motivations and overview}
\label{sec:setup-overview}

\subsection{Action and motivations}
\label{action-motivations}
In this paper, we use the model-independent language of the effective field theory of inflationary fluctuations \cite{Creminelli:2006xe,Cheung:2007st,Senatore:2010wk,Noumi:2012vr} to study the interactions between
the curvature perturbation $\zeta$ and a heavy scalar degree of freedom with a mass not far from the Hubble scale. We work in the decoupling limit in which the gravitationally induced interactions are ignored, we take the background spacetime as rigid de Sitter space, and we neglect subdominant deviations from scale-invariance. In this setup, one can consider that $\zeta=-H \pi$ where $\pi$ is the Goldstone boson that non-linearly realises the broken time diffeomorphism during inflation, and that $\pi$ enjoys a shift symmetry.
Up to first order in derivatives and cubic order in the field, the Lagrangian for the $\pi$ sector takes the standard form:
\begin{align}
    S_\pi =\int d\e\,d^3 \bfx  \,a^2  \epsilon H^2 \Mpl^2\,&\left[\frac{1}{c_s^2}\left(\pi'^2-c_s^2 (\partial_i \pi)^2\right) -\dfrac{1}{a}\left(\dfrac{1}{c_s^2}-1\right) \left(\pi'(\partial_i \pi)^2+\frac{A}{c_s^2} \pi'^3\right)+\dots\right]\,,
    \label{S-EFT}
\end{align}
where $\epsilon=-\dot{H}/H^2$, $c_s$ is the sound speed of $\pi$ and $A$ is a Wilson coefficient naturally of order one. The derivative self-interactions in $\pi'(\partial_i \pi)^2$ and $\pi'^3$ give rise to well known shapes of the bispectrum maximum near equilateral configurations (that can be approximated by the so-called equilateral and orthogonal templates \cite{Creminelli:2005hu,Senatore:2009gt}), and with the characteristic amplitude $\fNL \sim 1/c_s^2-1$. 
In addition, we consider the interaction of $\pi$ with an additional scalar degree of freedom $\sigma$ with a generic mass $m$ and with the following free action:
\begin{align}
    S^{(2)}_\sigma&=\int d\e d^3\bfx\, a^2 \left(\dfrac{1}{2}\sigma'^2-\dfrac{c_\sigma^2}{2}(\partial_i \sigma)^2-\dfrac{1}{2}m^2 a^2 \sigma^2\right)\,.
    \label{S2-sigma}
\end{align}
In general $\sigma$ can have a non-trivial speed of sound $c_\sigma$. However, one can always redefine spatial coordinates that absorb this and consider that $c_\sigma=1$, at the expanse of generating additional $c_\sigma$ dependence in other parts of the action. While this can be done straightforwardly and would not change the applicability of our analysis (concerning the $\pi-\sigma$ interactions of interest in this paper, this would correspond to having $c_s$ being the ratio between the propagation speeds of $\pi$ and of $\sigma$), this would clutter the equations, so we stick to $c_\sigma=1$ in the following. The dominant interactions between $\pi$ and $\sigma$ at low energies were classified in  \cite{Noumi:2012vr} (see e.g.~eqs 62-65 there) and here, we focus on the following operators giving rise to single-exchange contributions to the bispectrum:
\begin{equation}
\label{interpisigma}
S_{\pi\sigma}=\int d\eta d^3\bfx\,a^2\,\left(\rho a\pi'_c\sigma+\dfrac{1}{\Lambda_1}\pi'^2_c\sigma+\dfrac{c_s^2}{\Lambda_2}(\partial_i \pi_c)^2\sigma \right)\,,
\end{equation}
where, for future convenience, we have introduced the canonically normalized field $\pi_c=\frac{\sqrt{2 \epsilon} H \Mpl}{c_s}\pi$.
It is important to notice that the scale $\Lambda_2$ is not arbitrary. Indeed, it only emerges from the unitary gauge operator $\propto \rho \delta g^{00} \sigma$, where $\delta g^{00} \rightarrow -2 \dot{\pi} - \dot{\pi}^2 + \frac{(\partial_i \pi)^2}{a^2} $ upon reintroducing $\pi$. Because of this, one has
\begin{align}
    \frac{1}{\Lambda_2}=-\dfrac{1}{2}\dfrac{1}{\sqrt{2\epsilon} c_s \Mpl}\dfrac{\rho}{H}\,.
    \label{eq:Lambda-rho-link}
\end{align}
By contrast, $\Lambda_1$ is not fixed by the non-linearly realised time-diffeomorphism invariance, as the corresponding interaction is not only generated by $\delta g^{00}\sigma$, but also by $(\delta g^{00})^2 \sigma$. The scale $\rho$ can be a priori arbitrary, but we will restrict to situations in which the corresponding quadratic interaction in $\pi'_c\sigma$ can be treated perturbatively, namely $\rho \lesssim m$ (see section \ref{sec:size-NG} for a quantitative discussion).\\

\noindent In this paper, we are interested in the imprints left by heavy fields on inflationary correlators. Following the EFT logic, when sufficiently heavy (a point we will elaborate on below), the $\sigma$ field can be integrated out in a standard way, i.e. one can replace $\sigma$ in the action by the low-energy solution to its equation of motion: neglecting the kinetic and gradient terms of $\sigma$, one finds $\sigma \approx \rho \pi'_c/(a m^2)$, upon which replacement the total action takes the form \eqref{S-EFT} with a redefined Wilson coefficient $\tilde{A}$ and a new speed of sound such that\footnote{Naturally, the coefficient of the $\pi'(\partial_i \pi)^2$  interaction being tied to the new speed of sound is verified only with the specific value of $\Lambda_2$ in \eqref{eq:Lambda-rho-link}, as both are consequences of the non-linearly realised time-diffeomorphism invariance.}
\beq
\frac{1}{\tilde{c}_s^2}=\frac{1}{c_s^2}\left(1+\frac{\rho^2}{m^2} \right)\,.
\label{new-speed-sound}
\eeq
However, while this single-field EFT correctly reproduces the bispectrum for generic triangular configurations, it fails to capture the non-perturbative effects of spontaneous particle production, notably giving rise to oscillations in the squeezed limit. Moreover, this description is accurate only if the kinetic and gradient terms of $\sigma$ are negligible compared to its mass term in the action \eqref{S2-sigma}, around the relevant time for the dynamics of $\pi$, i.e. around sound horizon crossing (considering that $\rho \lesssim m$ and hence $\tilde{c}_s \approx c_s$). This is valid only for $m \gg H/c_s$. When the sound speed is small, this leaves an interesting parameter space $H \lesssim m \lesssim H/c_s$ in which the standard local EFT \eqref{S-EFT} (with redefined parameters) fails to reproduce the impact of massive fields even in equilateral configurations. However, as we will see in section \ref{non-localEFT}, one can still integrate out $\sigma$ in this regime (if its mass is not too close to the Hubble scale), albeit in a non-standard way that results in a single-field effective description that is non-local in space. In both cases though, the single-field description misses again the cosmological collider oscillations characteristic of a heavy field of mass $m\geq 3/2 H$. In the rest of this paper, we will consider a heavy field, leaving the study of lighter fields for future work. Our exact bootstrap analysis will be valid for any such mass and any sound speed $c_s$. However, equipped with our analytical results, and given that the situation with $m \gg H/c_s$ resembles the well understood one with $c_s=1$ for generic kinematical configurations (albeit with appreciable differences in the squeezed limit), we will mostly focus on the theoretical understanding and phenomenological implications of the opposite regime of parameter space with $m \ll H/c_s$, unique to a low speed of sound.\\

\noindent An additional motivation for concentrating on this regime comes from the following important consideration: when the sound speed is low, the cutoff the EFT action \eqref{S-EFT} becomes close to the Hubble scale, so that too heavy fields can not be self-consistently included in the description.\footnote{We thank Luca Santoni for discussions about this point.} More quantitatively, the cut-off energy scale of the EFT of inflation is given by (see e.g. \cite{Cheung:2007st,Baumann:2011su}) 
\begin{align}
    \Lambda=\dfrac{1}{ (2 \pi{\cal P}_\zeta)^{1/4}}\frac{c_s}{(1-c_s^2)^{1/4}} H\,,
\label{eq:cutoff} 
\end{align}
implying that the massive field $\sigma$ can be described in the EFT only if
\begin{align}
    m<\Lambda\simeq 100 \, \frac{c_s}{(1-c_s^2)^{1/4}} H \,.
    \label{bound-m}
\end{align}
Hence, one finds that for $c_s \leq 0.1$, i.e. for the bulk of the low sound speed parameter space, a self-consistent description requires that $m <H/c_s$: heavier fields exceed the cutoff scale, and should have been integrated out in the first place.
Let us now add some cautionary words: 
the Planck constraints on non-Gaussianity give a lower bound on the speed of sound (assuming a pure $c_s$-theory): $c_s \geq 0.021 (95 \% \textrm{CL})$ \cite{Planck:2019kim} (not far from the value at which the cutoff \eqref{eq:cutoff} approaches $H$ and the theory becomes useless). Hence, in the following, when we take the limit $c_s \to 0$ in some analytical formulae, this should be taken as a formal limit. In practice, one can check that such formulae are very accurate as soon as $c_s \lesssim 0.1$, and are therefore fully applicable for theories that are indeed observationally relevant. When it will come to numerical examples, our benchmark situation will be $c_s=0.1$, but we also consider $c_s=0.01$. In that case, the bound \eqref{bound-m} leaves barely room for a heavy field, of mass $m\geq 3/2 H$, to be coupled to the pure $c_s$-EFT of inflation, and some of our plots should then be taken for mere illustrative purposes of relevant trends.

\subsection{Qualitative picture}
\label{Qualitative}
Before exploring it in detail in this paper, one can understand in simple terms why the regime $m <H/c_s$ is interesting, both theoretically and phenomenologically, and anticipate on its most salient features. For a given $k$-mode, there exists two relevant times for the dynamics:
\begin{itemize}
    \item[$\star$]
Event (1): sound horizon crossing for $\pi$, such that $k/a= H/c_s$, and at which the uncoupled $\pi$ freezes;

 \item[$\star$]
Event (2): ``mass crossing'' for $\sigma$, such that $k/a=m$, before which the uncoupled $\sigma$ behaves as a quantum massless field in its vacuum, and after which it decays and oscillates.
\end{itemize}

\noindent For $m> H/c_s$, event (2) occurs before event (1), whereas the opposite is true for $m< H/c_s$. In that situation, there exists a window of time during which $\pi$, already outside its sound horizon, quantum mechanically interacts with the $\sigma$ field still following the Bunch-Davies behaviour.
This unusual situation leads to a growth of the power spectrum of $\pi$ during that interval of $-\log(\frac{m}{H} c_s)$ \textit{e}-folds, a growth that is stopped after event (2) and the decay of $\sigma$. This IR ``divergence'', regulated by the mass of $\sigma$, will show up as an unusual logarithmic dependence of the $\pi$ correlators on the combination $\frac{m}{H} c_s$. The exact parameter dependence can not be found without the full computation that we make in this paper, but this intuitive picture does capture the correct physics. \\

\begin{figure}
    \centering
    \includegraphics[scale=0.40]{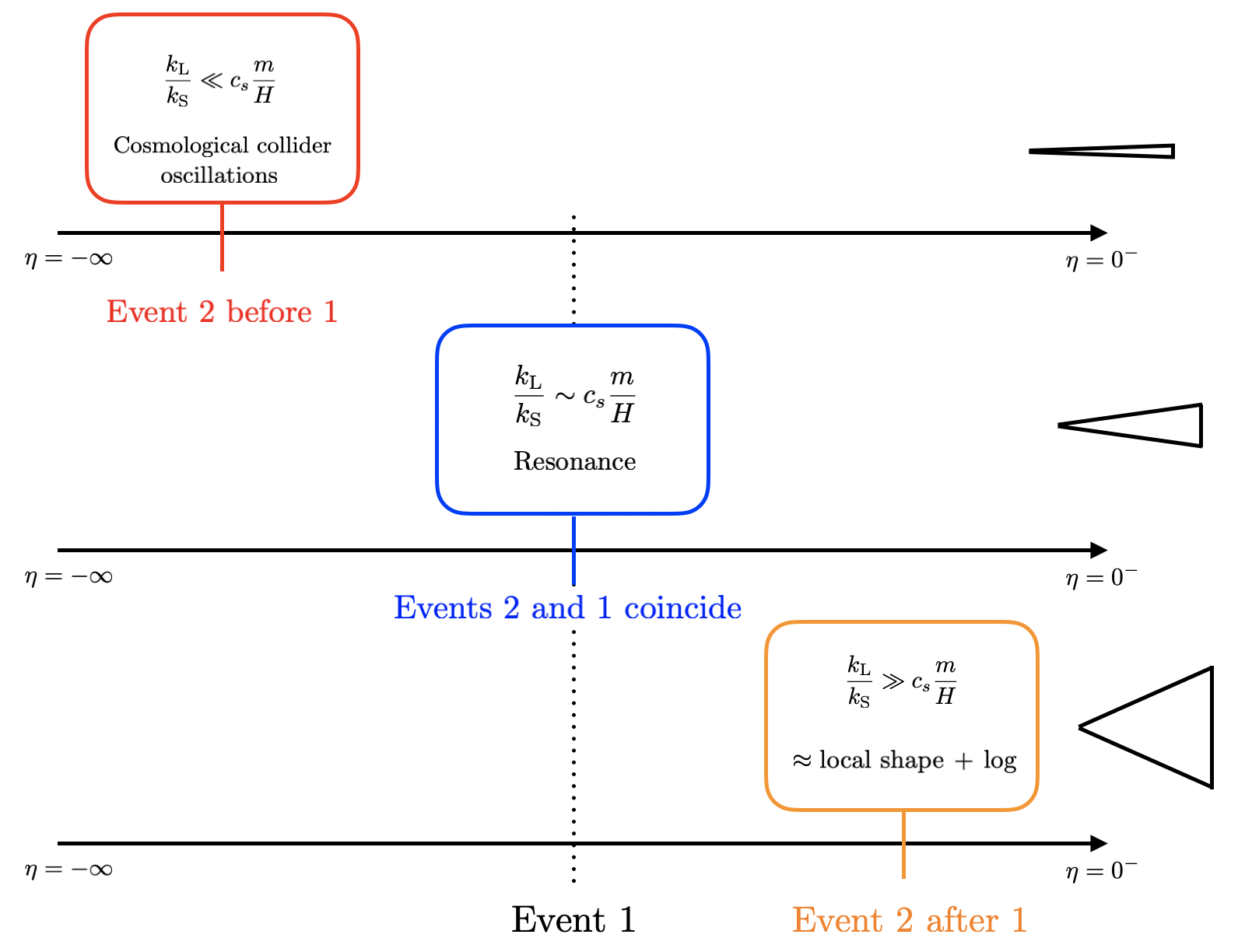}
    \caption{Schematic representation of the different kinematical regimes of the bispectrum. Event 1 is the sound horizon crossing of the short mode $k_\textrm{S}$, and event 2 is the mass crossing of the long mode $k_\textrm{L}$.}
    \label{fig:overview-bispectrum}
\end{figure}

\noindent This comparison of the relevant timescales is also useful to understand the different regimes of the bispectrum depending on how squeezed the corresponding triangle is. Let us consider for definiteness an isosceles triangle with $k_3=k_\textrm{L}< k_1= k_2 =k_\textrm{S}$. As our results will confirm, the relevant timescales to compare are now the ones of sound horizon crossing of the short mode $k_\textrm{S}$, and of mass crossing of the long mode $k_\textrm{L}$, still called events (1) and (2) for simplicity (see figure \ref{fig:overview-bispectrum}). In the usual situation with $c_s=1$, event (2) always occurs before event (1). This results in the squeezed limit of the bispectrum probing the super-Hubble oscillations of the massive field, manifesting as the cosmological collider oscillations. In contrast, for $c_s m/H \ll 1$, event (1) can occur before event (2), even for some squeezed triangles, resulting in three qualitatively different regimes for the bispectrum.
\begin{itemize}
\item The usual regime of the cosmological collider oscillations, with (2) before (1), now becomes pushed to what one may call ultra-squeezed configurations with $k_\textrm{L}/k_\textrm{S} \ll c_s m/H$ (top situation in figure \ref{fig:overview-bispectrum}).\footnote{We will give more refined estimates in section \ref{sec:CC-oscillations} as for when the cosmological collider oscillations actually dominate the signal, see Eq.~\eqref{eq:refined-estimates-cc}} 
\item Instead, for $k_\textrm{L}/k_\textrm{S} \gg c_s m/H$, (1) occurs before (2) (bottom situation in figure \ref{fig:overview-bispectrum}), resulting in a completely different signal, bearing resemblances with the local shape, albeit with the IR divergence described above also showing up as a logarithmic dependence in the number of \textit{e}-folds $-\log(\frac{k_\textrm{S}}{k_\textrm{L}}\frac{m}{H} c_s)$ between (1) and (2). 
\item Eventually, when the two characteristic times coincide (middle situation in \ref{fig:overview-bispectrum}), for $k_\textrm{L}/k_\textrm{S} \sim c_s m/H$, the shape of the bispectrum exhibits ``bump''-like features that we will call resonances (with details depending on the cubic interactions), characteristic of the low-speed collider.
\end{itemize}
Note that the argument for the presence of resonances for $k_\textrm{L}/k_\textrm{S} \sim c_s m/H$ is analogous to the one explaining that standard EFT shapes peak in equilateral configurations: in that case, the only characteristic time is sound horizon crossing for $\pi$, and the shape of the bispectrum is maximised in kinematical configurations for which the three characteristic times of the momenta coincide, i.e. in the equilateral limit. Indeed, if one of the mode is still inside the sound horizon, its rapid oscillations average out and leave a small signal, whereas the derivative interactions become inefficient outside the sound horizon. In our two-field situation, in addition to sound horizon crossing for $\pi$, another characteristic time enters the problem as we have explained, mass crossing for $\sigma$, also delineating the regimes before which it rapidly oscillates, and after which it decays. For the same reason as above, one thus expects the signal to be maximised for triangular configurations at which these characteristic times coincide.
These qualitative arguments will be explicitly confirmed quantitatively in what follows, notably in section \ref{4pt-from-EFT} within the single-field non-local EFT, but this simple physical picture guarantees
in a model-independent manner the robustness of the existence of ``resonances'' in squeezed configurations when $\pi$ interacts with a supersonic heavy field, for instance when considering other interactions leading to more complicated diagrams.
These resonances are expected to gradually disappear as the mass of the exchanged field diminishes, with the shape eventually becoming close to the local shape for a massless field. However, note that even in that case, resonances were already noticed in \cite{RenauxPetel:2011uk}, albeit in an approximate and much simpler computation, for the ``quantum'' contribution to the bispectrum (see e.g. section 4.3 and fig.~8 in \cite{RenauxPetel:2011uk}), with the characteristic time of the exchanged field being simply Hubble crossing, hence resulting in a resonance for $k_\textrm{L}/k_\textrm{S} \sim c_s$ for the reasons described above.

\subsection{Comments on UV completions}

It is perfectly legitimate to consider our starting point action \eqref{S-EFT}-\eqref{interpisigma} as our theory of interest, and compute observables within its framework, which is what we will do in the rest of this paper. But just like our theory can be (approximately) described by the local EFT action with redefined parameters for $m \gg H/c_s$, it is interesting to see how our setup itself may emerge as a low-energy effective description of some more fundamental theory. One such possible UV completion can be found by considering a 3-field model with $\pi$ propagating at the speed of light coupled to two interacting massive fields, also propagating at unit speed, and with quadratic Lagrangian
\begin{equation}
{\cal L}/a^3 = -\frac12(\partial_\mu \pi_c)^2+\tilde{\rho}\, \dot \pi_c \F_1 -\frac12(\partial_\mu \F_1)^2-\frac12(\partial_\mu \F_2)^2 -\frac12 M_1^2 \F_1^2 -\frac12 M_2^2 \F_2^2 - M_{12}^2 \F_1 \F_2\,.
\label{3field-beginning}
\end{equation}
Such a Lagrangian is commonplace in explicit realizations of inflation. For instance, it describes fluctuations in nonlinear sigma models, where $\tilde{\rho}$ is related to the deviation of the background trajectory from a geodesic in field space, i.e. it describes turns in a multifield landscape, and the $\F_1$ field directly coupled to $\pi_c$ corresponds to the fluctuation in the direction of the acceleration of the background fields orthogonal to the instantaneous velocity (see e.g.~\cite{Gordon:2000hv,GrootNibbelink:2000vx,GrootNibbelink:2001qt,Langlois:2008mn,Achucarro:2010da}).\footnote{Other (derivative) interactions between the entropic fields $\F_{1,2}$ are also present in general in these models, see e.g.~\cite{Pinol:2020kvw}.} Such an origin is not at all needed for our discussion though, and the action \eqref{3field-beginning} can be considered on its own at the level of the EFT of fluctuations only.\footnote{On a different note, the reduced sound speed of the curvature perturbation in our setup needs not emerge as an effective description at low energy, but it can be a ``fundamental'' property of the inflationary scenario formulated at the level of the full fields driving inflation, like in single- and multi-field DBI inflation \cite{Alishahiha:2004eh,Langlois:2008wt,Langlois:2008qf}. More generally, the existence of different sound speeds is a generic property of multifield scenarios with higher derivative terms, and some of its consequences have been studied in various works \cite{Langlois:2008mn,Gao:2008dt,RenauxPetel:2008gi,Langlois:2008wt,Langlois:2008qf,Arroja:2008yy,Langlois2009,RenauxPetel:2009sj,Mizuno:2009cv,Mizuno:2009mv,Gao:2009gd,Cai:2009hw,Gao:2009at,RenauxPetel:2011dv,RenauxPetel:2011uk}, although under the assumption that the couplings between fields propagating at different speeds is negligible around the times of sound horizon crossings, i.e. in a very simplified context that does not take into account the crucial aspects studied in this work.} In a two-field setup, i.e. with $\F_2=0$, the Lagrangian \eqref{3field-beginning} provides one with a typical UV completion of the EFT of inflation upon integrating out the massive field $\F_1$ (see e.g.~\cite{Tolley:2009fg,Cremonini:2010ua,Achucarro:2010da,Baumann:2011su} for early works and \cite{Garcia-Saenz:2019njm,Pinol:2020kvw} for recent applications).
The same logic follows here, but the interaction between the entropic fields $\F_{1,2}$, through the off-diagonal mass term in \eqref{3field-beginning}, plays an important role. To better understand this, it is useful to introduce the mass eigenstates $\sigma_{1,2}$ in terms of which the action \eqref{3field-beginning} reads
\begin{align}
{\cal L}/a^3 &= -\frac12(\partial_\mu \pi_c)^2+ \tilde{\rho}\,\dot \pi_c (\cos(\theta) \sigma_1+\sin(\theta) \sigma_2) -\frac12(\partial_\mu \sigma_1)^2-\frac12(\partial_\mu \sigma_2)^2 -\frac12 m_1^2 \sigma_1^2 -\frac12 m^2 \sigma_2^2 \nonumber \\
&- \frac{\tilde{\rho}}{2 \sqrt{2 \epsilon} H \Mp} \frac{(\partial_i \pi_c)^2}{a^2} (\cos(\theta) \sigma_1+\sin(\theta) \sigma_2)
\end{align}
with $\theta$ the angle of the rotation matrix between the ``flavor'' ($\F_{1,2}$) and the mass ($\sigma_{1,2}$) basis, such that the ``portal'' field $\F_1$ equals $\cos(\theta) \sigma_1+\sin(\theta) \sigma_2$, where one uses the terminology introduced in Ref.~\cite{Pinol:2021aun}, and one can choose $m_1$ larger than $m_2$. Here we have reinstored in the second line the unavoidable cubic terms in the Lagrangian that are fixed by the non-linearly realised time-diffeomorphism invariance, keeping in mind that the quadratic interaction in $\tilde{\rho} \dot{\pi}_c \F_1$ comes from a term $\propto \tilde{\rho} \delta g^{00} \F_1$ in the unitary gauge, see the discussion in section \ref{action-motivations}. Let us now consider a situation with a hierarchy $m_1 \gg  m \geq 3/2 H$ such that one can integrate out the heaviest mass eigenstate $\sigma_1$ while keeping $\sigma_2$ in the low-energy description. Paralleling the discussion in section \ref{action-motivations} and upon the replacement $\sigma_1 \to \tilde{\rho} \cos(\theta)/m_1^2\,\dot \pi_c$, this leads to our starting point action \eqref{S-EFT}-\eqref{interpisigma} (upon the redefinition $\pi_c \to c_s \pi_c$ so that $\pi_c=\frac{\sqrt{2 \epsilon} H \Mpl}{c_s}\pi$ still holds, and with the identification $\sigma_2=\sigma$), with parameters
\begin{equation}
\label{uv-parameters}
\frac{1}{c_s^2}=1+\frac{\tilde{\rho}^2 \cos^2(\theta)}{m_1^2}    \quad \textrm{and} \quad \rho=\tilde{\rho}\, c_s \sin(\theta)\,. 
\end{equation}
Note that a small sound speed requires $\tilde{\rho}^2 \gg m_1^2$ and that the applicability of the resulting EFT necessitates $m_1^2 \gg H^2/c_s^2$ \cite{Cremonini:2010ua,Baumann:2011su}, i.e. $m_1^2/H^2 \gg \tilde{\rho}^2/m_1^2 \cos^2(\theta) \gg 1$. In this UV completion, Eq.~\eqref{uv-parameters} entails that $\rho \simeq m_1 \tan(\theta)$, where remember that a perturbative treatment of the quadratic coupling demands $\rho \lesssim m$, while one has $m_1 \gg m$ in the first place for $\sigma$ to be consistently kept in the EFT (recall the bound \eqref{bound-m}). This has a clear physical interpretation: for a generic mixing angle, the portal field $\F_1$ is a linear combination of the mass eigenstates $\sigma_{1,2}$ with similar weights, and the coupling between $\pi$ and the portal field cannot at the same time generate a small speed of sound, while leaving a weak coupling between $\pi_c$ and $\sigma$. Instead, this can be realised for a small mixing angle $\theta \ll 1$, as the portal field is then mostly aligned with the heaviest mass eigenstate responsible for the low sound speed, leading to a reduced strength of the coupling $\rho$ between $\pi_c$ and the ``misaligned'' field $\sigma$.\\

\noindent Naturally, the effective theory stemming from integrating out the $\sigma_1$ field misses the associated particle production effects in the squeezed limit. The full cosmological collider signal from such many-field theories has been computed recently \cite{Pinol:2021aun} and exhibits a rich structure, especially for comparable masses or/and generic mixing angles, notably resulting in modulated oscillations with several frequencies (see also \cite{Aoki:2020zbj}). But in our situation of interest here with a hierarchy $m_1 \gg m$, the exponential suppression of the particle production effects as a function of the mass entails that the full (many-field) cosmological collider signal is, for practical purposes, indistinguishable from the one computed in the two-field effective field theory involving $\pi$ and $\sigma$ only.\footnote{If the mixing angle $\theta$ is so small that the cosmological collider signal originating from $\sigma_2$ is similar to the one originating from $\sigma_1$, the whole cosmological collider signal becomes uninterestingly small, as well as the effects studied in this paper, whose amplitudes are governed by the size of the coupling $\rho$.}
It would be interesting to study if the correlation functions studied in this work with the EFT \eqref{S-EFT}-\eqref{interpisigma} as a starting point faithfully reproduce the ones of the UV completion discussed here in the entire range of triangular configurations. The answer to such a question would anyway depend on the specific type of UV completion considered, and in the following, we content ourselves with characterising primordial correlators within our setup.

\section{Cosmological collider bootstrap and the speed of sound} 
\label{WS-bootstrap}
\subsection{Mode functions and diagrammatic rules}
In this section we recap the standard \textit{in-in} formalism which will be used later for writing the bulk integral expressions for the cosmological correlators of interest in this paper. Of course, following the cosmological bootstrap philosophy, we will not directly evaluate these time integrals and use instead the bootstrap techniques to directly solve for the boundary correlators that these bulk integrals represent.\\ 

\noindent First of all, for future reference, we quote the positive frequency and negative frequency mode functions for $\pi_c$ and $\sigma$:
\begin{align}
\pi^{\pm}_c(k,\e)&=\dfrac{i H}{\sqrt{2c_s^3k^3}}(1\pm i c_sk\e)\exp(\mp ic_sk\e)\,,
\\
 \sigma_+(k,\eta)&=\dfrac{\sqrt{\pi} H}{2}\exp(-\pi \mu/2)\,\exp(i\pi/4) (-\eta)^{3/2} H^{(1)}_{i\mu}(-k\eta)\,, \\
  \sigma_-(k,\eta)&=\dfrac{\sqrt{\pi} H}{2}\exp(\pi \mu/2)\,\exp(-i\pi/4) (-\eta)^{3/2} H^{(2)}_{i\mu}(-k\eta)\,,
\end{align}
where
\begin{align}
    \mu=\sqrt{\dfrac{m^2}{H^2}-\dfrac{9}{4}}\,,
\end{align}
$H^{(1)}_{i\mu}$ and $H^{(2)}_{i\mu}$ are the Hankel functions of order $i \mu$ and of respectively the first and second type, and we recall that we consider heavy fields with $\frac{m}{H}\geq \frac{3}{2}$ in this paper.\\

\begin{figure}
    \centering
    \includegraphics[scale=0.52]{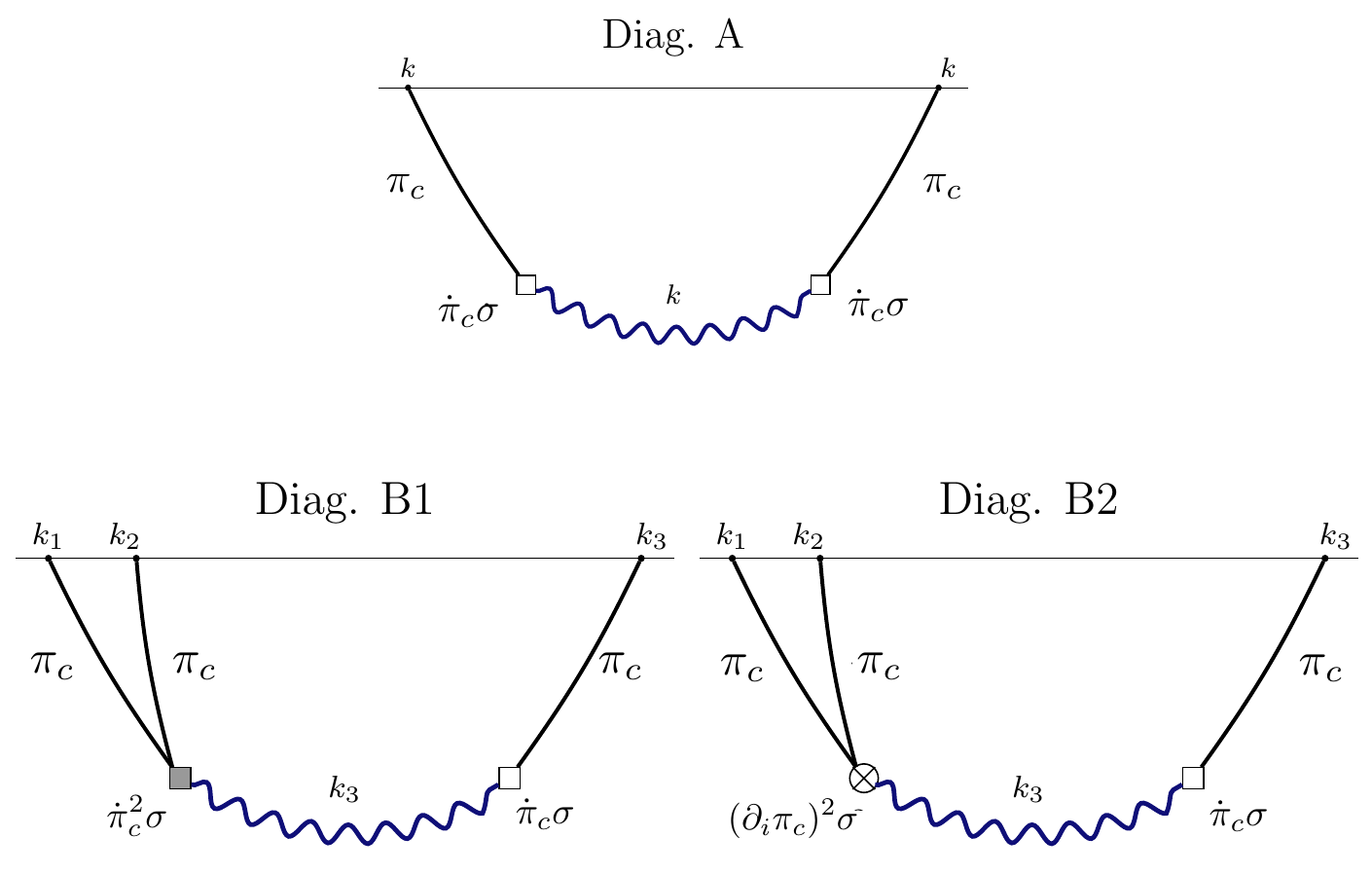}
    \caption{In this work we study in detail the three depicted single-exchange  diagrams for the two- and three-point correlation functions of $\pi_c$. The white rectangle represents the linear mixing operator $\dot{\pi}_c\sigma$, whereas the gray rectangle and the crossed circle stand for the $\dot{\pi}_c^2\sigma$ and $(\partial_i \pi_c)^2\sigma$ vertices, respectively. We also obtain the results for the three corresponding single-exchange four-point functions, with interactions either $\dot{\pi}_c^2\sigma$ or $(\partial_i \pi_c)^2\sigma$ at each vertex.}
    \label{feyab}
\end{figure}

\noindent
Having selected the interaction in \eqref{interpisigma}, we set out to calculate the 
correlation functions of $\pi$ mediated by $\s$. Using the \textit{in-in} approach \cite{Weinberg:2005vy}, the $n$-point function can be written as 
\begin{align}
\label{formalInIn}
    &\Big\langle \hat{\pi}(\bfk_1,\e_0)\dots \hat{\pi}(\bfk_n,\e_0)\Big\rangle=\\ \nonumber 
    &\Big\langle \bar{T}\left(e^{+i\int_{-\infty(1+i\epsilon)}^{\e_0}d\e\,H_{\text{int}}(\e)}\right)\, \hat{\pi}(\bfk_1,\e_0)\dots \hat{\pi}(\bfk_n,\e_0)\,T\left(e^{-i\int_{-\infty(1-i\epsilon)}^{\e_0}d\e\,H_{\text{int}}(\e)}\right)\Big\rangle_I\,,
\end{align}
where $\e_0$ is the end of inflation conformal time, and the subscript $I$ indicates that the operators and the vacuum are in the interaction picture, and $T$ ($\bar{T}$) denotes the time-order (anti-time-order) operation. To leading order in the couplings, the two-point function of $\pi$ induced by $\s$ is dominated by the diagram on the top of fig.~\ref{feyab}, which we refer to Diagram A hereafter. As for the three point function, 
two possible diagrams arise (see fig.~\ref{feyab} below): they are formed by the exchange of the particle $\sigma$ between the left and the right vertex. For both diagrams, the right vertex is given by the linear mixing term $\dot{\pi}_c \sigma$, while the left vertex is either $\dot{\pi}_c^2\s$ (Diagram B1) or $(\partial_i \Pc)^2\s$ (Diagram B2). 
By expanding the formal \textit{in-in} expression for the correlator in Eq.~\eqref{formalInIn}, the Feynmann rules for the diagrams can be summarised in the following steps: 
\begin{itemize}
    \item each vertex is labeled as ``+" or ``-", so a diagram with $N$ vertices entails $2^N$ contributions. Plus (minus) vertices come with a factor of ``$+i$" (``$-i"$). Each vertex is associated with a conformal time ($\e_i, i=1,\dots, N$) which is integrated over.
    \item an internal line (with momentum $s$) that connects two vertices is assigned an appropriate propagator, depending on the label of its vertices. Such bulk-to-bulk propagators (corresponding to $\s$) come in four different types that are defined by: 
    \begin{align}
    G_{++}(s,\e,\e') &=\s_-(s,\e')\s_+(s,\e)\theta(\e-\e')+\s_-(s,\e)\s_+(s,\e')\theta(\e'-\e)\,,\\ 
    G_{+-}(s,\e,\e') &=\s_+(s,\e')\s_-(s,\e)\,, \\
    G_{--}(s,\e,\e') &=\s_+(s,\e')\s_-(s,\e)\theta(\e-\e')+\s_+(s,\e)\s_-(s,\e')\theta(\e'-\e)\,,\\
    G_{-+}(s,\e,\e')&=\s_-(s,\e')\s_+(s,\e)\,,
    \end{align}
    where $\e$ and $\e'$ correspond to the conformal times of the vertices at each end (for real arguments, $G_{--}=G^*_{++}$ and $G_{-+}=G^*_{+-}$). 
    \item lines that connect a plus vertex (minus vertex) to the boundary, contribute a bulk-to-boundary propagator $\pi_c^{-}(k,\e)\pi_c^{+}(k,\e_0)$ ($\pi_c^{+}(k,\e)\pi_c^{-}(k,\e_0)$). 
    \item vertices with spatial derivatives come with a factor of $i\bfk$, where $\bfk$ is the momentum of the field that carries the derivative. As for a time derivative, the operator $\partial_\e$ act on the corresponding mode function, which might be either in the bulk-to-bulk or the bulk-to-boundary propagator that enters the vertex.\footnote{Notice that the time derivative does not act on the step function $\theta(\e-\e')$ since, in contrast to reference \cite{Chen:2017ryl} for instance, we are using the canonical version of the \textit{in-in} formalism where (in presence of operators with time derivatives) the interaction part of the Hamiltonian is not opposite to the interaction part of the Lagrangian (see \cite{Abolhasani:2022twf} for a related discussion).} 
\end{itemize}
\subsection{Conformally coupled field and the weight-shifting operators}
\begin{figure}
    \centering
    \includegraphics[scale=1.2]{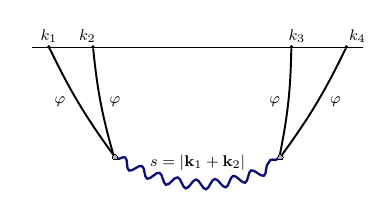}
    \caption{The four-point exchange diagram of $\vpi$ mediated by a massive scalar.}
    \label{fig:vpi-corr}
\end{figure}

The correlators of the conformally coupled (cc) scalar in dS space exhibit a simpler algebraic structure than the correlators of massless and massive fields. This is the direct result of the simplicity of its mode function:
\begin{align}
    \vpi_\pm(k,\e)=-\dfrac{H}{\sqrt{2k}}\e\,\exp(\mp ik\e)\,.
\label{modefunction-cc}   
\end{align}
Furthermore, the objects of primary interest in cosmology, namely the correlators of massless fields in dS can be obtained by acting with bespoke weight-shifting operators on the correlators of the conformally coupled field $\vpi$ (aka the ``cc field"). Using this method, all the exchange diagrams of the four-point function of a massless scalar field mediated by a massive field (including spinning ones) were computed in recent years \cite{Arkani-Hamed:2018kmz,Baumann:2019oyu}. The weight-shifting operators can be systematically derived by leveraging the dS $\text{SO}(4,1)$ isometry group. Nevertheless, regardless of the dS boost symmetry, the map between the correlators of the conformally coupled and the massless fields can be understood in terms of a set of relations between the corresponding mode functions (and derivatives thereof) \cite{Arkani-Hamed:2018kmz}.   
For example, the mode function $\Pc$ is related to $\vpi$ via a straightforward operation: 
\begin{align}
\label{pic}
    \Pc^\pm(k,\e)=\Pc(k,\e_0)\dfrac{1}{\e}(1-k\partial_k)h_\pm(c_sk,\e)\,.
\end{align}
where we have defined 
\begin{align}
    h_\pm(k,\e)&\equiv\e_0 \dfrac{\vpi_\pm(k,\e)}{\vpi_\pm(k,\e_0)}= \e \exp(\mp ik\e)\,,\\ 
    \Pc(k,\e_0)&\equiv \Pc^\pm(k,\e_0)=\dfrac{H}{(2c_s^3k^3)^{1/2}}\,.
\end{align}
For future references we also define $\vpi(k,\e_0)\equiv \vpi^\pm(k,\e_0)=H\e_0/(2k)^{1/2}$. 
An analogous equation to \eqref{pic} holds for the first derivative of $\Pc$, i.e. 
\begin{align}
\label{pprime}
  \partial_\e\Pc^\pm(k,\e)=\Pc(k,\e_0)c_s^2k^2\,h_\pm(c_sk,\e)\,,
\end{align}
and higher derivatives of $\Pc$ can be similarly expressed by virtue of its equation of motion. 
We will see in the remainder of this section that using these relations all the single-exchange diagrams of $\pi$, irrespective of the nature of the vertices, can be obtained by applying appropriate boundary operators on the four-point exchange diagram of $\vpi$ depicted in Figure \ref{fig:vpi-corr}, in which the intermediate field $\sigma$ interacts with $\vpi$ via the simple cubic interaction $g\,\vpi^2\,\sigma$. \\
 
\begin{figure}
    \centering
    \includegraphics[scale=0.5]{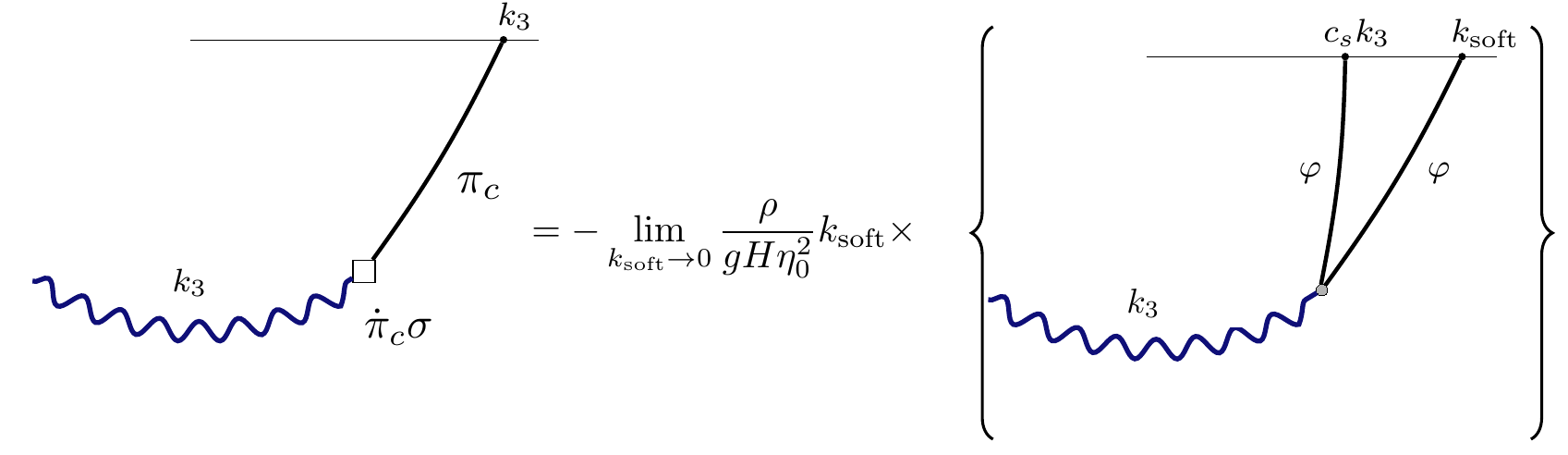}
    \includegraphics[scale=0.5]{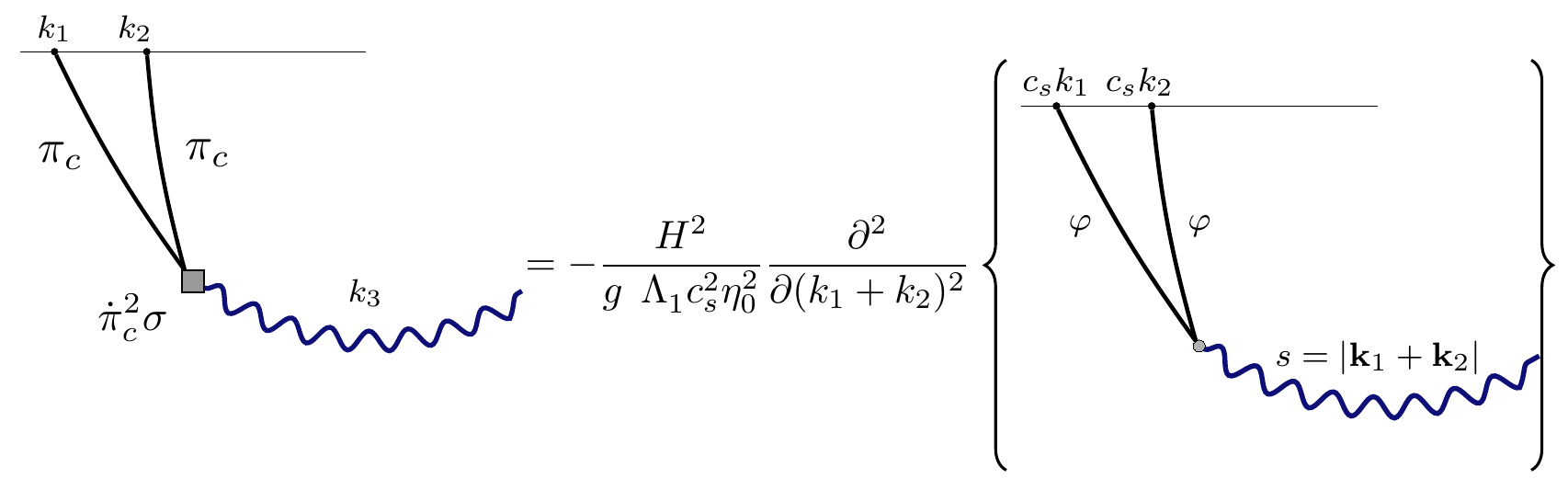}
    \includegraphics[scale=0.5]{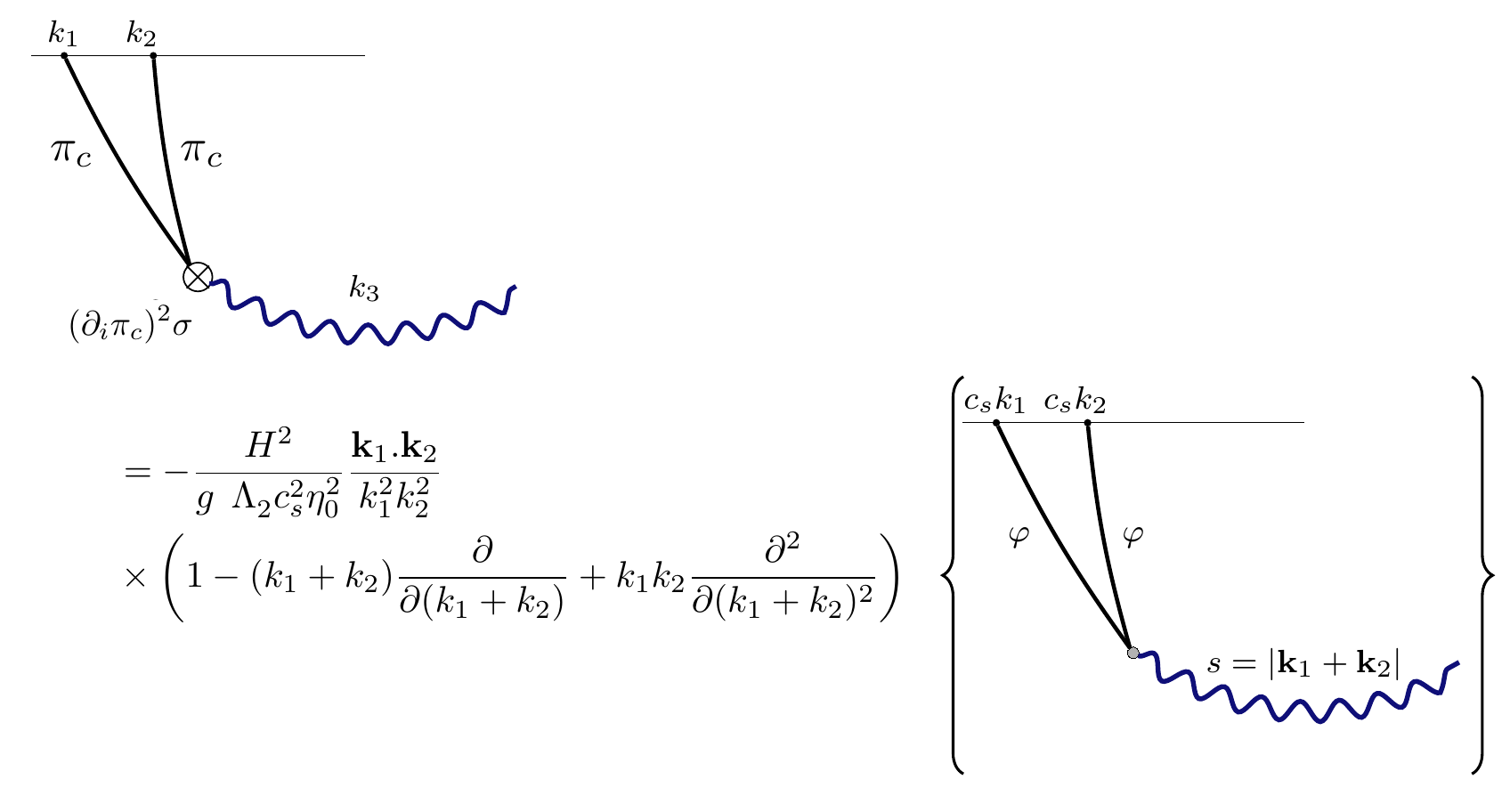}
    \caption{In this table we have collected the relationships between the building blocks of the correlators of $\pi_c$ and those of the four-point function of $\vpi$. By multiplying the operators that act on each vertex in Diagrams A, B1 and B2 and for the trispectra, one can deduce the weight shifting operators that relate the full diagrams to $\hat{F}$, as presented by Equations \eqref{diagmA}, \eqref{diagb1b2} and \eqref{trispectrum}}.
    \label{fig:building-blocks}
\end{figure}

\noindent
We begin by explicitly writing down the contribution of the exchange diagram depicted in Figure \ref{fig:vpi-corr} to  four-point correlator of $\vpi$ evaluated at the end of inflation $\e=\e_0$. Following a similar notation to Appendix B of \cite{Arkani-Hamed:2018kmz}, the answer is given by
\begin{align}
    \left\langle \vpi(\bfk_1,\e_0)\vpi(\bfk_2,\e_0)\vpi(\bfk_3,\e_0)\vpi(\bfk_4,\e_0)\right\rangle'=\dfrac{\eta_0^4\,H^2}{2k_1 k_2 k_3 k_4}\,\,F(k_1,\dots, k_4;s)+{t-}\,\,\text{and}\,{u-}\text{channels}\,,
\end{align}
in which 
\begin{align}
    F &=F_{++}+F_{+-}+F_{-+}+F_{--}\,,
\end{align}
where 
\begin{align}
\label{bulk-def-F++}
    F_{\pm\pm}(k_1,\dots,k_4; s)&=-\dfrac{g^2}{2 H^2}\int_{-\infty(1\mp i\epsilon)}^{\eta_0}\dfrac{d\e\,}{\e^2} \int_{-\infty(1\mp i\epsilon)}^{\eta_0}\dfrac{d\e'\,}{\e'^2}e^{\pm i(k_1+k_2)\e}\,e^{\pm i(k_3+k_4)\e'}\, G_{\pm\pm}(s,\e,\e')\,,\\ \label{Fpppm}
     F_{\pm\mp}(k_1,\dots,k_4;s)&=\dfrac{g^2}{2 H^2}\int_{-\infty(1\mp i\epsilon)}^{\eta_0}\dfrac{d\e\,}{\e^2} \int_{-\infty(1\pm i\epsilon)}^{\eta_0}\dfrac{d\e'\,}{\e'^2}e^{\pm i(k_1+k_2)\e}\,e^{\mp i(k_3+k_4)\e'}\,G_{\pm\mp}(s,\e,\e')\,.
\end{align}
Above, different components of the $s-$channel diagram are written in terms of four ``energy'' variables $\lbrace k_1,k_2,k_3,k_4,s\equiv |\bfk_1+\bfk_2|\rbrace$.\\
It is noteworthy that, for physical values of energies (namely $\lbrace k_a,s\rbrace \subset \mathbb{R}^+$), $F_{--}$ and $F_{+-}$ are given by the complex conjugates of $F_{++}$ and $F_{-+}$. Moreover, dilatation symmetry implies that the correlators of $\vpi$ scale as 
\begin{align}
    \langle \vpi(\lambda \bfk_1)\dots \vpi(\lambda \bfk_n)\rangle'=\dfrac{1}{\lambda^{2n-3}}\langle \vpi( \bfk_1)\dots \vpi( \bfk_n)\rangle'\,.
\end{align}
As a result, $F_{\pm\pm}$ and $F_{\pm\mp}$ can be expressed as
\begin{align}
    F_{\pm\pm}(k_a;s)=\dfrac{1}{s}\hat{F}_{\pm\pm}(u,v)\,,\quad F_{\mp\pm}(k_a;s)=\dfrac{1}{s}\hat{F}_{\pm\mp}(u,v)\,,
\end{align}
from which it follows that 
\begin{align}
    F=\dfrac{1}{s}\hat{F}(u,v)\,,\quad \hat{F}=\hat{F}_{++}+\hat{F}_{--}+\hat{F}_{+-}+\hat{F}_{-+}\,.
\end{align}
where we have defined the energy ratios 
\begin{align}
    u\equiv \dfrac{s}{k_1+k_2}\,,\qquad v\equiv \dfrac{s}{k_3+k_4}\,. 
\end{align}
For physical configurations, the triangle inequality implies that 
\begin{align}
    0\leq u \leq 1\,,\qquad 0\leq v\leq 1\,,\qquad \text{physical configurations}. 
\end{align}
However, relating our diagrams to $F$ will incorporate the \textit{analytic continuation} of $F$ as a function of $k_a$($a=1,\dots,4$) and $s$ (or equivalently $\hat{F}$ as a function of $u$ and $v$) in a domain that 
should at least cover all the real and positive values of $u$ and $v$ (especially the region defined by $u>1$ and $v>1$).\\
\\
\noindent  The single-exchange diagrams of $\pi$ can be related to the soft limit of the quantity $F$ defined above by means of appropriate weight-shifting operators. Using the relationships \eqref{pprime} and \eqref{pic} we infer that 
\begin{itemize}
    \item using \eqref{pprime} inside the \textit{in-in} expressions of all diagrams, the quadratic vertex 
    \begin{align}
    \nonumber
    \e^{-3}\partial_\e\Pc^{\pm}(k,\e)\sigma_\pm(k,\e)
    \end{align}
    can be related to the cubic vertex
    \begin{align}
    \nonumber
    \e^{-4}\vpi_\pm(c_s k,\e)\sigma_\pm(k,\e)\vpi_\pm(  k_{\text{soft}},\e)\,.
    \end{align}
    Above, the momentum of one the external cc fields is taken to zero ($k_{\text{soft}}\to 0$). The mere purpose of this soft cc field in the cubic vertex is to contribute a factor of $\e$ to the \textit{in-in} expression, hence adjusting the power of conformal time in the quadratic vertex.  It is also crucial that the energy of the other external cc field is re-scaled with $c_s$ while the energy of the intermediate field $\s$ is left intact. Notice that the prescription above and others that follow go the same for all combinations of positive and negative frequencies in the product of the fields. 
    \item in Diagram B1, the left vertex (in momentum space) gives the following contribution to the \textit{in-in} time integral: 
    \begin{align}
        \e^{-2}\partial_\e\Pc^\pm(k_1,\e)\partial_\e\Pc^\pm(k_2,\e)\sigma_\pm(|\bfk_1+\bfk_2|,\e)\,. 
    \end{align}
    This term is proportional to 
    \begin{align}
        \dfrac{\partial^2}{\partial (k_1+k_2)^2} \left(\e^{-4}\vpi_\pm(c_s k_1,\e)\vpi_\pm(c_s k_2,\e)\s_\pm(|\bfk_1+\bfk_2|,\e)\right)\,,
    \end{align}
    where here, the derivative operator generates a factor of $\e^2$ and raises the power of $\e^{-4}$ in the vertex $\vpi^2\s$ to $\e^{-2}$ in the vertex $\Pc'^2\sigma$. 
    \item in Diagram B2 the left vertex (in momentum space) takes the following form: 
        \begin{align}
        \bfk_1. \bfk_2 \e^{-2}\Pc^\pm(k_1,\e)\Pc^\pm(k_2,\e)\sigma_\pm(|\bfk_1+\bfk_2|,\e)\,. 
    \end{align}
    Using \eqref{pic}, this can be recast into
    \begin{align}
       \bfk_1. \bfk_2 (1-k_1\partial_{k_1})(1-k_2\partial_{k_2})\left(\e^{-4}\vpi_\pm(c_s k_1,\e)\vpi_\pm(c_s k_2,\e)\s_\pm(|\bfk_1+\bfk_2|,\e)\right)\,,
    \end{align}
    up to an energy dependent prefactor. Another simplification occurs in that the term in parenthesis depends on $k_{1,2}$ only through the combination $(k_1+k_2)$. Therefore $\partial_{k_1}=\partial_{k_2}=\partial_{(k_1+k_2)}$ (notice that $s=|\bfk_1+\bfk_2|$ is an independent variable), and consequently we can write
    \begin{align}
        (1-k_1\partial_{k_1})(1-k_2\partial_{k_2})=\left(1-(k_1+k_2)\partial_{k_1+k_2}+k_1 k_2 \partial^2_{k_1+k_2}\right)\,.
    \end{align}
\end{itemize}
The resulting relationships between the building blocks of the $\pi$ and $\vpi$ correlators are depicted in Figure \ref{fig:building-blocks}, where we have included the appropriate powers of external energies and prefactors. Converting $\Pc$ to the curvature perturbation $\zeta$, given by
\begin{align}
    \zeta=-H\pi=-\dfrac{c_s }{\sqrt{2 \epsilon} \Mpl}\pi_c\,,
\end{align}
we finally arrive at the relationships below between the four-point function $F$ and our desired correlators:\\
\textbf{Power spectrum. Diagram A}. The correction to the power spectrum of $\zeta$ is extracted from the double soft limit of the four-point function of the cc field: 
\begin{equation}
\dfrac{\Delta P_{\zeta}(k)}{P_{\zeta}(k)}=\dfrac{\rho^2}{g^2 H^2}(c_s k)\lim_{k_{\text{soft}}\to 0}F(c_s k,k_{\text{soft}},c_s k,k_{\text{soft}};k) \,,
\end{equation}
i.e.
\begin{tcolorbox}[colframe=white,arc=0pt]
\begin{equation}
\label{diagmA}
    \dfrac{\Delta P_{\zeta}}{P_{\zeta}}=\dfrac{c_s\rho^2}{g^2 H^2}\hat{F}\left(u=\dfrac{1}{c_s},v=\dfrac{1}{c_s}\right)\,,
\end{equation}
\end{tcolorbox}
where $P_\zeta(k)$ is the standard vacuum contribution to the scalar power spectrum
\begin{align}
    P_\zeta=2\pi^2 \dfrac{{\cal P}_\zeta }{k^3}\,,\qquad {\cal P}_\zeta=\dfrac{H^2}{8\pi^2 \epsilon c_s \Mpl^2}\,.
    \label{eq:standard-power-spectrum}
\end{align}
Notice that the arguments of $\hat{F}$ are bigger than unity for $c_s<1$. Therefore, evaluating the right-hand side above already involves an analytic continuation outside the physical domain of momenta for the seed correlator.  \\
\textbf{Bispectrum. Diagrams B1-B2}. The corresponding bispectra are related to the soft limit of $F$ followed by an appropriate weight-shifting operator. They are given by 
\begin{subequations}
\begin{align}
   B^{\text{B1}}_\zeta &=\left(-\dfrac{4\pi^3 \rho}{c_s^{1/2} g^2\Lambda_1 }\right)\,\dfrac{{\cal P}_\zeta^{3/2}}{k_1 k_2 k_3}\lim_{k_{\text{soft}}\to 0}\dfrac{\partial^2}{\partial (k_1+k_2)^2}F(c_sk_1,c_sk_2,c_s k_3,k_{\text{soft}};k_3)+{t-}\,\,\text{and}\,{u-}\text{channels}\,,\\ \nonumber
   B^{\text{B2}}_\zeta &=\left(-\dfrac{4\pi^3\rho}{c_s^{1/2} g^2\Lambda_2}\right)\dfrac{{\cal P}_\zeta^{3/2}}{k_1^3 k_2^3 k_3}\bfk_1.\bfk_2\,\\ \nonumber
   & \times \lim_{k_{\text{soft}}\to 0}\left(1-(k_1+k_2)\dfrac{\partial}{\partial (k_1+k_2)}+k_1 k_2 \dfrac{\partial^2}{\partial (k_1+k_2)^2}\right) F(c_sk_1,c_sk_2,c_s k_3,k_{\text{soft}};k_3)\\ 
   & +{t-}\,\,\text{and}\,{u-}\text{channels}\,.
\end{align}
\end{subequations}
It useful to write the final result in terms of $\hat{F}$ and its partial derivatives, i.e. 
\begin{tcolorbox}[colframe=white,arc=0pt]
\begin{subequations}
\label{diagb1b2}
\begin{align}
B^{\text{B1}}_\zeta(k_1,k_2,k_3)&=\dfrac{\alpha_1}{g^2 k_1 k_2 k_3 (k_1+k_2)^3}\Big(2\partial_u+u\partial_u^2\Big)\hat{F}(u,v)+ {t-}\,\,\text{and}\,{u-}\text{channels}\,.  \\ \nonumber 
&\\ \nonumber 
B^{\text{B2}}_\zeta(k_1,k_2,k_3)&=\dfrac{\alpha_2 \,\bfk_1.\bfk_2}{g^2 k_1^3 k_2^3 k_3^2}\left(1+u\partial_u+\dfrac{c_s^2 k_1 k_2}{k_3^2}u^3[2\partial_u+u\partial_u^2]\right)\hat{F}(u,v)\\ 
&+ {t-}\,\,\text{and}\,{u-}\text{channels}\,.
\end{align}
\end{subequations}
\noindent with 
\begin{align}
    u=\dfrac{k_3}{c_s(k_1+k_2)}\,,\qquad v=\dfrac{1}{c_s}\,,
    \label{def:u-v}
\end{align}
\end{tcolorbox}
\noindent and where we have defined  
\begin{align}
    \alpha_1=-\dfrac{4\pi^3 \rho}{c_s^{3/2}\Lambda_1}{\cal P}_\zeta^{3/2}\,,\qquad \alpha_2=-\dfrac{4\pi^3\rho}{c_s^{1/2}\Lambda_2}{\cal P}_\zeta^{3/2}\,.
\end{align}
\\
\textbf{Trispectrum}.
It is also immediate to combine the building blocks relationships of fig.~\ref{fig:building-blocks} to compute the three different four-point correlation functions of $\zeta$ mediated by the exchange of $\sigma$, built out of our two cubic vertices. We obtain the trispectra $T_\zeta=\langle \zeta(\bfk_1)\zeta(\bfk_2)\zeta(\bfk_3) \zeta(\bfk_4)\rangle'$, with obvious notations:
\begin{tcolorbox}[colframe=white,arc=0pt]
\begin{subequations}
\label{trispectrum}
\begin{align}
&T_\zeta^{\dot{\pi}_c^2\sigma-\dot{\pi}_c^2\sigma}= \dfrac{\beta_{1,1} k_{12}}{g^2 k_1 k_2 k_3 k_4 (k_1+k_2)^3 (k_3+k_4)^3}\Big(2\partial_u+u\partial_u^2\Big) \Big(2\partial_v+v\partial_v^2\Big)\hat{F}(u,v) + 2\, \textrm{perm.} \\
\nonumber 
\\ \nonumber
&T_\zeta^{\dot{\pi}_c^2\sigma-(\partial_i \pi_c)^2\sigma}= \dfrac{\beta_{1,2}}{g^2 k_1 k_2 k_3 k_4 (k_1+k_2)^3} \dfrac{\bfk_3.\bfk_4}{k_3^2 k_4^2} \nonumber \\
&\times \Big(2\partial_u+u\partial_u^2\Big) \left(1+v\partial_v+\dfrac{c_s^2 k_3 k_4}{k_{12}^2}v^3[2\partial_v+v\partial_v^2]\right)\hat{F}(u,v)
 + 5\, \textrm{perm.} 
 \\
\nonumber 
\\ \nonumber  
&T_\zeta^{(\partial_i \pi_c)^2\sigma-(\partial_i \pi_c)^2\sigma}= \dfrac{\beta_{2,2}}{g^2 k_1 k_2 k_3 k_4 k_{12}} \dfrac{\bfk_1.\bfk_2}{k_1^2 k_2^2} \dfrac{\bfk_3.\bfk_4}{k_3^2 k_4^2} \\
& \vspace{-2em} \times   \left(1+u\partial_u+\dfrac{c_s^2 k_1 k_2}{k_{12}^2}u^3[2\partial_u+u\partial_u^2]\right) \left(1+v\partial_v+\dfrac{c_s^2 k_3 k_4}{k_{12}^2}v^3[2\partial_v+v\partial_v^2]\right)\hat{F}(u,v)
 + 2\, \textrm{perm.} \nonumber
\end{align}
\end{subequations}
\noindent where here 
\begin{align}
    u=\dfrac{k_{12}}{c_s(k_1+k_2)}\,,\qquad v=\dfrac{k_{12}}{c_s (k_3+k_4)}
    \label{def:u-v-trispectrum}
\end{align}
\end{tcolorbox}
with
\begin{align}
    \beta_{1,1}=\dfrac{8 \pi^4 {\cal P}_\zeta^{2} H^2 }{c_s^{4}\Lambda_1^2}\,,\qquad \beta_{1,2}=\dfrac{8 \pi^4 {\cal P}_\zeta^{2} H^2 }{c_s^{3}\Lambda_1 \Lambda_2}
      \,, \quad \beta_{2,2}= \dfrac{8 \pi^4 {\cal P}_\zeta^{2} H^2 }{c_s^{2}\Lambda_2^2}\,.
\end{align}

\noindent Eventually, note that our results hold for any value of $c_s$ including values larger than unity
(remember that up to a rescaling of spatial coordinates, $c_s$ can be considered as the ratio between the propagation speeds of $\pi$ and of $\sigma$). However, this regime does not require extra theoretical work, as the observable correlation functions are then mapped to the seed four-point correlation function $\hat{F}$ with arguments $(u,v)$ inside the unit disk (see Eqs.~ \eqref{diagmA}, \eqref{def:u-v} and \eqref{def:u-v-trispectrum}), which has been computed in \cite{Arkani-Hamed:2018kmz}. A straightforward but interesting consequence is that due to the stretching between k-space and $(u,v)$ space by the sound speed, the usual cosmological collider oscillations, only present for the bispectrum in the squeezed limit for $c_s=1$, can extend to the whole triangular configurations up to equilateral ones, see \cite{Pimentel:2022fsc} for related plots. This has a clear physical origin in terms of the characteristic timescales discussed in section \ref{Qualitative}, as the first situation in fig.~\ref{fig:overview-bispectrum} then occurs for all triangles.

\subsection{Bootstrap toolkit}

In this subsection, we describe the bootstrap tools that we adopt in this work in order to deduce the four-point function $\hat{F}(u,v)$ in our region of interest. 
This contains a summary of already known features but also new results on their own.

\subsubsection{Analyticity and polology}
\label{Analyticity}
The singularities of the cosmological correlators have demonstrated constraining power in dictating their entire structures  \cite{Arkani-Hamed:2018kmz,COT,Jazayeri:2021fvk,Pajer:2020wxk, Baumann:2021fxj}. For Bunch-Davies initial conditions, these singularities are absent for physical configurations, and this by itself is an indispensable input for the bootstrap program. However, two general types of poles appear once the correlators  are 
analytically continued in their kinematical arguments: 
\begin{itemize}
    \item the \textit{total energy pole} is defined by the following hyperplane in the space of energy variables, 
    \begin{align}
        k_T=E_1+\dots+E_n=0\,,\quad E_i\equiv c_i k_i\,,
    \end{align}
    where $E_i=c_i k_i$ are the energies of the external fields with $c_i$'s standing for the speed of propagation for each external field in the correlation function. Near the singularity, the correlator behaves as 
    \begin{align}
        \text{correlator}\propto \dfrac{1}{k_T^p}\,,
    \end{align}
    where $p$ is fixed by dimensional analysis \cite{Pajer:2020wxk} 
    \begin{align}
    \label{degree}
        p=1+\sum_{\alpha}(\Delta_\alpha-4)\,.
    \end{align}
    Here $\alpha$ runs over all vertices in the diagram, and $\Delta_\alpha$ is the energy dimension of the operator that acts at the vertex (for the exceptional cases of $p=0$ and $p=-1$ the singularity behaves as $\log(k_T)$ and $k_T \log(k_T)$, respectively.)
    The residue of the total energy pole is proportional to the scattering amplitude associated with the same diagram in flat space \cite{Raju:2012zr,Maldacena:2011nz, COT}.
    \footnote{An exceptions to this rule arises when the leading order scattering amplitude of the theory vanishes. See \cite{Grall:2020ibl} for some discussions about this point, in the context of the DBI theory.}
    \item The \textit{subdiagram (partial) energy poles} are associated with the total energy of the subdiagrams that emerge after cutting an internal line in the original graph. The residue of such a pole is proportional to the amplitude that each subdiagram defines. The degree of the singularity is determined by the same formula as Equation \eqref{degree}. 
\end{itemize}
We are going to review the singularity structure for individual components of the correlator $F$, namely $F_{++}$ and $F_{+-}$ ($F_{--}$ and $F_{-+}$ are not independent quantities). For simplicity, we only analytically continue in the external energies $k_a$ and maintain $s$ as real and positive. The advantage of looking at the $++$ and $+-$ components separately is that the analytic continuation of each is already defined by the time integrals in Equation \eqref{bulk-def-F++} in certain domains in the complex plane. Moreover, as we discuss shortly, a cutting rule can only be stated for the $++$ part and not for the whole correlator.\\

\noindent The domains of analyticity of the formulae \eqref{bulk-def-F++}-\eqref{Fpppm} are determined by the convergence of the time integrals in the ultraviolet (i.e. at $\e\to -\infty$ limit), and they are given by
\begin{align}
   & F_{++}: \lbrace (k_a,s)|\,\text{Im}(k_a)<0, s>0\rbrace\,, \\ \nonumber
   & F_{+-}: \lbrace (k_a,s)|\,\text{Im}(k_{1,2})<0,\text{Im}(k_{3,4})>0 ,s>0\rbrace\,,
\end{align}
or equivalently
\begin{align}
\nonumber
    &\hat{F}_{++}: \lbrace (u,v)|\,\text{Im}(u)>0, \text{Im}(v)>0\rbrace \\ \nonumber
    & \hat{F}_{+-}: \lbrace (u,v)|\,\text{Im}(u)>0, \text{Im}(v)<0\rbrace\,.
\end{align}
We begin by $F_{+-}$, which is simply the product of two three-point functions 
\begin{align}
\label{factorized}
    F_{+-}=\dfrac{1}{2} f_3(k_1-i\epsilon,k_2-i\epsilon,s)f_3^*(k_3-i\epsilon,k_4-i\epsilon,s)\,,\qquad k_{a}>0, s>0\,.
\end{align}
Above, we have defined
\begin{align}
\label{threepoint}
    f_3(k_1,k_2,s)=\dfrac{1}{\sqrt{s}}\hat{f}_3(u)\equiv \dfrac{i g}{H}\int_{-\infty}^{0}\dfrac{d\e}{\e^2}e^{i(k_1+k_2)\e}\s_-(s,\e)\,,
\end{align}
across the area $\lbrace (k_1,k_2,s)|\text{Im}(k_{1,2})<0, s>0\rbrace$ (or $\text{Im}(u)>0$). 
This three-point function
\footnote{The reason we call $f_3$ a three-point is that it is proportional to the wavefunction coefficient $\psi_{\vpi\vpi\s}$ in the late time wavefunction of the universe \cite{Arkani-Hamed:2018kmz,Anninos:2014lwa}, this quantity might be thought as a three-point function associated with a putative dual theory that lives on the boundary. The actual three-point correlation function $\langle\vpi\vpi\s\rangle$ is instead proportional to $\text{Re}[f_3/\sigma_+(s,\e_0)]$.}
corresponds to the one that emerges after cutting the internal line of the four-point exchange diagram. It will be sufficient to bootstrap this three-point function, and $F_{+-}$ will simply follow from \eqref{factorized}. \\

\noindent  The sole singularity of the three-point function $f_3$ is the total energy pole located at 
\begin{align}
    E_L=k_1+k_2+s=0\,, \qquad \text{or}\qquad u=-1\,.
\end{align}
This will contribute as a (left) partial energy pole to the full correlator $F$. There will equally be a right partial energy pole located at 
\begin{align}
    E_R=k_3+k_4+s=0\,, \qquad \text{or}\qquad v=-1\,.
\end{align}
The singularity of $f_3$ can be easily revealed by the direct inspection of the time integral: singularities can only arise in the UV part of the integral, where the integrand is exponentially suppressed unless the external energies sum to zero. Near the limit $E_L\to 0$, the integral is dominated by its behaviour at large conformal time, where 
the mode function $\s_-$ can in effect be replaced by $\s_-\to -\dfrac{H}{\sqrt{2s}}\e \exp(i s\e)$, 
and the behaviour of $f_3$ in the vicinity of $E_L=0$ is found to be 
\begin{align}
\label{flatspace}
\lim_{u\to -1+i\epsilon}f_3(k_1,k_2,s)=-\dfrac{i g}{\sqrt{2s}}\log(1+u)\,.
\end{align}
The coefficient $i g$ in front is the three-particle amplitude $\vpi\vpi\to\s$. Also, the degree of the divergence agrees with Eq.~\eqref{degree} because the diagram has a single vertex with a relevant operator $\vpi^2\s$ resulting in $p=0$.\\

\begin{figure}
    \centering
    \includegraphics[scale=0.4]{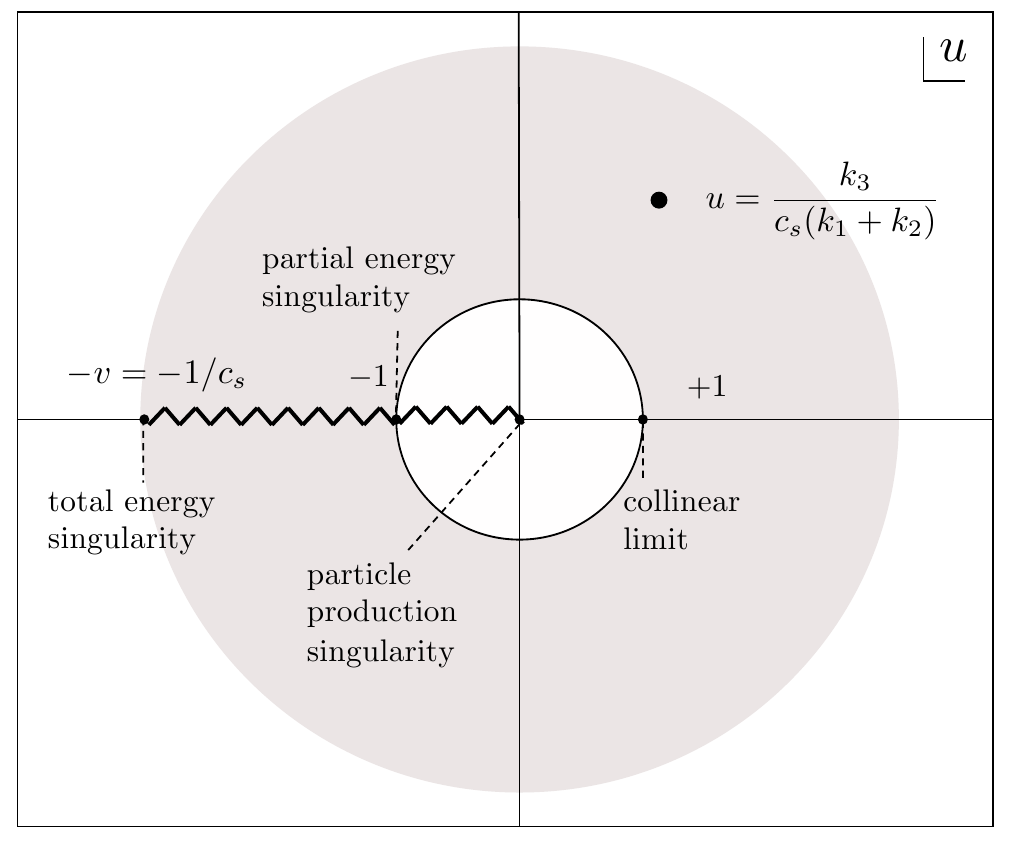}
    \caption{The analytical structure of $\hat{F}_{++}$ as a function of complex $u$. Motivated by the fact that in the bispectra formulae \eqref{diagb1b2} we set $v=1/c_s$, we have taken $v$ to be a positive and bigger than $1$ parameter. Moreover, in contrast with the ordinary case of $c_s=1$, these equations also imply that $u=k_3/c_s(k_1+k_2)$ is allowed to lie outside the unit disk. The shaded region is the annulus where the expression \eqref{fullFpp} is valid. There are three branch points, located on the real axis at $u=0$ (due to the particle production effect discussed in Section \ref{asymptlimitseed}), at $u=-1$ (due to the partial energy singularity \eqref{partial}) and finally at $u=-v$ (due to the total energy singularity \eqref{totalenergy}). The four-point function should have no divergence near the collinear limit ($u=1$), see Section \ref{fixinghom}.}
    \label{fig:analF++}
\end{figure}

 \noindent  The analytical structure of $f_3$ alongside the boundary equations and the cutting rules, discussed in future sections, will form enough ingredients to pinpoint $F_{++}$ as well (see Figure \ref{fig:analF++} where the analytical structure of $\hat{F}_{++}$ is represented). So, even though we will not directly need them, for completeness we briefly review the divergences of $F_{++}$. 
Near $k_T=0$ (or equivalently $u+v=0$), the double-time integral is dominated by the regime where both vertices are evaluated at infinite past. As a result, the time integral simplifies to \cite{Arkani-Hamed:2018kmz} 
\begin{align}
    \label{totalenergy}
    \lim_{u+v\to 0+i\epsilon} \hat{F}_{++}=\dfrac{g^2}{v^2-1}(u+v)\log(u+v)\,.
\end{align}
The right-hand side is proportional to the $s-$channel two-to-two scattering of $\vpi$ exchanged by $\s$ (which is given by $A_{\text{flat}}=\frac{1}{s_{\text{flat}}}=\frac{1}{(k_3+k_4)^2-s^2}$) and the degree of divergence corresponds to $p=-1$ in Equation \eqref{degree}. \\

\noindent  As for the partial energy pole, the residue is totally fixed by unitarity (see the discussion below). But it could also be seen directly at the level of the time integral that near $E_L=0$, the integral is dominated by the $\e\to -\infty$ limit, and $F_{++}$ reduces to
\begin{align}
\label{partial}
    \lim_{u\to -1+i\epsilon} F_{++}=-\dfrac{i g}{2\sqrt{2s}}\log(1+u)\,f_3^*(-k_3^*,-k_4^*,s)\,.
\end{align}
We see that in this case $F_{++}$ factorises into the product of the three-particle amplitude $i g$ and a three-point correlator (with deformed arguments
\footnote{As a technical side, notice that $-k_{3,4}^*$ are within the domain of analyticity of $f_3$, namely the lower half complex plane of the $k_{3,4}$ space.}). A similar factorisation occurs near $E_R=0$ (i.e. $v=-1$). 
\subsubsection{Locality: boundary differential equations}
It was pointed out in \cite{Arkani-Hamed:2018kmz} that the Ward identities associated with the dS boost symmetries imply two \textit{boundary differential equations} for the four-point function $\hat{F}$ 
\begin{align}
    &{\cal O}(u,\partial_u)\hat{F}(u,v)=g^2\frac{u\,v}{2(u+v)}\,,\\ \nonumber
    & {\cal O}(v,\partial_v)\hat{F}(u,v)=g^2\frac{u\,v}{2(u+v)}\,,
\end{align}
where 
\begin{align}
    {\cal O}(u,\partial_u)\equiv \left[ u^2(1-u^2)\partial_u^2-2u^3\partial_u+\left(\mu^2+\frac{1}{4}\right)\right]\,.
\end{align}
The RHS of the equations above corresponds to the contact correlator $\vpi^4$, which emerges after the operator ${\cal O}(u,\partial_u)$ has turned the exchange interaction into a contact one (see \eqref{btob} below).\\

\noindent  Here we outline the direct derivation of these boundary equations based on locality, without any reference to the dS boost symmetry---in this case, locality is synonymous of the fact that the four-point function is induced by the propagation of $\s$ in the bulk of spacetime. Therefore, we begin with the bulk differential equations that govern the bulk-to-bulk propagators in the \textit{in-in} formalism \cite{Arkani-Hamed:2018kmz}, i.e. 
\begin{align}
\label{btob}
&\left[\del_\e^2-\dfrac{2}{\e}\del_\eta+s^2+\dfrac{m^2}{\e^2H^2}\,\right]G_{\pm\pm}(s,\eta,\eta')=(\eta' H)^2 \delta(\eta-\eta')\,,\\ \nonumber
& \left[\del_\e^2-\dfrac{2}{\e}\del_\eta+s^2+\dfrac{m^2}{\e^2H^2}\,\right]G_{\pm\mp}(s,\eta,\eta')=0\,,
\end{align}
where $s$ is the energy of the exchanged field. These bulk equations can be converted into boundary equations for $F$ by trading the derivatives with respect to time for the derivative with respect to momentum, when acted on the plane wave $\exp(ic_sk\eta)$, namely 
\begin{equation}
    \e \partial_\e\,\exp(ic_s\,k\e)=k\partial_k\,\exp(ic_s k\e)\,.
\end{equation}
After performing a number of integration by parts one arrives at 
\begin{align}
\label{boundaryeq1}
     &{\cal O}(u,\partial_u)\,\hat{F}_{\pm\pm}(u,v)=g^2\frac{u\,v}{2(u+v)}\,,\\ \label{boundaryeq2}
     & {\cal O}(u,\partial_u)\,\hat{F}_{\pm\mp}(u,v)=0\,,
\end{align}
plus the same copies of equations with operator ${\cal O}(v,\partial_v)$ substituted on the left-hand side. It follows from the above equations that the full correlator $\hat{F}$ will satisfy the same equation as the first line above. The second equation holds for the three-point function $\hat{f}_3$ as well. \\

\noindent  The differential equations \eqref{boundaryeq1} and \eqref{boundaryeq2} can be solved for $\hat{F}_{++}$ and $\hat{F}_{+-}$(or $\hat{f}_3$) by supplementing enough initial conditions in specific limits of the kinematics. As emphasised in Subsection \ref{Analyticity}, one such constraint derives from forbidding singularities in physical configurations, in particular at the collinear limit $u=1$ (or $v=1$). To illustrate this point, let us begin by solving the homogeneous equation \eqref{boundaryeq2} for the three-point function. The most general ansatz is given by
\begin{align}
\label{ansatz-f3}
    \hat{f}_3(u)=A_+ f_+(u)+A_- f_-(u)\,,
\end{align}
where $f_\pm$ are two linearly independent solutions to the homogeneous boundary equation:
\begin{align}
\label{newbaseplus}
   &\qquad  f_+(u)= {}_2 F_1\left(\frac{1}{4}-\frac{i \mu }{2},\frac{1}{4}+\frac{i \mu
   }{2};\frac{1}{2};\frac{1}{u^2}\right)\,,\,\,\,\\ 
  &\qquad  f_-(u)=\dfrac{2}{u} \times\,{}_2 F_1\left(\frac{3}{4}-\frac{i \mu }{2},\frac{3}{4}+\frac{i \mu
   }{2};\frac{3}{2};\frac{1}{u^2}\right)\,\,\, \label{newbaseminus}
\end{align}

\noindent Each of these two functions exhibit branch point singularities at $u=\pm 1$. Near the collinear singularity $u=+1$ we have
\begin{align}
\label{fpmcollinear}
    f_+\to -\dfrac{\sqrt{\pi}}{\Gamma(1/4+i\mu/2) \Gamma(1/4-i\mu/2)}\log(1-u)\,,\,\, f_-\to -\dfrac{\sqrt{\pi}}{\Gamma(3/4+i\mu/2) \Gamma(3/4-i\mu/2)}\log(1-u)\,.
\end{align}
Asking the cancellation of this logarithmic singularity in the linear combination \eqref{ansatz-f3} and matching onto the flat space limit ($u\to -1+i\epsilon$) in \eqref{flatspace}, we arrive at 
\begin{align}
    \hat{f}_3(u)=\dfrac{ig }{2\sqrt{2\pi }}\bigg(\Gamma\left(1/4+i\mu/2\right) \Gamma\left(1/4-i\mu/2\right) f_+(u)-\Gamma\left(3/4+i\mu/2\right) \Gamma\left(3/4-i\mu/2\right) f_-(u)\bigg)\,.
    \label{explicit-f3}
\end{align}
It is worth mentioning that the choice of basis \eqref{newbaseplus}-\eqref{newbaseminus} is particularly convenient for expanding $\hat{f}_3$ (and, as we will see later, the homogeneous part of $\hat{F}_{++}$), because they are both fully analytic across the region $\text{Im}(u)>0$ (the only branch cut of $f_\pm$ stretches from $-1$ to $+1$, i.e. it lies on the boundary of the analytical domain\footnote{A different set of basis functions $F_{\pm}(u)$ was used in \cite{Arkani-Hamed:2018kmz} with branch cuts positioned on the positive imaginary axis. The introduced basis functions $f_\pm(u)$ better suit us because we are working with specific continuations of $\hat{F}_{++}(u,v)$ and $\hat{f}_3$ that are entirely analytical on the upper-half plane of the complex space $(u,v)$ (see Section 4.2 of \cite{Goodhew:2021oqg} for a related discussion).}). 
\subsubsection{Unitarity: cosmological cutting rules} 
Recently, the implications of perturbative unitarity for the structure of cosmological correlators were studied in a series of works \cite{COT, Meltzer:2020qbr, Cespedes:2020xqq, Melville:2021lst,Goodhew:2021oqg}. It was shown that the unitarity of the evolution translates into an infinite number of constraints on the coefficients of the perturbative wavefunction of the universe. These constraints appear as a set of \textit{cutting rules} that recursively relate the discontinuity of a Feynman diagram to a linear combination of the products of the discontinuity of its subdiagrams. Instead of the wavefunction coefficients, below we derive a cutting rule directly for our object of interest, namely $\hat{F}_{++}$. \\

\noindent Underlying the derivation of the Cosmological cutting rules are the Hermitian analyticity of the bulk-to-boundary propagator, which for our setup is the simple fact that
\begin{align}
    \vpi_+^*(k,\e)=\vpi_+(-k^*,\e),\qquad k\in \mathbb{C}\,,
\end{align}
where $\vpi$ is the conformally coupled mode function \eqref{modefunction-cc}, and the factorisation of the imaginary part of the bulk-to-bulk propagator in the wavefunction picture. This last property can be easily converted into a statement about $G_{++}$:
\begin{align}
    G^*_{++}(s,\e,\e')+G_{++}(s,\e,\e')=\sigma_-(s,\e)\sigma_+(s,\e')+\eta \leftrightarrow \eta'\,.
\end{align}
Together with the time-integral definition of $F_{++}$ \eqref{bulk-def-F++}, this property enables us to write down the following cutting rule: 
\begin{tcolorbox}[colframe=white,arc=0pt]
\begin{equation}
\label{cuttingrule}
    \hat{F}_{++}(u,v)+\hat{F}^*_{++}(-u^*,-v^*)=-\dfrac{1}{2}\,\hat{f}_3(u)\hat{f}_3^*(-v^*)-\frac{1}{2} \hat{f}_3(v) \hat{f}_3^*(-u^*)\,,
\end{equation}
\end{tcolorbox}
\noindent valid within the upper half of the complex plane of $(u,v)$. \footnote{Note that $-u^*$ and $-v^*$ lie in the upper half of the complex plane as well.}
This relation has the anticipated format: a specific linear combination of the analytically continued four-point exchange diagram factorises into the sum over the product of its constituent three-points. This cutting rule will serve as an essential ingredient in solving the boundary equation for $\hat{F}_{++}$. As a corollary, it follows from the above equation that near the left partial energy pole, namely $u=-1$, $\hat{F}_{++}$ reduces to \eqref{partial}. This is so because near the singularity the second term on the LHS is finite, hence negligible, while on the right-hand side only the first term diverges. 

\section{Seed four-point function}
\label{seedfourpoint}

\noindent Having established the relationships \eqref{diagmA} and \eqref{diagb1b2}, our task now reduces to finding the four-point function $F(k_a;s)$ (or equivalently $\hat{F}(u,v)$). 
This correlator was bootstrapped in \cite{Arkani-Hamed:2018kmz} by means of locality and consistent factorisation of the four-point function on its partial energy poles (namely when $k_1+k_2+s$ and $k_3+k_4+s$ are simultaneously sent to zero). However, the final analytical result presented in the aforementioned paper contains a power series expansion which schematically looks like 
\begin{align}
    \hat{F}(u,v)\supset \sum_{m=0}^{\infty}\sum_{n=0}^{\infty} c_{m,n} u^{2m+1}\,(u/v)^n\,, \quad |u|<|v|\,,
\end{align}
where $c_{m,n}$'s are mass dependent constants (a similar expression holds for the opposite regime, i.e. $|u|>|v|$, upon replacing $u\leftrightarrow v$). This expansion is perfectly convergent when $u$ and $v$ are both inside the unit circle, i.e. $|u|\leq 1, |v|\leq 1$; this includes all physical configurations of the quadrilateral (i.e. $0\leq u\leq 1$, $0\leq v\leq 1$). In this work, we are interested in the opposite case: upon analytic continuation of $\hat{F}$, our setup probes values of $u$ (and $v$) larger than unity, hence beyond the unit disk and where the above series expansion is no longer applicable.\footnote{Along the lines of Appendix C of \cite{Arkani-Hamed:2018kmz}, one can pursue an alternative approach by resumming the power series inside the unit circle and analytically continue to the whole complex plane. In practice, however, this will involve Kampé de Fériet functions 
$^{p+q}F_{r+s}\left(\left\lbrace \begin{array}{l} a_1,\dots a_p:b_1,b'_1\dots b_1,b'_q \\ c_1,\dots c_r:d_1,d'_1,\dots d_s,d'_s
\end{array}\right\rbrace,x,y\right)$ 
which are defined as power series of their arguments $x,y$ only within the unit disk $|x|<1,|y|<1$ (we were unable to find an asymptotic expansion for these functions outside this region in the literature). For the purpose of our computation one needs to go beyond the unit disk of $x$ and we found it more insightful to solve the bootstrap equations there from first principles.} We also take a different pathway in our derivation as compared to \cite{Arkani-Hamed:2018kmz}. We use the boundary equation supplemented with the cosmological cutting rule to solve for $\hat{F}_{++}$ and subsequently arrive at $\hat{F}$, whereas the earlier derivation was based on factorisation in the $u\to -1, v\to -1$ limit and regularity at the junction $u=v$ for the full correlator $\hat{F}$. 

\subsection{Ansatz for the particular solution}

As we saw before, the correlators of $\pi$  are written as the outcomes of the action of the weight-shifting operators on the seed four-point function $\hat{F}(u,v)$ with arguments that can be outside or inside the respective unit disks (i.e. $|u|<1, |v|<1$). This is clear from the substitutions $u\to \frac{s}{c_s(k_1+k_2)}$ and $v\to \frac{s}{c_s(k_3+k_4)}$ made in \eqref{diagmA} and \eqref{diagb1b2}.
In fact, one can probe an interesting limit by sending $c_s\to 0$ while keeping the energies $(k_a,s)$ fixed.\footnote{Note that this should be thought as a formal limit. Physically speaking, however, the EFT of inflation becomes strongly coupled when $c_s\to 0$, see the discussion below Eq.~\eqref{eq:cutoff}.} This is equivalent to sending $u$ and $v$ to infinity, where the boundary equations for $\hat{F}_{++}$ simplify to: 
\begin{align}
\nonumber
   & -u^4\partial_u^2 \hat{F}_{++}-2u^3 \partial_u\hat{F}_{++}=\dfrac{g^2}{2}\dfrac{u v}{u+v}\,\\
   &-v^4\partial_v^2 \hat{F}_{++}-2v^3 \partial_v\hat{F}_{++}=\dfrac{g^2}{2}\dfrac{u v}{u+v}\,,\qquad |u|\gg 1, |v|\gg 1\,,
\end{align}
The most general solution to the above equations, after imposing the symmetry $u\leftrightarrow v$, is given by
\begin{align}
\label{asymptotic}
    \hat{F}_{++}=-\dfrac{g^2}{2}\left(\frac{1}{u}+\dfrac{1}{v}\right)\log(\dfrac{u+v}{u v})+a\left(\dfrac{1}{u}+\dfrac{1}{v}\right)+\dfrac{b}{u v}+c\,,\quad u,v\to \infty\,,
\end{align}
where $a,b$ and $c$ are constant, and in retrospect one can check that dropping the $\mu$-dependent terms in the boundary equations was indeed consistent for $u \gg \mu$. 
Another feature of the solution is the appearance of branch points at $u=0$ and $u=-v$. For $v\in \mathbb{R}^+ +i\epsilon$, the branch cut falls within the interval $[-v,0]$, in $u$-space. It is noteworthy that the behaviour near the branch point $u=-v$ is dictated by the total energy singularity of $\hat{F}_{++}$ in Eq.~\eqref{totalenergy}. In contrast, since the above relation was derived for large $u,v$ it is not applicable near $u=0$.   
\\

\noindent For finite values of $u$ and $v$, the Taylor expansion of the RHS of the boundary equation, namely 
\begin{equation}
    {\cal O}(u,\partial_u) \hat{F}_{++}={\cal O}(v,\partial_v) \hat{F}_{++}=
    \begin{cases}
    & \frac{g^2}{2}\sum_{n=0}^\infty (-1)^n\,\frac{u^{n+1}}{v^{n}}\qquad |u|<|v| \\
    &\\
    & \frac{g^2}{2}\sum_{n=0}^\infty (-1)^n\,\frac{v^{n+1}}{u^{n}}\qquad |u|>|v|
    \end{cases}
    \,,
\end{equation}
suggests the following ansatz for a particular solution 
\begin{align}
\label{ansatz}
    \hat{F}_{p}=\sum_{m,n=0}^{\infty}\left( a_{m n}+b_{m n}\,\log(u) \right)u^{-m}\,\left(\frac{u}{v}\right)^n\,,\qquad 1<|u|<|v|\,.
\end{align}
Above, since we are expanding across an annulus (see Figure \ref{fig:analF++}), we allow for both positive and negative powers of $u$, i.e. $m$ might be bigger than $n$ or not. 
The restriction to non-negative integers $m$ and the addition of the logarithmic term are both motivated by the asymptotic limit of the first term in \eqref{asymptotic}:
\begin{align}
   \lim_{1 \ll |u|<|v|} \hat{F}_{++}=\dfrac{g^2}{2}\left(\dfrac{1}{u}+\dfrac{1}{v}\right) \left[\log(u)+\sum_{n=0}^\infty \dfrac{(-1)^{n}}{n}\left(\dfrac{u}{v}\right)^n \right]\,.
\end{align}
In the next section we solve for the series coefficients $a_{mn}$ and $b_{mn}$ hence finding $\hat{F}_{++}(u,v)$ inside the indicated domain. One can then easily extend the solution to the opposite side (i.e. $1<|v|<|u|$) by virtue of the symmetry under the exchange of $u$ and $v$.

\subsection{Series coefficients and resummation}

Plugging the ansatz \eqref{ansatz} inside the boundary equation leads to a set of recursive relations for the series coefficients $a_{mn}$ and $b_{mn}$ that can be solved. We go straight to the final answer here and leave the details of the derivation to Appendix \ref{detailsderivation}. To express the result, it proves useful to switch to a new set of coefficients defined by
\begin{align}
  B_{k,n}\equiv b_{(n-k) n},\,\quad A_{k,n}\equiv a_{(n-k)n}\,,\quad -\infty<k\leq n\,,
\end{align}
with which we write
\begin{align}
\label{ansatz2}
    \hat{F}_{p}=\sum_{n=0}^{\infty}\sum_{k=-\infty}^{n} \bigg( A_{k,n}+B_{k,n}\,\log(u) \bigg)\,\dfrac{u^k}{v^n}\,,\qquad 1<|u|<|v|\,.
\end{align}
The only non-vanishing elements of the matrices $A_{k,n}$ and $B_{k,n}$ can be found in Eqs.~\eqref{B-2ln}-\eqref{A-2l-1n}. We show in Appendix \ref{detailsderivation} that the logarithmic piece in the particular solution \eqref{ansatz2} resums to
\begin{align}
\label{Bpart}
    \sum_{n=0}^{\infty}\sum_{k=-\infty}^{n} B_{k,n}\log(u)\dfrac{u^k}{v^n}=\dfrac{g^2}{4}\bigg(f_+(u)f_-(v)+f_-(u)f_+(v)\bigg)\log(u)\,.
\end{align}
Furthermore, the first contribution can be repackaged into
\begin{align}
    \label{simplifiedsum}
    \sum_{n=0}^{\infty}\sum_{k=-\infty}^{n} A_{k,n}\dfrac{u^k}{v^n} &=
    \dfrac{1}{8\pi^2}\cosh(\pi\mu)\Gamma(3/4+i\mu/2) \Gamma(3/4-i\mu/2)\,f_-(v)\sum_{l=0}^\infty p_l\dfrac{1}{u^{2l}}\\
    \nonumber
  &+\dfrac{1}{8\pi^2}\cosh(\pi\mu)\Gamma(1/4+i\mu/2) \Gamma(1/4-i\mu/2)\,f_+(v)\sum_{l=0}^\infty q_l\,\dfrac{1}{u^{2l+1}}\\ 
  \nonumber
    & +\sum_{l=0}^\infty Y_l(u,v)\dfrac{1}{v^{2l+1}}\,,
\end{align}
where 
\begin{align}
   p_l &=g^2\dfrac{2^{2l-1}\Gamma\left(\frac{1}{4}+l+\frac{i\mu}{2}\right)\Gamma\left(\frac{1}{4}+l-\frac{i\mu}{2}\right)}{\Gamma(1+2l)}\,
   \left(H_{2l}+H_{-\frac{3}{4}+\frac{i\mu}{2}}-H_{-\frac{3}{4}+l+\frac{i\mu}{2}}+(\mu \to -\mu)\right)\,,\\ \nonumber
   q_l &=g^2\dfrac{2^{2l} \Gamma\left(\frac{3}{4}+l+\frac{i\mu}{2}\right) \Gamma\left(\frac{3}{4}+l-\frac{i\mu}{2}\right)}{\Gamma(2l+2)}\,\left(-1+H_{2l+1}+H_{-\frac{1}{4}+\frac{i\mu}{2}}-H_{-\frac{1}{4}+l+\frac{i\mu}{2}}+(\mu \to - \mu)\right)\,,
\end{align}
and
\begin{align}
  \label{Ycoef}
 Y_l(u,v)&= -g^2\frac{\,4^l\Gamma \left(\frac{5}{4}+l +\frac{i \mu }{2}\right) \Gamma \left(\frac{5}{4}+l -\frac{i \mu }{2}\right) }{  \Gamma \left(\frac{5}{4}+\frac{i \mu
   }{2}\right)  \Gamma \left(\frac{5}{4}-\frac{i \mu
   }{2}\right) \Gamma (2 l+3)}\dfrac{u}{v}\\ \nonumber  
  &+g^2\dfrac{\sqrt{\pi}u^2}{8v^2}\Gamma\left(\frac{7}{4}+l+\frac{i \mu }{2}\right) \Gamma\left(\frac{7}{4}+l-\frac{i \mu }{2}\right) \,
   _5\tilde{F}_4\left( \begin{array}{l}
   1,1,\frac{3}{2},l-\frac{i \mu }{2}+\frac{7}{4},l+\frac{i \mu
   }{2}+\frac{7}{4}\\ 
   \frac{7}{4}-\frac{i \mu }{2},\frac{i \mu
   }{2}+\frac{7}{4},l+2,l+\frac{5}{2}
   \end{array}
;\frac{u^2}{v^2}\right)\\ \nonumber
     & -g^2 \dfrac{\sqrt{\pi}u^3}{8v^3}
\Gamma\left(\frac{9}{4}+l+\frac{i \mu }{2}\right) \Gamma\left(\frac{9}{4}+l-\frac{i \mu }{2}\right)  \,
   _5\tilde{F}_4\left(
   \begin{array}{l}
    1,\frac{3}{2},2,l-\frac{i \mu }{2}+\frac{9}{4},l+\frac{i \mu
   }{2}+\frac{9}{4}      \\
    \frac{9}{4}-\frac{i \mu }{2},\frac{i \mu
   }{2}+\frac{9}{4},l+\frac{5}{2},l+3     
   \end{array}
   ;\frac{u^2}{v^2}\right)\,,
\end{align}
with $_5\tilde{F}_4$ the regularized hypergeometric function, which is regular on the entire unit circle for the parameters here.
\begin{tcolorbox}[toggle enlargement=none, colback=white]
\underline{\textit{Series coefficients}}. For $n=\text{odd}$, the only non-zero components are given by, 
\small
\begin{align}
\label{B-2ln}
      B_{-2l,n} &=\frac{g^2}{8\pi^2}\,\frac{2^{2l+n}\cosh(\pi\mu)}{\Gamma(2l+1)\Gamma(n+1)}\,\Gamma\left(\frac{1}{4}+l+\frac{i\mu}{2}\right) \Gamma\left(\frac{1}{4}+l-\frac{i\mu}{2}\right) \\ \nonumber
    &\qquad \times  \Gamma\left(\frac{1}{4}+\frac{n}{2}+\frac{i\mu}{2}\right) \Gamma\left(\frac{1}{4}+\frac{n}{2}-\frac{i\mu}{2}\right)\,,\,\,l\geq 0\,, \\ \nonumber
       &\\  
           A_{2l,n}&=\frac{g^2}{4}\frac{2^{n-2l}\Gamma(2l)}{\Gamma(n+1)}\frac{\Gamma(\frac{1}{4}+\frac{n}{2}+\frac{i\mu}{2}) \Gamma(\frac{1}{4}+\frac{n}{2}-\frac{i\mu}{2})}{\Gamma(\frac{3}{4}+l+\frac{i\mu}{2}) \Gamma(\frac{3}{4}+l-\frac{i\mu}{2})}\,,\quad 1\leq l\leq \frac{n-1}{2}\,, \\ \nonumber
          &\\   
      A_{-2l,n} &=-\frac{g^2}{16 \pi^2}\frac{\cosh(\pi \mu)2^{2l+n}}{\Gamma(2l+1)\Gamma(n+1)}\Gamma\left(\frac{1}{4}+l+\frac{i\mu}{2}\right) \Gamma\left(\frac{1}{4}+l-\frac{i\mu}{2}\right) \\ \nonumber
    & \times  \Gamma\left(\frac{1}{4}+\frac{n}{2}+\frac{i\mu}{2}\right) \Gamma\left(\frac{1}{4}+\frac{n}{2}-\frac{i\mu}{2}\right) \left(-H_{2l}-H_{-\frac{3}{4}+\frac{i\mu}{2}}+H_{-\frac{3}{4}+\frac{i\mu}{2}+l}+(\mu \to -\mu)\right)\,,\,\, l\geq 0\,,
\end{align}
where $H_\nu$ are Harmonic numbers.\\ 

\noindent Similarly, for $n=\text{even}$ we find:  
\small
\begin{align}
\label{Boddeven}
 B_{-2l-1,n} &=\dfrac{g^2}{4\pi^2 }\dfrac{2^{2l+n}\cosh(\pi\mu)}{\Gamma(2l+2)\Gamma(n+1)}
\Gamma\left(\frac{3}{4}+l+\frac{i\mu}{2}\right) \Gamma\left(\frac{3}{4}+l-\frac{i\mu}{2}\right) \\ \nonumber
    & \times \Gamma\left(\frac{1}{4}+\frac{n}{2}+\frac{i\mu}{2}\right) \Gamma\left(\frac{1}{4}+\frac{n}{2}-\frac{i\mu}{2}\right) \,,\quad l \geq 0\,,\\ \nonumber 
 &\\ 
 A_{2l+1,n} &=-\dfrac{g^2}{4}\dfrac{\,2^{2l+n}l \Gamma(2l)}{\Gamma(n+1)}\dfrac{\Gamma(\frac{1}{4}+\frac{n}{2}+\frac{i\mu}{2})\Gamma(\frac{1}{4}+\frac{n}{2}-\frac{i\mu}{2})}{\Gamma(\frac{5}{4}+l+\frac{i\mu}{2}) \Gamma(\frac{5}{4}+l-\frac{i\mu}{2})}\,, \quad 0<l\leq \frac{n-2}{2}\,,\\ \nonumber
 &\\ 
  A_{1,n} &=-\dfrac{g^2}{8}\dfrac{2^{n}}{\Gamma(n+1)}\dfrac{\Gamma(\frac{1}{4}+\frac{n}{2}+\frac{i\mu}{2}) \Gamma(\frac{1}{4}+\frac{n}{2}-\frac{i\mu}{2})}{\Gamma(\frac{5}{4}+\frac{i\mu}{2}) \Gamma(\frac{5}{4}-\frac{i\mu}{2})}\,,\\ \nonumber
  & \\   
  \label{A-2l-1n}
  A_{-2l-1,n} &=-\frac{g^2}{8\pi^2}\frac{2^{2l+n}\cosh(\pi\mu)}{ \Gamma(2l+2)\Gamma(n+1)}
\Gamma\left(\frac{3}{4}+l+\frac{i\mu}{2}\right) \Gamma\left(\frac{3}{4}+l-\frac{i\mu}{2}\right) \Gamma\left(\frac{1}{4}+\frac{n}{2}+\frac{i\mu}{2}\right)\\ \nonumber
 & \times  \Gamma\left(\frac{1}{4}+\frac{n}{2}-\frac{i\mu}{2}\right)\left(1-H_{2l+1}-H_{-\frac{1}{4}+\frac{i\mu}{2}}+H_{-\frac{1}{4}+l+\frac{i\mu}{2}}+(\mu \to -\mu)\right)\,,\,\, l \geq 0\,.
 \end{align}
\end{tcolorbox}
\noindent The merit of the expression \eqref{simplifiedsum} is that, unlike the original series \eqref{ansatz2}, the dependence on $u/v$ is fully resummed. This will be especially useful in computing the power spectrum, for which we set $u=v=1/c_s$, or for the $t-$ and $u-$channel contributions to the squeezed limit bispectrum (with $k_3\to 0$) where $u/v$ approaches unity. 

\subsection{Fixing the homogeneous solution}  
\label{fixinghom}

Having derived the particular solution to the boundary equation for $\hat{F}_{++}$, we now move to the freedom in adding to it any solution of the homogeneous differential equations. It is crucial to observe that the particular solution derived above cannot describe the entire $\hat{F}_{++}$ for a few reasons: $(i)$ once continued to the $|u|>|v|$ region, $\hat{F}_p$ is not smooth at $u=v$, $(ii)$ $\hat{F}_p$ is plagued by a spurious pole at $u=1$, and $(iii)$ it does not satisfy our cutting rule \eqref{cuttingrule}. Below we demonstrate that imposing regularity at $u=1$ and the cutting rule totally determines the homogeneous solution. Therefore, the regularity of the final answer at $u=v$ will be an automatic output.\\

\noindent \textit{\textbf{Cutting rule}}.  Incorporating the homogeneous solutions to the boundary equations ${\cal O}(u,\partial_u) \hat{F}_{h}={\cal O}(v,\partial_v) \hat{F}_{h}=0$, the most general ansatz for $F_{++}$ becomes:
\begin{align}
\label{fullFpp}
    \hat{F}_{++}(u,v)=\sum_{m,n}\bigg(a_{m,n}+b_{m,n}\log(u)\bigg)\dfrac{u^{n-m}}{v^n}+\sum_{\pm\pm} \beta_{\pm\pm}f_{\pm}(u)f_{\pm}(v)\,,\quad 1<|u|<|v|\,,
\end{align}
where $\beta_{\pm\pm}$ are four free parameters that we will identify later. It will be sufficient to exploit the cutting rule \eqref{cuttingrule} across the following domain: 
\begin{equation}
{\cal D}\equiv \left\lbrace (u,v) |u=\text{Re}(u)+i\epsilon,v=\text{Re}(v)+i\epsilon, 1<|u|<|v|\right\rbrace\,,
\end{equation}
within which the $f_\pm(u)$ basis functions display the following properties:  
\begin{align}
\label{fpmprop}
    &f_+^*(-u+i\epsilon)=f_+(u+i\epsilon)\,,\quad f_-^*(-u+i\epsilon)=-f_-(u+i\epsilon)\,,\quad \text{Im}(f_\pm(u))=0\,. 
\end{align}
Using these equalities together with the expression for $f_3$ in \eqref{explicit-f3}, the cutting rule can be recast into 
\begin{align}
   & \hat{F}_{++}(u-i\epsilon,v-i\epsilon)+\hat{F}^*_{++}(-u-i\epsilon,-v-i\epsilon)=\\ \nonumber
& -\dfrac{g^2}{8\pi}\left(\Gamma(1/4+i\mu/2)^2 \Gamma(1/4-i\mu/2)^2 f_+(u)f_+(v)-\Gamma(3/4+i\mu/2)^2 \Gamma(3/4-i\mu/2)^2 f_-(u)f_-(v)\right)\,.
\end{align}
It can be viewed that all the $a_{mn}$ elements in $\hat{F}_{++}$ disappear from the LHS of the cutting rule above. The logarithmic pieces partially cancel against each other, leaving behind a residual term that survives due to the simple fact that 
\begin{align}
\nonumber
    \log^*(-u+i\epsilon)=-i\pi+\log(u)\,. 
\end{align}
Putting everything together and equating the coefficients of the $f_\pm(u)f_\pm(v)$ terms on both sides of the cutting rule, we find 
\begin{align}
    &\text{Im}(\beta_{+-})=\text{Im}(\beta_{-+})=-\dfrac{\pi g^2}{8}\,,\\ 
     &\text{Re}(\beta_{++})=-\dfrac{g^2}{16\pi} \Gamma(1/4+i\mu/2)^2 \Gamma(1/4-i\mu/2)^2\,,\\ 
    & \text{Re}(\beta_{--})=\dfrac{g^2}{16\pi} \Gamma(3/4+i\mu/2)^2 \Gamma(3/4-i\mu/2)^2\,.
    \end{align}
The real parts of $\beta_{-+}$ and $\beta_{+-}$ and the imaginary parts of $\beta_{++}$ and $\beta_{--}$ are so far arbitrary. They will be dictated by requesting the regularity of $\hat{F}_{++}$ in the collinear limit. 
\\

\noindent \textit{\textbf{Cancellation of the collinear singularity}}. The ansatz \eqref{fullFpp} exhibits a spurious pole at $u=1$ unless we appropriately tune the parameters $\beta_{\pm\pm}$. The potential singularity stems from $(i)$ the logarithmic divergence in the basis functions $f_\pm$ given in Eq.~\eqref{fpmcollinear}, and $(ii)$ the last term in \eqref{simplifiedsum} involving $A_{k,n}$ elements with $k\geq 0$. Near $u=1$, the latter behaves as
\begin{align}
\label{logpar}
    \sum_{m,n} a_{mn}\dfrac{u^{n-m}}{v^n}\sim\log(u-1) \left(c_1(\mu)f_-(v)+c_2(\mu)f_+(v)\right)\,,
\end{align}
where
\begin{align}
    & c_1(\mu)=-\frac{1}{2} g^2\dfrac{\cosh(\pi\mu)}{8\pi^{3/2}}\Gamma(3/4+i\mu/2) \Gamma(3/4-i\mu/2)\,(H_{-3/4+i\mu/2}+\log(2)+(\mu \to -\mu))\,, \label{c1}\\ 
    & c_2(\mu)=-\frac{1}{2} g^2\dfrac{\cosh(\pi\mu)}{8\pi^{3/2}}\Gamma(1/4+i\mu/2) \Gamma(1/4-i\mu/2) \,(-1+H_{-1/4+i\mu/2}+\log(2)+(\mu \to -\mu))\,. \label{c2}
\end{align}
Asking the cancellation of the logarithmic divergence in the particular solution \eqref{logpar} against the one in the homogeneous part (the last term in \eqref{fullFpp}) we arrive at
\begin{align}
       \text{Re}\beta_{+-}&=-\dfrac{\pi g^2}{8\cosh(\pi\mu)}-\dfrac{g^2}{8}\bigg(H_{-3/4+i\mu/2}+\log(2)+(\mu \to -\mu)\bigg)\,,\\ 
    \text{Re}\beta_{-+}&=\dfrac{\pi g^2}{8\cosh(\pi\mu)}-\dfrac{g^2}{8}\bigg(-1+H_{-1/4+i\mu/2}+\log(2)+(\mu \to -\mu)\bigg)\,,\\  
\Im(\beta_{--})&=\dfrac{\pi g^2}{8} \frac{\Gamma(\frac{3}{4}+\frac{i\mu}{2}) \Gamma(\frac{3}{4}-\frac{i\mu}{2})}{\Gamma(\frac{1}{4}+\frac{i\mu}{2}) \Gamma(\frac{1}{4}-\frac{i\mu}{2})}\,, \\
\Im(\beta_{++})&=\dfrac{\pi g^2}{8}\frac{\Gamma(\frac{1}{4}+\frac{i\mu}{2}) \Gamma(\frac{1}{4}-\frac{i\mu}{2})}{\Gamma(\frac{3}{4}+\frac{i\mu}{2}) \Gamma(\frac{3}{4}-\frac{i\mu}{2})}\,.
\end{align}
In summary, we identified all real and imaginary components of the free parameters $\beta_{\pm\pm}$ by imposing the cutting rule and the regularity at the collinear limit. Since there is no more free parameter in \eqref{fullFpp}, it must have all the other properties that $\hat{F}_{++}$ is supposed to possess. Specifically, we demonstrate in Appendix \ref{singularity-structure-appendix} that our solution has the anticipated singularities when the total or the partial energies vanish. 

\subsection{Full correlator}

For real values of energies, i.e. $u,v\in \mathbb{R}^+$, the full correlator $\hat{F}$ is twice the real part of the sum of $F_{++}$ \eqref{fullFpp} and $F_{+-}$ \eqref{factorized}. Using also \eqref{explicit-f3} and after some algebra, the final answer simplifies to
\begin{tcolorbox}[colframe=white,arc=0pt]
\begin{align}
\label{correlatorfull}
    \hat{F} &=2\sum_{m,n} a_{m,n}\dfrac{u^{n-m}}{v^n} +\dfrac{g^2}{2}\bigg(f_+(u)f_-(v)+f_-(u)f_+(v)\bigg)\log(u) \\ \nonumber 
    & +\dfrac{g^2}{4\pi} 
    \Gamma^4\left(\frac{3}{4}+\frac{i\mu}{2}\right) \Gamma^4\left(\frac{3}{4}-\frac{i\mu}{2}\right)\,f_-(u)f_-(v)   
    \\ \nonumber 
    &
   -\frac{g^2}{4}\bigg(\log(2)-1+H_{-\frac{1}{4}+\frac{i\mu}{2}}+(\mu \to -\mu) \bigg)f_-(u)f_+(v)\\ \nonumber
    &     -\frac{g^2}{4}\left(\log(2)+ H_{-\frac{1}{4}+\frac{i\mu}{2}}+(\mu \to -\mu) \right) f_+(u)f_-(v)
    \,,\quad u,v\in \mathbb{R}\,,\quad 1<u<v\,
\end{align}
\end{tcolorbox}
\noindent where the first term has the convenient (partial) resummation \eqref{simplifiedsum}. Given that $f_\pm(u)$ do not display any discontinuity across the interval $(1,\infty)$, we did not have to specify the $i\epsilon$ prescription within their arguments above.\\

\noindent In addition to \eqref{correlatorfull}, we also need the correlator within the unit disk. This region becomes of particular interest when we evaluate the three-point function of $\pi$ in the ultra-squeezed limit $\frac{k_\L}{2k_\S}\ll c_s$. Within the domain of $u<v<1$, $\hat{F}$ was given in \cite{Arkani-Hamed:2018kmz} as a double series in powers of $u$ and $u/v$, plus a specific homogeneous solution to the boundary equation. Below, we quote the expression for $\hat{F}$, except that here we expand the homogeneous solution $g_h(u,v)$ in terms of our basis functions $f_\pm$
\begin{align}
\label{fullF}
      \hat{F}= \sum_{m,n=0}^\infty c_{m,n}u^{2m+1}\,\left(\frac{u}{v}\right)^n+\dfrac{\pi g^2}{2 \cosh (\pi \mu)}g_h(u,v)\, \quad 0<u<v<1\,.
\end{align}
where
\begin{align}
    g_h(u,v)&=-\dfrac{i}{2}\cosh(\pi\mu)\dfrac{\Gamma(\frac{1}{4}+\frac{i\mu}{2}) \Gamma(\frac{1}{4}-\frac{i\mu}{2})}{\Gamma(\frac{3}{4}+\frac{i\mu}{2}) \Gamma(\frac{3}{4}-\frac{i\mu}{2})}f_+(u+i\epsilon)f_+(v+i\epsilon)\\ \nonumber
    &+\left(1-\dfrac{i}{2}\cosh(\pi\mu)\right)\dfrac{\Gamma(\frac{3}{4}+\frac{i\mu}{2}) \Gamma(\frac{3}{4}-\frac{i\mu}{2})}{\Gamma(\frac{1}{4}+\frac{i\mu}{2}) \Gamma(\frac{1}{4}-\frac{i\mu}{2})}f_-(u+i\epsilon)f_-(v+i\epsilon)\\ \nonumber  
    &+\dfrac{i}{2}\cosh(\pi\mu)f_+(u+i\epsilon)f_-(v+i\epsilon)-\left(1-\dfrac{i}{2}\cosh(\pi\mu)\right)f_-(u+i\epsilon)f_+(v+i\epsilon)\,, \\ \nonumber
    c_{mn}&=\dfrac{(-1)^n (n+1)(n+2)\dots (n+2m)}{[(n+\frac{1}{2})^2+\mu^2][(n+\frac{5}{2})^2+\mu^2]\dots [(n+\frac{1}{2}+2m)^2+\mu^2]}\,.
\end{align}
Two remarks are in order about this result. First,  due to the discontinuity of the basis functions across $u,v\in [0,1]$, the $i\epsilon$ term had to be inserted inside the arguments of $f_\pm$. Secondly, the expression \eqref{fullF} can be readily used even if $v$ was bigger than one, as long as $u$ remains less than $v$ hence guaranteeing the convergence of the series. In summary, the formula \eqref{fullF} together with \eqref{correlatorfull} defines the correlator $\hat{F}(u,v)$ across the entire span of the $(u,v)$ space.

\subsection{Asymptotic limits of the seed correlator}
\label{asymptlimitseed}

\subsubsection*{Large energy ratios}
In the $z\gg \text{max}\lbrace 1,\mu\rbrace$ regime, $f_\pm$ can be approximated by the series expansion
\begin{align}
\label{TaylorfP}
       f_+(z)&=1+\dfrac{1}{z^2}\left(\frac{\mu ^2}{2}+\frac{1}{8}\right)+\dots\,, \\ 
   f_-(z) &=\frac{2}{z}+\dfrac{1}{z^3}\left(\frac{\mu ^2}{3}+\frac{3}{4}\right)+\dots\,, \label{TaylorfM}
\end{align}
Inserting the above expressions in \eqref{correlatorfull} and keeping only the leading order terms in $u^{-1}$ and $v^{-1}$, we arrive at
\begin{tcolorbox}[colframe=white,arc=0pt]
\begin{align}
\label{Flimit}
    \hat{F}_{\text{asymp}}&=g^2\left(\dfrac{1}{u}+\dfrac{1}{v}\right)\left(\log(\frac{1}{C(\mu)}\dfrac{u v}{u+v})+1-\gamma_E\right),\quad u,v\in \mathbb{R}^+\,, u,v \gg \text{max}\lbrace 1,\mu\rbrace 
\end{align} 
\end{tcolorbox}
\noindent where we have defined
\begin{align}
   C(\mu)=2\exp(\frac12H_{-1/4+i\mu/2}+\frac12H_{-1/4-i\mu/2}-\gamma_E)\,.
\end{align}
\noindent This simple formula for $\hat{F}_{\text{asympt}}$ will subsequently generate analytical expressions for the correlators of $\pi$ (associated with diagrams A, B1 and B2) in the $c_s\ll 1$ regime. Notice that the usual non-analyticities due to particle production, which enter the correlator through oscillatory factors such as $u^{\pm i\mu}$, are absent in $\hat{F}_{\text{asympt}}$. This owes to the fact that we are expanding the correlator around $u,v=\infty$, while $u/v$ is held fixed. In contrast, the branch cut attributed to the particle production is visible only in the vicinity of the origin (i.e. $u$ or $v$ equal to zero), as we will review shortly.\\ 

\noindent It can be verified that there is no contribution to $\hat{F}$ at order $1/u^2$ or $1/v^2$, and the first correction to \eqref{Flimit} arises at order ${\cal O}(\frac{1}{u v})$, given by 
\begin{align}
\label{NLOF}
    \Delta \hat{F}_2=g^2\dfrac{\Gamma(\frac{3}{4}+i\mu) \Gamma(\frac{3}{4}-i\mu)}{\pi u v}\,. 
\end{align}
As long as $u$ and $v$ are large, this correction remains small since it does not grow with $\mu$ (in fact,  $\Gamma(\frac{3}{4}+i\mu) \Gamma(\frac{3}{4}-i\mu)< \Gamma(\frac{3}{4})^2$ for real $\mu$).
However, the NNLO term, namely the cubic order terms in inverse powers of $u$ and $v$, eventually dominates over $\hat{F}_{\text{asymp}}$ for $\mu\gtrsim \text{max}\lbrace u,v\rbrace$, invalidating the asymptotic formula above (see Equation \eqref{deltaF3} in Appendix \ref{appendix:F-asymptotic}). \\

\begin{figure}
    \centering
    \includegraphics[scale=0.5]{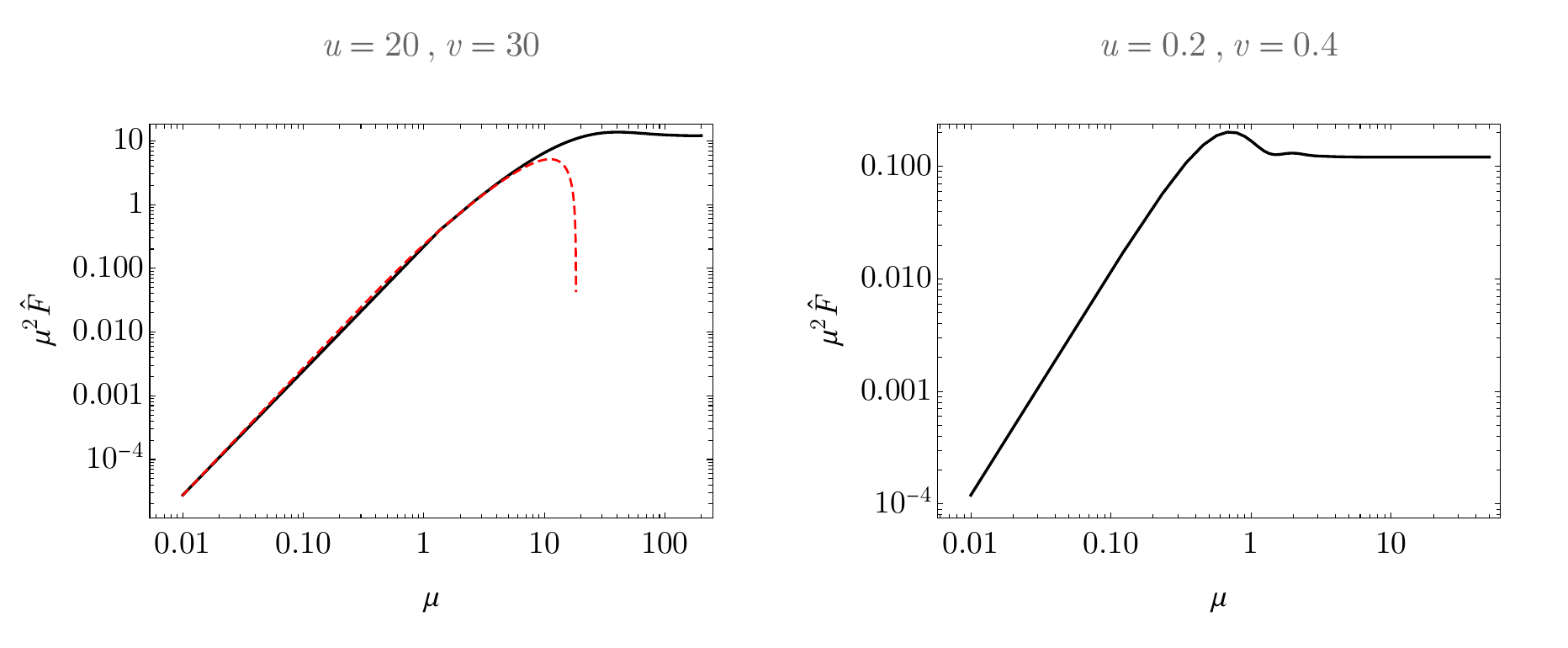}
    \caption{The four-point function $\hat{F}(u,v)$ multiplied by mass squared as a function of $\mu$, for fixed values of $u$ and $v$. \textit{Left}: The dashed line indicates $F_{\text{asympt}}$ \eqref{Flimit}. The $1/\mu^2$ behaviour of $\hat{F}(u\gg 1,v\gg 1)$ is not reached before $\mu$ grows larger than $\text{max}(u,v)$. \textit{Right}: In contrast to the previous case, when $u$ and $v$ are both within the unit disk, $\hat{F}$ starts to decay as $1/\mu^2$ for $\mu\gtrsim 1$. In both diagrams we have normalized to $g=1$.}
    \label{fig:fmu12}
\end{figure}

\noindent Notice that the mass of $\s$ enters the asymptotic correlator \eqref{Flimit} only through the $C(\mu)$ factor, which goes as
\begin{align}
    \lim_{\mu\gg 1} C(\mu)=\mu-\dfrac{1}{24\mu}+{\cal O}(\mu^{-2})
    \label{C-mu-large-mass}
\end{align}
in the large mass limit. Actually, this behaviour is accurate already for $\mu \gtrsim 1$, while for smaller values, $C(\mu)$ deviates from this behaviour to monotonously reach the constant value $0.68$ at $\mu=0$. All in all, one can qualitatively remember that for all masses $m \geq 3/2 H$ as relevant here, $C(\mu)$ can be thought of as simply $ \approx m/H$.
Therefore, according to \eqref{Flimit}, intermediate heavy fields that are still lighter than $\text{max}\lbrace u,v\rbrace \times H$ induce four point-functions that vary with the mass but only logarithmically --- they are not suppressed by the inverse power of mass squared, nor by the Boltzmann factor $\exp(-\pi\mu)$ that characterises the particle production effects in dS space. For very heavy particles, the $\mu^{-2}$ decline in the correlator is expected from an EFT standpoint: once $\sigma$ is integrated out (at tree-level and at leading order in derivatives to yield a local EFT) it can impact the correlators of $\vpi$ only through the quartic EFT operator $\frac{g^2}{H^2(\mu^2+9/4)}\vpi^4$, leading to the anticipated $\mu^{-2}$ decay.  
In fact, we recover this behaviour in the large mass limit once we include the corrections to $\hat{F}_{\text{asympt}}$ which become important for $\mu\gg \text{max}\lbrace u,v\rbrace $, as illustrated in Figure \ref{fig:fmu12}. In contrast, for the ordinary case of $u$ and $v$ both lying within the unit disk, according to Figure \ref{fig:fmu12} the correlator starts to decay as $\mu^{-2}$ as soon as $\mu$ grows larger than unity.
Naturally, these observations, formulated in term of the seed four-point correlator, matches the discussion on the two qualitatively different regimes of the exchanged field being lighter or heavier than $H/c_s$ in section \ref{action-motivations}, and that will be further elaborated upon in sections \ref{sec:correlators} and \ref{non-localEFT}.

\subsubsection*{Small energy ratio $u\ll 1$}
The ultra-squeezed configuration of the bispectrum, such that $\frac{k_\L}{k_\S}\ll c_s$, can be deduced from $\hat{F}$ by sending $u\to 0$ while keeping the second argument at the fixed value $v=1/c_s$. This limit corresponds to a collapsed quadrilateral with two of its adjacent sides approaching zero. In this kinematical limit, the four-point $\hat{F}$ becomes entirely dominated by the homogeneous solution $g_h$ in \eqref{fullF}, and is given by
\begin{align}
\label{squeezedF}
    \lim_{u\to 0, v=1/c_s}\,\hat{F} &=\dfrac{\pi g^2}{2\cosh(\pi\mu)}\sum_{\pm}\xi_\pm(c_s,\mu)\,u^{\frac{1}{2}\pm i\mu}\,,
\end{align}
where 
\begin{align}
\label{def-xi}
\xi_-&=\dfrac{1}{\pi}\dfrac{ 2^{-\frac{3}{2}+i\mu}}{\Gamma(\frac{1}{2}+i\mu)}(i \sinh(\pi \mu)-1) \Gamma(i\mu)\\ \nonumber
     &\times  \left[
    \Gamma\left(3/4+i\mu/2 \right)  \Gamma\left(3/4-i\mu/2 \right) f_-(c_s^{-1}+i\epsilon)-\Gamma\left(1/4+i\mu/2\right) \Gamma\left(1/4-i\mu/2\right) f_+(c_s^{-1}+i\epsilon)\right]\,, \\ \nonumber
    \xi_+&=\xi_- (\mu\to -\mu)\,.
    \end{align}
In the asymptotic form \eqref{squeezedF}, the non-analytic dependence on the energy ratio $u$ through the oscillatory phases $u^{\pm i\mu}$ is the famous hallmark of particle production in de Sitter spacetime. We will discuss the dependence of this signal on $c_s$ and $\mu$ in Section \ref{sec:CC-oscillations}.

\section{Inflationary correlators and the low speed collider}
\label{sec:correlators}

\subsection{Power spectrum}

\begin{figure}
    \centering
    \includegraphics[scale=0.54]{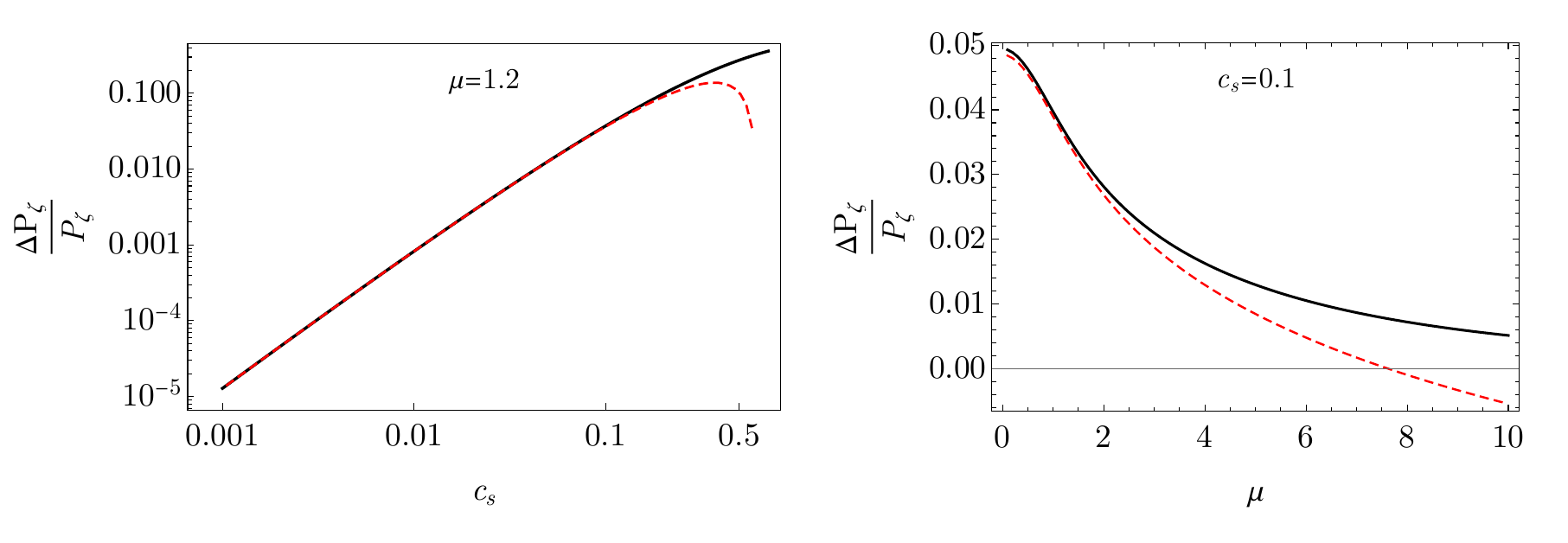}
    \caption{\textit{Left}: The fractional shift in the power spectrum of $\zeta$ (due to the exchange of $\s$) as a function of the speed of sound. The black curve is the exact result \eqref{diagmA}, and the dashed red is the asymptotic behaviour
    \eqref{powerasymp} for a small enough mass $c_s m/H \ll 1$.
    \textit{Right}: The same quantities for different values of $\mu$. As expected, for masses in Hubble units of order $1/c_s$ and heavier (in this case $\mu\sim 10$) the small sound speed results starts to deviate from the exact one. In both diagrams we have set $\rho/H=1$.}
    \label{fig:power}
\end{figure}

The power spectrum for arbitrary $c_s$ is analytically given by Equation \eqref{diagmA} and is plotted in Figure \ref{fig:power} as a function of $c_s$ and $\mu$. In the regime $c_s m/H \ll 1$, the fractional shift in the power spectrum of $\zeta$ induced by the exchange of $\s$ follows from \eqref{diagmA} and \eqref{Flimit}:
\begin{align}
\label{powerasymp}
    \dfrac{\Delta P_\zeta}{P_\zeta}&=\dfrac{2\rho^2 c_s^2}{H^2}\left(\log(\frac{1}{2c_s C(\mu)})+1-\gamma_E\right)\quad \textrm{for} \quad c_s \frac{m}{H} \ll 1\,.
\end{align}
The fractional correction to the power spectrum has been computed numerically in \cite{Lee:2016vti} from the bulk picture, and our analytical formula agrees with the results there (figure 5). Moreover, it has been observed to vanish more rapidly than $c_s$ in the low sound speed limit. Our analytical result \eqref{powerasymp} confirms this, and additionally provides one with the corresponding $c_s$-dependence analytically, including the unusual logarithmic dependence. Just like the asymptotic limit of the four-point seed function \eqref{Flimit}, the correction to the power spectrum \eqref{powerasymp} decreases with $m/H$ only logarithmically. Nevertheless, as we discussed in Section \ref{asymptlimitseed}, this growth eventually turns into the $1/\mu^2$ fall off behaviour once we consider the order one corrections to \eqref{Flimit} (and the subsequent correction to the above formula) that arise for $\mu\gtrsim 1/c_s$ (see Figure \ref{fig:power}). \\

\noindent The logarithmic mass dependence of the power spectrum is a sign that, within the mass range $1\ll m/H\ll 1/c_s$, the intermediate heavy field cannot be integrated out to yield a local action with terms suppressed by powers of $1/m^2$. At first glance, this might seem at odds with the ordinary EFT reasoning: the energy scale at which the cosmological correlators are generated is Hubble, and we expect heavier degrees of freedom to be irrelevant. This argument has an important caveat that we can integrate out only those fields that are non-relativistic at the time of sound-horizon crossing. Fluctuations of $\pi$ at this time have energies of order Hubble. However, the gradient energy of the $\sigma$ fluctuations with the same spatial scale is of order $H/c_s$. This shows that every degree of freedom lighter than $H/c_s$ should be kept in the EFT as a dynamical field simply because, at the sound-horizon crossing, its gradient energy is comparable to its mass, namely it is relativistic. The origin of the logarithmic mass dependence of the power spectrum will be transparent within the non-local EFT studied in section \ref{non-localEFT}.

\subsection{Bispectrum}

\subsubsection{Generic configurations}

Acting with the weight-shifting operators given by \eqref{diagb1b2} on the asymptotic form of the four-point function $\hat{F}(u,v)$ in \eqref{Flimit}, one can obtain the bispectra associated with Diagrams B1 and B2, up to leading order in $c_s$. After summing over all three channels (in other words, symmetrising among $k_1$, $k_2$ and $k_3$), we find 
\begin{tcolorbox}[colframe=white,arc=0pt]
\begin{align}
     B^{\text{B1}}_{\text{low speed}}
     (k_1,k_2,k_3)&=(-\alpha_1 c_s^2)\dfrac{e_2^2-2e_1 e_3}{e_1 e_3^3}=(-\alpha_1 c_s^2) \dfrac{\sum_{i< j} k_i^2 k_j^2}{(k_1 k_2 k_3)^3 (k_1+k_2+k_3)}\,, \label{asymptB1} \\ 
     B^{\text{B2}}_{\text{low speed}}(k_1,k_2,k_3)&=\left(\dfrac{\alpha_2 c_s}{2}\right)\dfrac{1}{e_3^3 e_1} \Big[ -(\gamma_E+1) e_1^4+(3+4\gamma_E)e_1^2 e_2 \label{asymptB2}\\
     & \quad -2e_2^2-(6\gamma_E+2) e_1 e_3-e_1\sum_{a=1}^3 k_a(e_1^2-2e_2-2k_a^2)\log(\frac{k_a}{c_s C(\mu)\,e_1})\Big]\nonumber
\end{align}
\end{tcolorbox}
\noindent where $e_i$'s are the symmetric polynomials, i.e.  
\begin{align}
\nonumber
    e_1=k_1+k_2+k_3\,,\qquad e_2=k_1 k_2+k_1 k_3+k_2 k_3\,,\quad e_3=k_1 k_2 k_3\,.
\end{align}

\noindent In order to derive the above formulae we assumed that, for all channels, the energy ratio $u$ (which is equal to $\frac{k_3}{c_s(k_1+k_2)}$ for the $s-$channel) is much bigger than $\mu$ and unity, whichever is the maximum. This condition is equivalent to: 
\begin{align}
    c_s \frac{m}{H} \ll \text{min}\bigg\lbrace \frac{k_1}{k_2+k_3},\frac{k_2}{k_1+k_3},\frac{k_3}{k_1+k_2}\bigg\rbrace\,. 
\end{align}
In other words, the asymptotic forms of the bispectra presented above are applicable only in the regime $c_s m/H \ll 1$ of particular interest, and for not too squeezed triangles, i.e. it holds for $k_\L/k_\S \gg c_s m/H$.
We will see later how the bispectrum behaves when these conditions are not met, first in the ultra-squeezed configurations in \ref{sec:CC-oscillations}, second in the complementary region of extended equilateral configurations ${\cal O}(1)c_s \frac{m}{H}\lesssim k_\L/k_\S \leqslant 1$, in sections \ref{sec:resonances} and \ref{non-localEFT}, covering all values of $c_s$ and of $m/H$.

\begin{figure}
    \centering
    \includegraphics[scale=0.53]{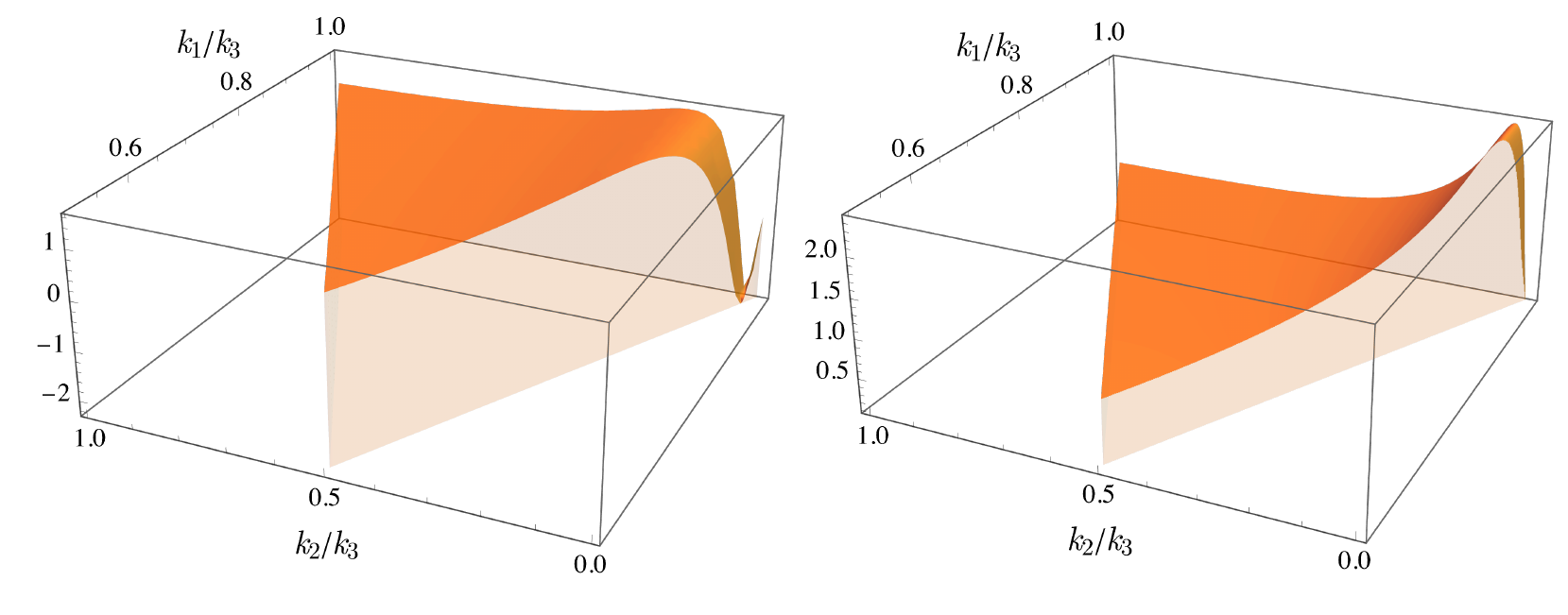}
    \caption{The shapes of the bispectra $S(k_1/k_3,k_2/k_3)$, normalized to unity in the equilateral limit ($k_1=k_2=k_3$), for Diagram B1 (\textit{left}) and Diagram B2 (\textit{right}). Here we have set $c_s=3\times 10^{-2}$ and $\mu=2$. Both the approximate local-shape behaviour Eqs.~\eqref{smallcsB1}-\eqref{smallcsB2}, and the resonances near squeezed configurations typical of the low speed collider are well visible.}
    \label{fig:Shapeasympt}
\end{figure}

Let us now highlight the interesting features of \eqref{asymptB1}-\eqref{asymptB2}: 
\begin{itemize}
    \item The low speed bispectrum associated with Diagram B1 does not depend on the intermediate mass at all. From the boundary point of view, this is so because the only mass-dependent combination in the asymptotic four-point function \eqref{asymptotic} is proportional to $(1/u+1/v)$ which gets annihilated by the weight-shifting operator $(2\partial_u+u\partial_u^2)$. In contrast, Diagram B2 in the $c_s\ll 1$ limit varies with mass through $C(\mu)$. 
    Nevertheless, much like the power spectrum, both diagrams start to decay like $1/\mu^2$ when the intermediate field becomes much heavier than $H/c_s$ (for which the $c_s\ll 1$ approximation above is not valid). The logarithmic mass dependence of the bispectrum finds a simple explanation in terms of the effective non-local single field theory which we discuss in detail in Section \ref{non-localEFT}. Equivalently, as we described intuitively in \ref{sec:setup-overview}, this logarithmic dependence can be seen as a consequence of an IR ``divergence'' between sound horizon crossing of the short mode and mass crossing of the long one. Given that $\partial_i\pi/a$ decays more slowly outside the sound horizon than $\dot{\pi}$ (like $\eta$ versus $\eta^2$), one can convince oneself from the bulk integrals that only diagram B2 is affected by that effect, explaining the difference in that respect between \eqref{asymptB1} and \eqref{asymptB2}.

    \item The $k_T=0$ singularity of the bispectrum directly follows from the total energy singularity of the four-point seed function (at $u=-v$). But since the above analytical formulae for the bispectra are very simple, as a non-trivial cross-check, here we directly look at their amplitude limit. Around the total energy pole ($k_T=c_s e_1=0$), both diagrams behave as 
    \begin{align}
    \label{total}
      \lim_{e_1\to 0}
     B^{\text{B1,B2}}_{\text{low speed}}\propto\dfrac{e_2^2}{e_1 e_3^3}\,.
    \end{align}
    This accords with the general relationship between the correlator and the associated amplitude given in Equation 4.35 of \cite{COT}, which reduces to  
    \begin{align}
    \label{BArel}
        \lim_{k_T\to 0}B_\zeta=\text{constant}\times \text{Re}\left(\dfrac{i A_3}{k_T^p k_1^2 k_2^2 k_3^2} \right)\,,
    \end{align}
    for the special case of the three-point function. Here $A_3$ stands for the three-particle amplitude due to the exchange of $\s$. To verify this equation, we first observe that in both cases the degree of the total energy singularity is unity (i.e. $p=1$). According to the formula \eqref{degree}, this simply follows from the dimensions of the vertices, namely $[\dot{\pi}_c\s]=3$ and $[\dot{\pi}_c^2 \s]=[(\partial_i \pi_c)^2\s]=5$. Second, 
    we need the flat space amplitudes associated with Diagrams B1 and B2 (summed over all channels), both of which simplify to
    \begin{align}
        \lim_{c_s\ll 1}A_3^{\text{B1,B2}}(k_1,k_2,k_3)=\text{constant}\times \dfrac{e_2^2}{e_3}\,,
    \end{align}
    at low speeds. Comparing this result with the right-hand side of \eqref{total} confirms the expected relation between the three-point function and the three particle amplitude \eqref{BArel}.  
    As an aside, notice that the proportionality of $A_3^{\text{B1}}$ and $A_3^{\text{B2}}$ was not an accident. It is due to the fact that the vertices $\dot{\pi}^2 \s$ and $(\partial_i \pi)^2 \s$ are related through the equation of motion of $\pi$ accompanied with the field redefinition $\pi\to \pi+\pi \sigma$. \\
    \item Unlike the ultra-squeezed regime (i.e. $k_\textrm{L}/k_\textrm{S}\ll c_s m/H$), equations \eqref{asymptB1}-\eqref{asymptB2} apply to the mildly-squeezed configurations, such that
    \begin{equation}
     c_s \frac{m}{H}\ll \frac{k_\L}{k_\S}\ll 1\,,\qquad \text{mildly-squeezed regime}\,. 
     \label{eq:mildly-squeezed}
    \end{equation}
    In this limit, one finds 
    \begin{align}
        \lim_{c_s m/H \ll \frac{k_\L}{k_\S}\ll 1}B^{\text{B1}}_{\text{low speed}}&=\dfrac{1}{2\pi} \left(\dfrac{c_s}{{\cal P}_\zeta}\right)^{\frac{1}{2}}\left(\dfrac{\rho}{\Lambda_{1}}\right)\,P_\zeta(k_\L) P_\zeta(k_\S)\left(1-\dfrac{k_\L}{2k_\S}+{\cal O}\left(\dfrac{k_\L^3}{k_\S^3}\right)\right) \label{smallcsB1}\\ 
        \lim_{c_s m/H \ll \frac{k_\L}{k_\S}\ll 1}B^{\text{B2}}_{\text{low speed}}&=\dfrac{-1}{2\pi} \left(\dfrac{c_s}{{\cal P}_\zeta}\right)^{\frac{1}{2}}\left(\dfrac{\rho}{\Lambda_{2}}\right) \label{smallcsB2}\\ 
        & \times P_\zeta(k_\L) P_\zeta(k_\S)\left(5+\dfrac{k_\L}{2k_\S}\left[19+4\gamma_E-4\log(\dfrac{k_\L}{4c_s\mu k_\S})\right]+{\cal O}\left(\dfrac{k_L^3}{k_S^3}\right)\right)\,. \nonumber
    \end{align}
     This behaviour, coinciding at leading order in $k_\L/k_\S$ with the one of the local shape, violates Maldacena's single-clock consistency condition \cite{maldacena2003non,Creminelli:2004yq}, implying that the asymptotic expressions \eqref{asymptB1}-\eqref{asymptB2} cannot be mimicked by any local cubic operator involving $\pi$ only. Naturally, this is consistent with the fact that for a small sound speed and in the window of mass $m/H \ll 1/c_s$, the heavy field $\sigma$ can be integrated out (see section \ref{non-localEFT}), albeit only to yield a Lagrangian that is \textit{non-local} in space.\\
     
     \noindent It is customary to define the shape function of the bispectrum, such that
      \begin{align}
    \label{shapefunctiondef}
    B_\zeta(k_1,k_2,k_3) = (2 \pi)^4 \frac{S(k_1,k_2,k_3)}{(k_1 k_2 k_3)^2} A_s^2
    \end{align}
    where $A_s$ denotes the amplitude of the curvature power spectrum, which in the scale invariant limit, and neglecting the small corrections from the exchange of the massive field,    simply coincides with ${\cal P}_\zeta$ in Eq.~\eqref{eq:standard-power-spectrum}. We will discuss the amplitudes of the non-Gaussian signals studied here in section \ref{sec:size-NG}, but their shapes, normalized to unity in the equilateral limit, can be seen in figure \ref{fig:Shapeasympt}. The behaviours \eqref{smallcsB1}-\eqref{smallcsB2}, implying a power-law growth proportional to $(k_S/k_L)$ in the mildly-squeezed regime, are readily visible. As we will discuss shortly, the corrections to \eqref{asymptB1}-\eqref{asymptB2} tame this growth as triangles become more squeezed, leaving behind  bump-like features around $k_L/k_S\sim  c_s m/H$, also well visible. 
\end{itemize}

\subsubsection{Ultra-squeezed configurations}
\label{sec:CC-oscillations}

In the ultra-squeezed limit $\frac{k_L}{2c_s k_S}\to 0$, the bispectra are found by acting with the weight-shifting operators \eqref{diagb1b2} on the asymptotic formula \eqref{squeezedF}. The overall behaviour of the ultra-squeezed bispectrum is not qualitatively different than that of the canonical case, in which the signal is characterised by oscillations in $\log(k_\L/k_\S)$ with frequency $\mu$ and an amplitude that decays as $(k_\L/k_\S)^{3/2}$. Nonetheless, the amplitude and the phases of the oscillations are subject to important modifications sensitive to $c_s$. The analytical formulae for the bispectra in this limit are given by: 
\begin{align}
\label{cosmocollider}
    \lim_{k_\L\ll 2c_s k_\S}B^{\text{B1},\text{B2}}_\zeta &=A^{\text{B1},\text{B2}}(c_s,\mu) P_\zeta(k_\L)P_\zeta(k_\S)\\ \nonumber
    & \times \left(\dfrac{k_\L}{k_\S}\right)^{\frac{3}{2}}\cos\left(\mu \log\left(\frac{k_\L}{2 c_s k_\S}\right)+\phi^{\text{B1},\text{B2}}(c_s,\mu)\right)\,,
\end{align}
where 
\begin{align}
    A^{\text{B1}}&=-\dfrac{\left((4\mu^2+5)^2-16\right)^{\frac{1}{2}}}{2^{5/2}\pi c_s {\cal P}_\zeta^{\frac{1}{2}}}\left(\dfrac{\rho}{\Lambda_1}\right)\dfrac{\pi}{\cosh(\pi\mu)}|\xi_+(c_s,\mu)|\,,\\ \nonumber
     A^{\text{B2}}&=\dfrac{(9\mu^2/4+1)^{\frac{1}{2}}}{\pi c_s {\cal P}_\zeta^{\frac{1}{2}}}\left(\dfrac{\rho}{\Lambda_2}\right)\dfrac{\pi}{\cosh(\pi\mu)}|\xi_+(c_s,\mu)|\,,\\ \nonumber
     \phi^{B_1}&=\text{Arg}[(-\mu^2+2i\mu+3/4)\xi_+(c_s,\mu)]\\ \nonumber
     \phi^{B_2}&=\text{Arg}[(1+2i\mu/3)\xi_+(c_s,\mu)]
\end{align}
and we recall that $\xi_+(c_s,\mu)$ is defined in Eq.~\eqref{def-xi}. These expressions simplify for large masses $\mu\gg 1$, as long as $c_s\mu\ll 1$. To see this, we use Eqs.~\eqref{TaylorfP}-\eqref{TaylorfM} to find
\begin{align}
    \lim_{1\ll \mu\ll c_s^{-1}}\xi_+=-\dfrac{2^{-3/2}(1-i)}{\mu}2^{-i\mu} e^{\pi\mu/2}\,,
\end{align}
For larger masses, i.e. $\mu\gtrsim c_s^{-1}$, one can check that a slightly modified version of the above formula, namely 
\begin{align}
\label{xilimit}
    \lim_{1\ll \mu,\, c_s\ll 1}\xi_+\approx -\dfrac{2^{-3/2}(1-i)}{\mu}2^{-i\mu} e^{(\frac{\pi}{2}-c_s)\mu}\,,
\end{align}
still captures the overall behaviour of $\xi_+$ as soon as $c_s \leqslant 0.1$. In conclusion, for $\mu\gg 1$ we find
\begin{align}
    \lim_{1\ll \mu,\, c_s\ll 1}A^{\text{B1}}&=-\dfrac{1}{2^{5/2} c_s {\cal P}_\zeta^{\frac{1}{2}}}\left(\dfrac{\rho}{\Lambda_1}\right)\mu\,e^{-(\frac{\pi}{2}+c_s)\mu}\,,\quad \lim_{1\ll \mu,\, c_s\ll 1}\phi^{\text{B1}}=-\dfrac{\pi}{4}-\mu \log(2) \label{A1phi1}\\ 
    \lim_{1\ll \mu,\, c_s\ll 1}A^{\text{B2}}&=\dfrac{3}{2}\dfrac{1}{c_s {\cal P}_\zeta^{\frac{1}{2}}}\left(\dfrac{\rho}{\Lambda_2}\right)\, e^{-(\frac{\pi}{2}+c_s)\mu}\,,\quad  \lim_{1\ll \mu,\, c_s\ll 1}\phi^{\text{B2}}=\dfrac{5\pi}{4}-\mu \log(2)\,. \label{A2phi2}
\end{align}
In \cite{Lee:2016vti}, it was noted that the conventional Boltzmann suppression factor $\exp(-\pi\mu)$ for $c_s=1$ was turned into $\exp(-\pi\mu/2)$ for a sufficiently low speed of sound. Our analysis both confirms this and gives the more accurate dependencies \eqref{A1phi1}-\eqref{A2phi2} in $\exp(-(\pi/2+c_s)\mu)$, with \eqref{xilimit} being very accurate already for $c_s \leqslant 0.1$ and $\mu \geq 5$.\\ 

\noindent As a final remark, notice that the ultra-squeezed limit formula \eqref{cosmocollider} receive corrections from a few sources: $(i)$ the particular solution to $\hat{F}(u,v)$ (i.e. the first term in \eqref{fullF}), $(ii)$ the subleading terms in the series expansion of $f_\pm$ around $u=0$, and $(iii)$ the $t-$ and $u-$channels that have been neglected in Eq.~\eqref{squeezedF}. At leading order in $u$ and for large $\mu$'s, $(i)$ is of order $u/\mu^2$ and is therefore always suppressed, as long as $u\lesssim 1$. 
However, corrections from point $(ii)$ are small only if $u\ll \mu^{-1/2}\lesssim 1$. 
The contribution of the other two channels to the three-point function scale as $P(k_\L)P(k_\S)(k_\L/k_\S)^2$, and hence are eventually subdominant in sufficiently squeezed configurations.\footnote{This is true only for the bispectrum. For $\hat{F}$, however, the other two channels contribute an offset given by 
    $\hat{F}_{\text{offset}}\sim -4g^2 c_s\left(\log(2\mu c_s)+\gamma_E-1 \right)$ to the oscillatory part of $\hat{F}$ in Equation \eqref{squeezedF}.} By and large, asking these corrections to be small refines the regime of the validity of \eqref{cosmocollider} by updating the upper bounds on $k_\L/k_\S$. For each diagram we find: 
\begin{align}
\label{eq:refined-estimates-cc}
    &\text{Diagram B1:}\qquad  \dfrac{k_L}{k_S}\ll \text{min}\lbrace 2c_s\mu^{-1/2}, 2c_s^{-3}\mu^2 e^{-(\pi+2c_s)\mu}\rbrace \,,\\ \nonumber
    &\text{Diagram B2:} \qquad \dfrac{k_L}{k_S}\ll \text{min}\lbrace 2c_s\mu^{-1/2}, 2c_s^{-3} e^{-(\pi+2c_s)\mu}\rbrace\,, 
\end{align}
where we have implicitly assumed $\mu\gtrsim 1$.

\subsubsection{Low speed resonances}
\label{sec:resonances}

The analytic expressions for the bispectra, given as inputs an arbitrary speed of sound, the mass of the intermediate field and the triangle configuration, are provided by Eq.~\eqref{diagb1b2} alongside \eqref{correlatorfull}-\eqref{fullF} and Eqs.~\eqref{Bpart}-\eqref{Ycoef}. Before discussing the amplitudes of the bispectra in section \ref{sec:size-NG}, here we concentrate on characterising their overall shapes. Hence, in figures \ref{fig:shapes-mu-one}, \ref{fig:shapes-others} and \ref{fig:largemu}, we plot the shape function (defined in\eqref{shapefunctiondef}) in isosceles configurations (i.e. $k_1=k_2$), normalized to unity for equilateral triangles, for various values of the parameters $c_s$ and $\mu$, for both diagrams B1 and B2. We cover all types of triangles by varying the ratio $\frac{k_3}{2k_1}$ from small values for squeezed triangles, through $1/2$ for equilateral configurations, to $1$ for flattened ones. When relevant and for comparison, we also plot the corresponding low speed signal \eqref{asymptB1}-\eqref{asymptB2} (orange dashed), the cosmological collider one \eqref{cosmocollider} (red dashed), as well as the standard local EFT one (blue dashed) that would result from integrating out the heavy field (see e.g.~the discussion around Eq.~\eqref{new-speed-sound}). The main characteristics of the signals are as follows: 
\begin{itemize}
    \item Let us first concentrate on the qualitatively new regime of particular interest, i.e. small $c_s$ and $c_s m/H$ (fig.~\ref{fig:shapes-mu-one}). There, the B2 shape is characterised by a pronounced ``resonance'' where it reaches its maximum, around $k_3/k_1\simeq c_s m/H$. This  bump of symmetric profile (in logarithmic scale) has an amplitude (compared to the equilateral configuration) that grows as $c_s$ or/and $m/H$ diminishes, with an enhancement factor $\sim {\cal O}(1)/(5 c_s\mu)$. Away from the resonance, one can see that the low speed result for larger values of $k_3/k_1$, and the cosmological collider one for smaller values, very well describe the signal. The latter remark also applies to the B1 shape, but its resonance signal is more complex: not only does it comprise a local positive maximum, attained for $k_3/k_1\simeq 2 c_s m/H$ and with an enhancement $\sim {\cal O}(1)/(10 c_s\mu)$; but it is also characterised by a second ``resonance'' for more squeezed triangles $k_3/k_1\simeq 0.5 c_s m/H$, at which the signal reaches a local (negative) extremum of similar amplitude. Eventually, the shape goes to zero between the two extrema at the intermediate value $k_3/k_1\simeq c_s m/H$.
    
    \item As one increases $\mu$, roughly above $0.1/c_s$ for Diagram B1 and $0.2/c_s$ for Diagram B2, the extremum of the shape B2 fades away, as well as the upward peak of shape B1. Instead, the downward peak of shape B1 remains distinctly visible even at larger values (see the plot $(\mu=5, c_s=0.1)$ in Fig.~\ref{fig:shapes-others}), before it eventually also fades away as $\mu$ approaches $1/c_s$.\footnote{For $0.6 \lesssim c_s \mu \lesssim 1$, one can check that the downward extremum turns upward (before fading away for larger values). However, this is an artefact of considering the shape normalized to its value in the equilateral limit. One can check indeed that the shape function there changes sign around $c_s \mu \simeq 0.6$, whereas the shape at the location of the (downward) resonance is always of the same sign (the one of $\alpha_1)$.} Eventually, note that even when the resonances have disappeared and the shapes monotonously decrease from equilateral to squeezed triangles (where the oscillations appear), 
    the shapes significantly differ from those of the local EFT cubic operators $\dot{\pi}^3$ (for Diagram B1) or $\dot{\pi}(\partial_i \pi)^2$ (for Diagram B2) if one has not reached the regime $\mu \gg 1/c_s$ (see e.g.~Fig.~\ref{fig:shapes-others} top right).  
\begin{figure}
    \centering
    \includegraphics[scale=0.8]{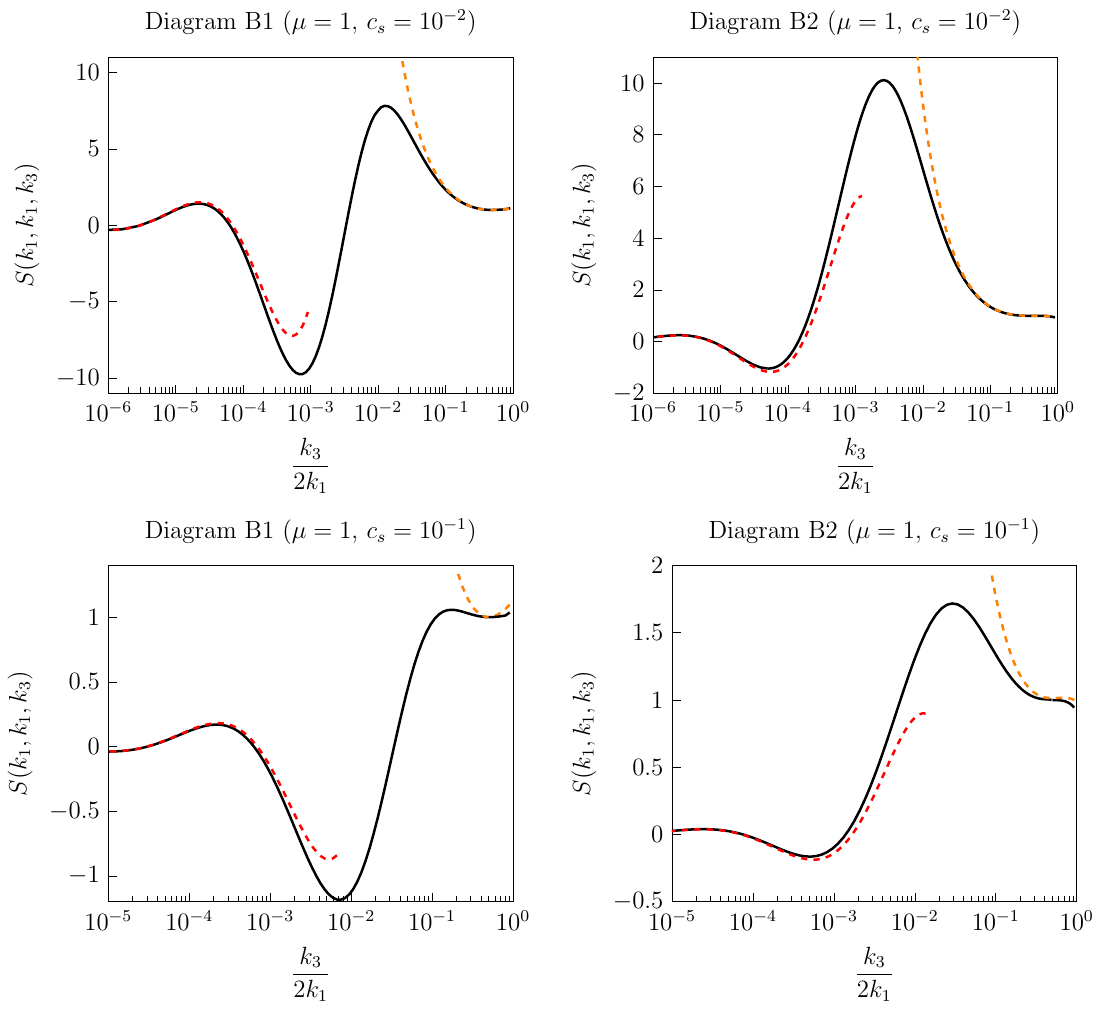}
    \caption{Shapes $S(k_1,k_1,k_3)$ of the bispectra for isosceles triangles, normalized to unity in equilateral configurations, for diagrams B1 (left) and B2 (right), and for $(\mu=1, c_s=0.01)$ (top) and $(\mu=1, c_s=0.1)$ (bottom). The dashed red and orange curves represent respectively the ultra-squeezed \eqref{cosmocollider} and the low speed signals \eqref{asymptB1}-\eqref{asymptB2}.}
    \label{fig:shapes-mu-one}
\end{figure}

     \item The features described above are characteristic of a subluminal speed of propagation for $\pi$. In other words, for a speed of sound close to unity, the two shapes monotonously decrease from unity (at $k_3/k_1=1$) before reaching the oscillatory phase (see fig.~\ref{fig:shapes-others}, bottom, for $c_s=0.7$).
    \item  Eventually, note that in the large mass limit $\mu\gg 1$, the character of the resonances becomes maximally distinct from that of the particle production effect in dS space. This is so because increasing the mass of the intermediate field makes the oscillations in the ultra-squeezed regime exponentially dim, irrespective of the value of $c_s$. Conversely, the resonances characteristic of the low speed collider survives as long as we keep $c_s\lesssim \mu^{-1}$ (see Fig.~\ref{fig:largemu}).
\end{itemize}

\begin{figure}
    \centering
    \includegraphics[scale=0.8]{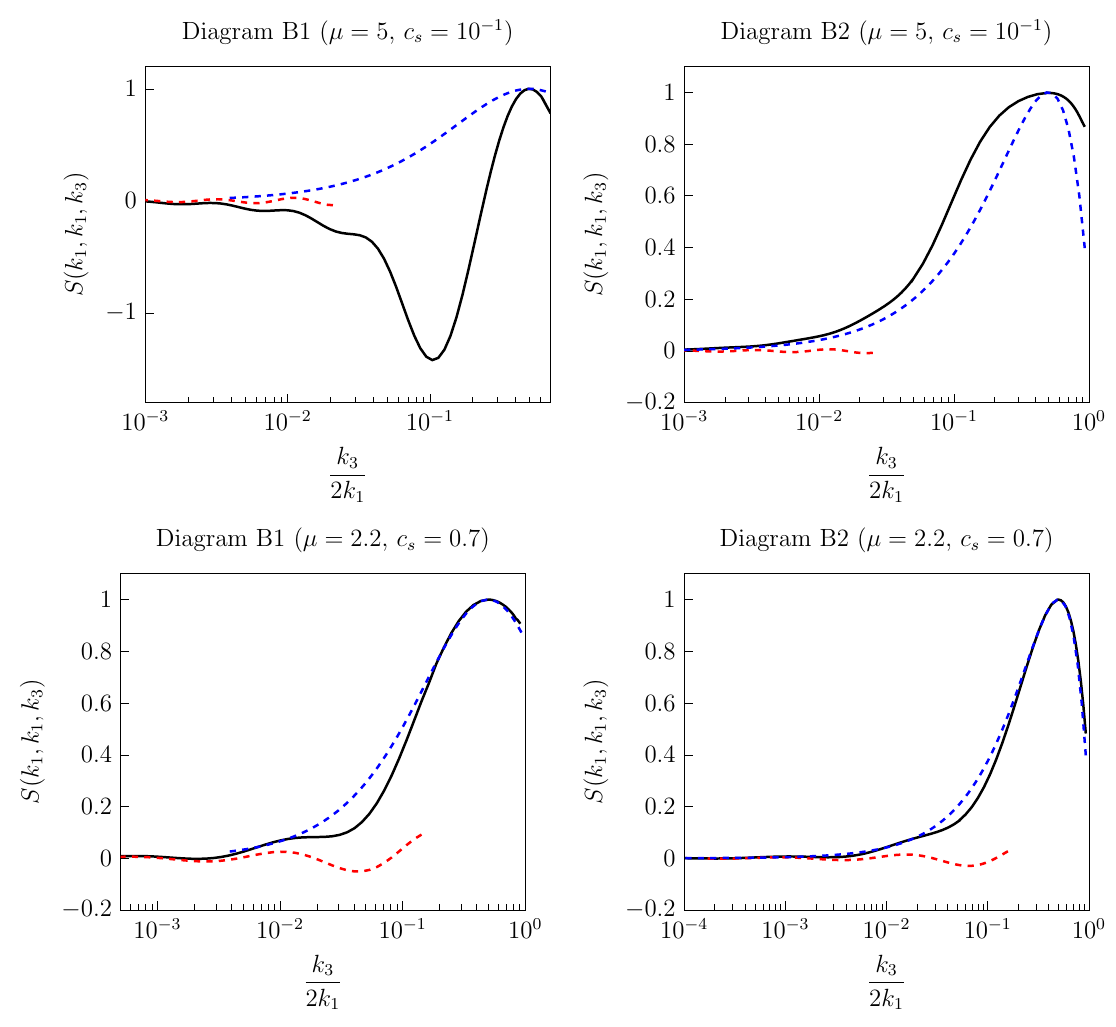}
    \caption{Shapes $S(k_1,k_1,k_3)$ of the bispectra for isosceles triangles, normalized to unity in equilateral configurations, for diagrams B1 (left) and B2 (right), and for $(\mu=5, c_s=0.1)$ (top) and $(\mu=2.2, c_s=0.7)$ (bottom). The dashed red and blue curves represent respectively the ultra-squeezed signal \eqref{cosmocollider} and the one that would result from the local EFT after integrating out the heavy field.}
    \label{fig:shapes-others}
\end{figure}

\begin{figure}
    \centering
    \includegraphics[scale=0.8]{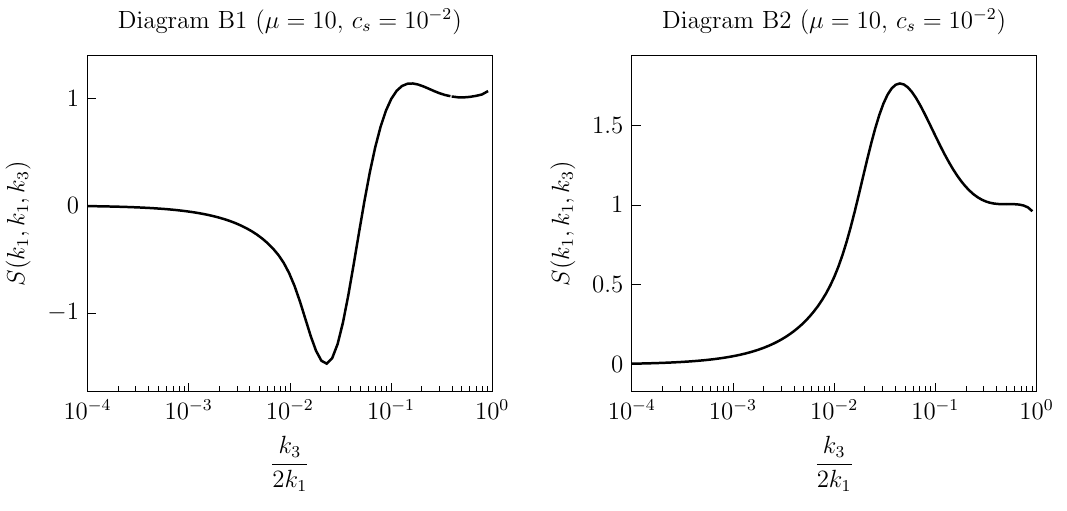}
    \caption{Shapes $S(k_1,k_1,k_3)$ of the bispectra for isosceles triangles, normalized to unity in equilateral configurations, for diagrams B1 (left) and B2 (right) and $(\mu=10, c_s=0.01)$.}
    \label{fig:largemu}
\end{figure}

\subsubsection{Size of non-Gaussianity and perturbativity}
\label{sec:size-NG}

So far, we have concentrated on characterising the shape of the bispectrum, but we now discuss its amplitude. As customary, a convenient overall measure of the bispectrum is the amplitude of the shape function in the equilateral configuration, more precisely, with the usual convention, the parameter $\fNL=\frac{10}{9}S(k,k,k)$. This provides a fair estimate of the signal for $m \gg H/c_s$, for which the shapes are maximal in equilateral configurations and the dependence on parameters anyway follows from the usual EFT treatment. For the new regime of interest  $m \ll H/c_s$, the shapes are maximal near the resonances, but the enhancement of the signal there compared to the equilateral limit is known, going as $(c_s m/H)^{-1}$ (see the previous section), so we first stick to the usual $\fNL$ parameter for simplicity, and concentrate on this most interesting regime.\\

\noindent From the definition of the shape \eqref{shapefunctiondef}, the expression \eqref{eq:standard-power-spectrum} for the leading-order power spectrum, and the results \eqref{asymptB1}-\eqref{asymptB2} for the bispectra, one then finds
\be
\fNL^{(1)}=\frac{5}{18 \pi} \frac{\rho}{\Lambda_1} \left( \frac{c_s}{{\cal P}_\zeta} \right)^{\frac12} \,, \,\, \fNL^{(2)}=\frac{5}{36 \pi} \frac{\rho}{\Lambda_2} \left( \frac{c_s}{{\cal P}_\zeta} \right)^{\frac12}
\Bigg[8-3 \gamma_E + 3\ln(\frac{1}{3 c_s C(\mu)}) \Bigg]\,, \textrm{for} \, \, c_s \frac{m}{H} \ll 1 \,.
\ee
for diagrams B1 and B2  respectively. Except for the mild and understood logarithmic dependence on $c_s C(\mu)$ for diagram B2, both thus share the parametric dependence $\fNL \sim \frac{\rho}{\Lambda_{1,2}} \left( \frac{c_s}{{\cal P}_\zeta} \right)^{\frac12}$. It is instructing to discuss first diagram B2, as remember that the scale $\Lambda_2$ suppressing the corresponding cubic interaction is determined by the quadratic coupling $\rho$ through the non-linearly realised symmetry of time-diffeomorphism invariance, i.e. both terms come from the interaction proportional to $\rho \delta g^{00} \sigma$ in the unitary gauge, see the discussion after Eq.~\eqref{interpisigma} and the explicit relation \eqref{eq:Lambda-rho-link}.\footnote{We are assuming that $c_\sigma=1$, so that no rescaling of the spatial coordinates is needed for our analysis to apply. It is straightforward to generalise our results to more general cases, but it would make the discussion more complex without much physical differences: we are interested in the qualitatively new regime in which the ratio between the speed of $\pi$ and the one of $\sigma$ is small. If we stick to subluminal propagation speeds, and consider the lower bound on the speed of propagation of $\pi$ coming from Planck constraints \cite{Planck:2019kim}, $c_s \geq 0.021 (95 \% \textrm{CL})$, or even simply the lower bound on $c_s$ to prevent strong coupling $c_s^2 \gg \sqrt{{\cal P}_\zeta}$, the speed of propagation of $\sigma$ can not appreciably deviate from the speed of light.} The latter can be simply rewritten in terms of the amplitude of the power spectrum as $\frac{H}{\Lambda_2} \left( \frac{c_s}{{\cal P}_\zeta} \right)^{\frac12}=-\pi \frac{\rho}{H}$, in such a way that $\fNL^{(2)}$ simplifies to
\be
\fNL^{(2)}=-\frac{5}{36} \left(\frac{\rho}{H}\right)^2 \Bigg[8-3\gamma_E+ 3\ln(\frac{1}{3 c_s C(\mu)}) \Bigg]\,,
\label{fNL2}
\ee
i.e. the amplitude of the non-Gaussian signal from diagram B2 is tied to the amplitude of the quadratic coupling.\footnote{From Eqs.~\eqref{trispectrum}, one can easily deduce that the dimensionless amplitude of the trispectrum in generic configurations parametrically reads $\tau_{\textrm{NL}} \sim \frac{1}{c_s^2}\left(\frac{\rho}{H} \right)^2$.} 

At first sight, this seems to limit the size of the bispectrum to tiny values, as one may think that the requirement of treating perturbatively the quadratic coupling  requires $(\rho/H)^2 \ll 1$. However, more room is actually left in our situation of interest. It is actually not completely straightforward to assess what is the correct perturbativity criterion. For instance, from the correction to the power spectrum coming from the exchange of $\sigma$ that scales as $\Delta P/P \sim (\rho/H)^2 c_s^2$ for $c_s m/H \ll 1$  (see Eq.~\eqref{powerasymp} and neglecting the logarithmic dependence), one may think that the bound is considerably weakened to $(\rho/H)^2 \ll 1/c_s^2$. Instead, if one uses the standard lore criterion that the quadratic mixing term should be negligible compared to the rest of the quadratic action around the relevant time characteristic of the dynamics of $\pi$, namely around sound horizon crossing, one straightforwardly finds instead a more stringent bound $(\rho/H)^2 \ll 1/c_s$. However, none of these reasonings are actually correct. Quite simply, the perturbative treatment of the quadratic coupling is warranted if and only if the uncoupled mode functions of $\pi$ and $\sigma$, which are taken as building blocks in the perturbative approach, faithfully reproduce the dynamics governed by the full quadratic action. If this is the case, then the use of these uncoupled mode functions will provide a correct approximation to the computation of both 2-point, but also 3-point and all higher-order correlation functions. However, one can numerically compute all correlation functions and assess the accuracy of the perturbative approach against exact (numerical) results, and one can check that this requires $\rho \lesssim m$ in the regime of interest. This ensures that the dynamics of $\sigma$ is not substantially modified by the coupling to $\pi_c$. One should keep in mind indeed the asymmetry between the two fields: the power spectrum of $\pi_c$ is much larger than the one of $\sigma$ at all times. For instance, around sound horizon crossing for $\pi_c$, when $\sigma$ is still sub-Hubble, $\pi_c \sim \frac{1}{\sqrt{c_s}} \sigma$, and the hierarchy is even bigger at later times when (the uncoupled) $\pi_c$ has frozen and $\sigma$ has further decayed. This hierarchy also intuitively explains why the correction to the power spectrum is smaller than just what the perturbativity bound would imply, $(\Delta P/P)/(\rho/m)^2 \sim (c_s m/H)^2 \ll 1$: even when one approaches the perturbative bound and the dynamics of $\sigma$ is substantially modified by the coupling to $\pi_c$, the effect on $\pi_c$ of much larger amplitude is comparatively much weaker.\\

\noindent As the discussion above points out once more, the existence of the two sound speeds and hence of different characteristic times (by contrast to only sound horizon crossing usually) is such that conventional back-of the envelope estimates do not hold. For instance, comparing the size of the cubic action compared to the quadratic one at sound horizon crossing would give $H/\Lambda_{1,2} \ll c_s$, which do not necessarily encode the correct criterion for treating the cubic interactions perturbatively. Such a proxy for the full computation is not needed though: we have computed the non-Gaussian signal, of size $\fNL \sim (\rho/H)^2 \lesssim (m/H)^2$, and it largely satisfies the perturbative criterion $\fNL \sqrt{{\cal P}_\zeta} \ll 1$.\footnote{We implicitly consider that natural values of $\Lambda_1$ are of order $\Lambda_2$, in which case the two cubic interactions are suppressed by the same scale.}
The non-Gaussian signal in the equilateral configuration can thus be observationally large, scaling like $\fNL \sim (m/H)^2 \gg 1$ when saturating the parametric bound for treating  $\rho$ perturbatively. However, one should keep in mind that the bispectrum signals we have computed add to the unavoidable one generated by the derivative self-interactions of $\pi$ in \eqref{S-EFT}, overall of equilateral type and of amplitude $1/c_s^2$. Our signal is hence a subdominant component to the total one in the equilateral configuration, although, as we will see below, it is actually not negligible when taking into account all numerical factors beyond the scalings here.\\

\begin{figure}
    \centering
    \includegraphics[scale=0.8]{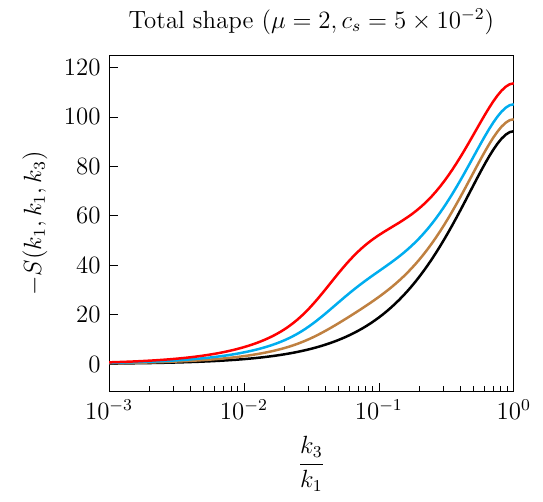}
    \caption{Shape $S(k_1,k_1,k_3)$ of the total bispectrum for isosceles triangles, found by adding the effects from the self-interactions in $\dot{\pi} (\partial_i \pi)^2$ and the effect of Diagram B2 (see the main text), for $c_s=0.05, \mu=2$ and $\rho/m=(0,2,3,4)$ (black, brown, blue, red).}
    \label{fig:total-signal}
\end{figure}

\noindent More importantly, as we have stressed, the signal induced by the interactions with the heavy field is enhanced near the resonances compared to the one in equilateral configuration, scaling like $\fNL^{\textrm{res}} \sim 1/(c_s m/H) (\rho/H)^2$. At the same time, the standard EFT shapes decrease in the squeezed limit like $k_\L/k_\S$, so that the ``contamination'' from the self-interactions of $\pi$, in the resonance region $k_\L/k_\S \sim c_s m/H$, is only $\fNL^{\textrm{contamination}} \sim  m/(H c_s)$. The ratio of the two is thus $\fNL^{\textrm{res}}/\fNL^{\textrm{contamination}} \sim (\rho/m)^2$, so that the two signals can become of the same amplitude there, leading to visible resonances. Beyond these instructive scalings, these features can be confirmed quantitatively by explicitly representing the total signal including all numerical factors, as can be seen in fig.~\ref{fig:total-signal} for the representative set of parameters $(c_s=0.05, \mu=2)$ and for various values of $\rho \sim \m$. For simplicity, there, we only showed the part of the signal that is entirely fixed by symmetries. Namely, for the self-interactions of $\pi$, we took into account only the one in $\dot{\pi} (\partial_i \pi)^2$, and for the interactions with $\sigma$, we similarly only considered the one in $\sigma (\partial_i \pi)^2$ (Diagram B2). We checked that the contribution from Diagram B1 does not change the picture for $\Lambda_1 \sim \Lambda_2$, although of course, the effect of the resonance can be made even more visible by considering smaller values of $\Lambda_1$, or/and fine-tuning the value of the Wilson coefficient $A$ in \eqref{S-EFT}. We note that the resonance signal from Diagram B2 is always negative (see \eqref{fNL2} and the shapes e.g.~in fig.~\ref{fig:shapes-mu-one}), just like the one from the $\dot{\pi}(\partial_i \pi)^2$ interaction, hence the fact that the total signal does not present a dip, but truly a bump-like feature. Given these results, it would naturally be interesting to further study our setup by treating the $\dot{\pi} \sigma$ interaction non-perturbatively, which we leave for future work.

\section{Non-local single field effective field theory}
\label{non-localEFT}

In the previous section we saw that the exchange of heavy supersonic particles between the curvature fluctuations leaves a characteristic imprint as resonances in the shape of the three-point correlation function as long as the massive field is lighter than $H/c_s$. Moreover, the variation of the two- and three-point functions with respect to the mass of the intermediate particle is only logarithmic. In this section, we show that both of these properties, and more generally the features of the signal that are not attributed to the non-perturbative particle production, can be explained by estimating the corresponding exchange diagram with a non-local contact interaction that emerges after integrating out the heavy field. We provide a number of numerical and analytical justifications in favor of such approximation for $\mu \gtrsim O(1)$.
At the same time, we demonstrate that the (non-local) single field description breaks down when $\mu\lesssim 1$ or when the three-point function configuration is ultra-squeezed.

\subsection{Mode function analysis} 
\label{modefunctionanal}

It is most convenient to look at the four-point seed function $\hat{F}(u,v)$ in order to establish the regime of  validity of the (non-local) single field picture. It goes without saying that the same single-field description will apply to the desired three-point functions. Nevertheless, it is sufficient for us to look back at the four-point function, and the three-points simply follow from the weight-shifting operators acted on $\hat{F}(u,v)$. To that end and at the cost of being pedantic, here we introduce a new field $\tilde{\vpi}$ that propagates with the speed of sound $c_s$ and has the same mass and cubic interaction as those of the $\vpi$ field, namely $m^2=2H^2$ and $g \tilde{\vpi}^2 \sigma$ (remember that $\vpi$ propagates at speed one). As a result, the four-point function of $\tilde{\vpi}$ due to the exchange of $\sigma$ is the same as $\hat{F}(u,v)$ where this time $u$ and $v$ are manifestly $c_s$-dependent, namely we have
\begin{align}
    u=\dfrac{s}{c_s(k_1+k_2)}\,,\quad v=\dfrac{s}{c_s(k_3+k_4)}\,.
\end{align}
Before studying the cosmological four-point correlator, let us first look at this field theory in the $H\to 0$ limit (with $m$ held fixed) by computing the two-to-two scattering of the $\tilde{\vpi}$ particles due to the tree-level exchange of $\sigma$.  The answer is given by
\begin{align}
{\cal A}\left({\tilde{\vpi}(\bfp_1)\tilde{\vpi}(\bfp_2)\to \tilde{\vpi}(\bfp_3)\tilde{\vpi}(\bfp_4)}\right)=\dfrac{g^2}{c_s^2(p_1+p_2)^2-|\bfp_1+\bfp_2|^2-m^2}+{t-}\,\,\text{and}\,{u-}\text{channels}\,,
\end{align}
where ${\cal A}$ is the scattering amplitude, and $(E_i=c_s p_i,\bfp_i)$ are the 4-momenta of the $\tilde{\vpi}$ particles (which are all taken to be incoming such that $\sum_i \bfp_i=\sum_i p_i=0$). 
Provided that the ratios
\begin{align}
\dfrac{c_s^2(p_1+p_2)^2}{|\bfp_1+\bfp_2|^2+m^2}\,,\,\,\dfrac{c_s^2(p_1+p_3)^2}{|\bfp_1+\bfp_3|^2+m^2}\,,\,\,\text{and}\,\,\,\dfrac{c_s^2(p_1+p_4)^2}{|\bfp_1+\bfp_4|^2+m^2}\,
\end{align}
are small, the resulting scattering amplitude at leading order in $c_s$ becomes
\begin{align}
\label{leading2to2}
{\cal A}_{2\tilde{\vpi}\to 2\tilde{\vpi}}\approx -\dfrac{g^2}{|\bfp_1+\bfp_2|^2+m^2}+{t-}\,\,\text{and}\,{u-}\text{channels}\,.
\end{align}
At this order in $c_s$, this amplitude arises from neglecting the time derivative of $\sigma$ 
compared to its spatial derivatives, which amounts to treating $\sigma$ as a non-dynamical field that can be solved in terms of $\tilde{\vpi}$, i.e. 
\begin{align}
\sigma\approx \dfrac{1}{-\nabla^2+m^2}g \tilde{\vpi}^2\,. 
\end{align}
Inserting $\sigma$ back inside the Lagrangian, we obtain a single field theory for $\tilde{\vpi}$ characterised by a non-local quartic interaction
\begin{align}
    {\cal L}_I=\dfrac{g^2}{2}\tilde{\vpi}^2\dfrac{1}{-\nabla^2+m^2}\tilde{\vpi}^2\,.
\end{align}
Obviously, this quartic non-local contact term generates the same 2-to-2 amplitude as \eqref{leading2to2}.\\

\noindent In fact, the corrections to \eqref{leading2to2} can be captured by adding an infinite tower of operators that are organized in powers of $\partial_t^2$.  These operators simply follow from solving $\sigma$ as 
\begin{align}
    \sigma=\dfrac{1}{-\Box+m^2}g\,\vpi^2=\dfrac{g}{-\nabla^2+m^2}\sum_{n=0}^{\infty}\left(\dfrac{-\partial_t^2}{-\nabla^2+m^2}\right)^n\,\tilde{\vpi}^2\,,
\end{align}
and plugging it back inside the action to find
\begin{align}
    {\cal L}_I=\frac{g^2}{2}\tilde{\vpi}^2\dfrac{1}{-\nabla^2+m^2}\sum_{n=0}^{\infty}\left(\dfrac{-\partial_t^2}{-\nabla^2+m^2}\right)^n\,\tilde{\vpi}^2\,.
\end{align}
\\
\noindent It might seem straightforward to use the same picture as above in the cosmological setting by neglecting the time derivatives of $\sigma$ in the exchange diagram of interest. However, the analysis gets more complicated due to the time dependence of the background. Before making further progress, it proves useful to switch to the following field variables
\begin{align}
\label{fieldredef}
    \Sigma\equiv a^2(\eta)\sigma\,,\quad \fI=a(\eta)\tilde{\vpi}\,,
\end{align}
in terms of which the Lagrangian becomes 
\begin{align}
\label{actionSigmapi}
    S&=\int d\eta \left(\dfrac{1}{2}\fI'^2-\dfrac{c_s^2}{2}(\partial_i \fI)^2+\dfrac{\eta^2 H^2}{2}\Sigma'^2-\dfrac{\eta^2 H^2}{2}(\partial_i \Sigma)^2-\dfrac{1}{2}(m^2-2H^2)\Sigma^2-g \fI^2 \Sigma\right)\,.
\end{align}
The virtue of these field redefinitions is that $\fI$ (aka the Sasaki-Mukhanov variable) behaves as a massless field in flat space at all times (as a virtue of its carefully chosen mass), while with the chosen rescaling of $\sigma$ the cubic term $f^2\Sigma$ is left with no explicit time dependence.  
In order to inspect under what circumstance the field $\Sigma$ can be regarded as non-dynamical,  it is useful to think of the exchange diagram \ref{fig:vpi-corr} as a contribution to the quartic part of the wavefunction of the universe at late times (i.e. $\eta_0\to 0$)
\footnote{For the purpose of the following discussion we do not need a thorough review of the wavefunction method. The interested reader can look at e.g. Appendix A of \cite{COT}.} 
, i.e. 
\begin{align*}
\Psi\lbrace \fI(\bfk),\eta_0\rbrace=\exp(-\dfrac{1}{2!}\int_\bfk \psi_2(\bfk) \fI(\bfk)\fI(-\bfk)-\dfrac{1}{4!}\left(\prod_{i=1}^4 \int_{\bfk_i}\right)\psi_4(\bfk_i)\fI(\bfk_1)\dots \fI(\bfk_4)+\dots )\,.
\end{align*}
Here $\fI(\bfk)$ is the boundary value of the field $\fI$ (in momentum space),  $\psi_n$'s are the so called wavefunction coefficients , and $\dots$ stand for higher order corrections to the wavefunction. To avoid cluttered notation, here we have neglected the dependence of $\Psi$ on the boundary value of $\Sigma$.  In the tree-level approximation, the wavefunction is given by 
\begin{align}
\Psi\lbrace \fI(\bfk)\rbrace=\exp(i S[\fI_{\text{cl}}(\bfk,\eta),\Sigma_{\text{cl}}(\bfk,\eta)])\,,
\end{align}
where $S$ is the action of the theory, with $\fI_{\text{cl}}$ and $\Sigma_{\text{cl}}$ satisfying the classical equations of motion, i.e. 
\begin{align}
    \dfrac{\delta S}{\delta \fI_{\text{cl}}}=\dfrac{\delta S}{\delta \Sigma_{\text{cl}}}=0\,, 
\end{align}
which have to be solved with the following boundary conditions
\begin{align}
\label{boundary}
& \fI_{\text{cl}}(\bfk,\eta_0)=\fI(\bfk)\,,\quad  \Sigma_{\text{cl}}(\bfk,\eta_0)=\Sigma(\bfk)\,,\\ \nonumber & \fI_{\text{cl}}(\bfk,-\infty(1-i\epsilon)=\Sigma_{\text{cl}}(\bfk,-\infty(1-i\epsilon)=0\,.
\end{align}
In the wavefunction approach, computing the contribution of the single-exchange diagram in Figure \ref{fig:vpi-corr} to $\psi_4(\bfk_1,\dots,\bfk_4)$ reduces to solving the equation of motion for $\Sigma$, which is
\begin{align}
\label{PDESigma}
\left[H^2\eta^2\partial_\eta^2+2H^2\eta \partial_\eta+H^2\eta^2(\bfq_1+\bfq_2)^2+(m^2-2H^2)\right]\Sigma=-g\,\fI(\bfq_1)\fI(\bfq_2)\exp(i c_s(q_1+q_2)\eta)\,, 
\end{align}
with the boundary conditions \eqref{boundary} (with $\Sigma(\bfk)=0$, since there is no external $\Sigma$ leg in the diagram). 
The source on the RHS of this equation is formed by the product of two factors of $\fI_{\text{cl}}$ in the free theory, i.e. 
\begin{align}
\fI_{\text{cl}}(\bfq_i,\eta)|_{\text{free}}=\fI(\bfq_i)\exp(ic_s q_i \eta)\,, \quad i=1,2\,.
\end{align}
Also, for the purpose of computing the $s-$channel contribution to $\psi_4$, the momenta $(\bfq_1,\bfq_2)$ should be equated with $(\bfk_1,\bfk_2)$ or $(\bfk_3,\bfk_4)$. The unique solution to \eqref{PDESigma} can be written as
\begin{align}
    \Sigma_{\text{sol}}=\Sigma(q_1,q_2,\eta)f(\bfq_1)f(\bfq_2)\,,
\end{align}
Subsequently, this solution should be inserted back inside the action $S_{\text{cl}}$ in order to compute $\psi_4$. $\Sigma_{\text{sol}}$ is quadratic in $f$, hence the quartic piece in the wavefunction originates from two contributions, namely the kinetic term of $\Sigma$ in \eqref{actionSigmapi} and the cubic interaction. The resulting four-point up to an unimportant prefactor is
\begin{align}
\label{psi4Sigma}
\nonumber
    \psi_4(\bfk_1,\dots,\bfk_4)&=\left(i\,g\int_{-\infty(1-i\epsilon)}^0 d\eta\, e^{ic_s(k_3+k_4)\eta}\,\Sigma(k_1,k_2,\eta)+ (\bfk_1,\bfk_2)\leftrightarrow (\bfk_3,\bfk_4)\right)\\ 
    & +{t-}\,\,\text{and}\,{u-}\text{channels}\,. 
\end{align}
Given the quartic wavefunction coefficient $\psi_4$, the correlator of $\tilde{\vpi}$ can be computed by the following relation \cite{Anninos:2014lwa} 
\begin{align}
    \langle \tilde{\vpi}(\bfk_1)\dots \tilde{\vpi}(\bfk_4)\rangle'=-2\left(\prod_{i=1}^4\,\langle \tilde{\vpi}(\bfk_i)\tilde{\vpi}(-\bfk_i)\rangle'\right)\text{Re}\psi_4(\bfk_1,\dots,\bfk_4)\,.
\end{align}
We are going to argue that when $c_s\ll 1$, for most part of the $(\bfq_1,\bfq_2)$ space it is a good approximation to ignore the time derivatives of $\Sigma$ in \eqref{PDESigma}. As a result, one finds
\footnote{We are forced to add a homogeneous solution to $\Sigma_0$ in order to comply with the boundary condition $\Sigma(q_1,q_2,\eta_0)=0$. However, it can be easily verified that the contribution of such piece to $\psi_4$ vanishes once $\eta_0$ is sent to zero for heavy fields.}
\begin{align}
\label{components}
\Sigma_0(q_1,q_2,\eta)&=-\dfrac{g}{H^2\eta^2 (\bfq_1+\bfq_2)^2+m^2-2H^2}e^{ic_s(q_1+q_2)\eta}\,.
\end{align}
The obvious sanity check is to ensure that the time derivative of $\Sigma_0$ are indeed negligible compared to the RHS of Eq.~\eqref{PDESigma}, which is to ask
\begin{align}
\label{negligible}
\dfrac{1}{g}\Big|(H^2\eta^2\partial_\eta^2+2H^2\eta \partial_\eta)\Sigma_0(q_1,q_2,\eta)\Big|\ll 1\,.
\end{align}
For the $s-$channel, $(q_1,q_2)$ can be set to $(k_1,k_2)$ or $(k_3,k_4)$. Therefore we arrive at two inequalities: 
\begin{align}
\label{ineq}
\Big|(\eta^2\partial_\eta^2+2\eta\partial_\eta)\dfrac{e^{ic_s(k_1+k_2)\eta}}{\eta^2s^2+\mu^2+1/4}\Big|\ll 1\,,\qquad \Big|(\eta^2\partial_\eta^2+2\eta\partial_\eta)\dfrac{e^{ic_s(k_3+k_4)\eta}}{\eta^2s^2+\mu^2+1/4}\Big|\ll 1\,, 
\end{align}
Switching to the dimensionless variable $x=s\eta$, one finds 
\begin{align}
\label{defrux}
 r(u,x)&=|(x^2\partial_x^2+2x\partial_x)\dfrac{e^{ix/u}}{x^2+\mu^2+1/4}|\ll 1\,,\\ \nonumber  r(v,x)&=|(x^2\partial_x^2+2x\partial_x)\dfrac{e^{ix/v}}{x^2+\mu^2+1/4}|\ll 1\,. 
\end{align}
In principle, we need these inequalities to hold for every $\eta$ in order for $\Sigma_0$ to be an accurate solution to \eqref{PDESigma}. However, the integrand of \eqref{psi4Sigma} (after Wick rotation) is exponentially small when either of the comoving scales $(k_1+k_2)$ and $(k_3+k_4)$ are deep inside the sound horizon (i.e. $c_s(k_1+k_2)\eta\gg 1$ and $c_s(k_3+k_4)\eta\gg 1$). Therefore, as far as computing the four-point function $\psi_4$ is concerned it is sufficient to ensure that the inequalities \eqref{ineq} are maintained around and after the sound-horizon crossing of the aforementioned scales, namely for $|x|\lesssim u\,(\text{or}\,v)$. Plotting $r(u,x)$ across the $|x|<u$ domain in Figure \ref{fig:r(u,x)}, one finds that (this can also be easily understood analytically):
\begin{itemize}
    \item for very heavy fields (i.e. $\mu\gg 1$), the ratio $r(u,x)$ is small for every $u>0$. This means that, for large masses, the non-local single field description is accurate in describing the exchange diagram irrespective of the four-point configuration (i.e. for any positive values of $u$ and $v$). 
    \item for $\mu\gtrsim 1$, $r(u,x)$ is negligible only if both $u$ and $v$ are much greater than one. In other words, the single-field picture fails to reproduce $\hat{F}(u,v)$ when $u$ or $v$ approach unity and $\mu$ is of order 1.  
    \item for barely heavy fields $m \simeq 3/2 H$, $r(u,x)$ becomes of order one when $x$ nears $1/2$. This is around mass crossing of the intermediate momentum, namely when $s\eta\sim  m/H\sim {\cal O}(1)$ (for the three-point function, which is related to $\hat{F}$ in the soft limit $k_4\to 0$, this time coincides with mass crossing of the $s-$channel $k_3$). 
\end{itemize}

\noindent The conclusion of the above observations is that the time derivative terms in \eqref{PDESigma} can be treated perturbatively when $\mu\gg 1$ or when $\mu\gtrsim 1$ is an order one number but $u$ and $v$ are much greater than one. \\ 
\begin{figure}
    \centering
    \includegraphics[scale=1.2]{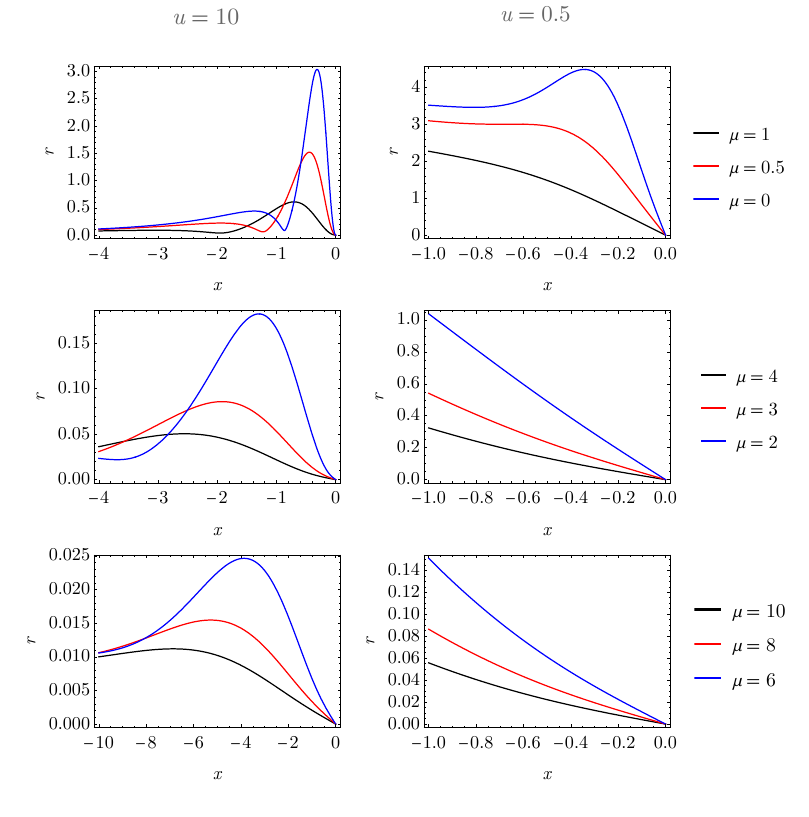}
    \caption{Plot of $r(u,x)$ \eqref{defrux} for various values of $u$ and $\mu$. The validity of the EFT requires $r \ll 1$ for $|x|\lesssim u$, see the text after Eq.~\eqref{defrux}.}
    \label{fig:r(u,x)}
\end{figure}

\noindent At the level of the action and zeroth order in time derivatives, the manipulations that led to \eqref{psi4Sigma} are nothing but solving the real space equation of motion for $\Sigma$ by neglecting its time derivatives and plugging the result back inside the action to find a non-local quartic operator. Assuming $r\ll 1$, one can even go beyond the leading order approximation and solve $\Sigma$ to full order in $\partial_\eta$, finally arriving at
\begin{align}
    \Sigma &=-\dfrac{g}{H^2}\dfrac{1}{\eta^2\partial_\eta^2+2\eta \partial_\eta-\eta^2\nabla^2+\mu^2+\frac{1}{4}} \fI^2\,,\\ \nonumber
    &=-\dfrac{g}{H^2}\dfrac{1}{-\eta^2\nabla^2+\mu^2+\frac{1}{4}}\sum_{n=0}^\infty\,\left[(\eta^2\partial_\eta^2+2\eta \partial_\eta)\dfrac{-1}{-\eta^2\nabla^2+\mu^2+\frac{1}{4}}\right]^n\,\fI^2\,.
\end{align}
Plugging this solution back inside the action yields the following non-local EFT for $\tilde{\vpi}$:
\begin{tcolorbox}[colframe=white,arc=0pt]
\begin{align}
\label{EFTnonlocal}
    {\cal L}_\textrm{EFT}&=a^2(\eta)\left[\dfrac{1}{2}\tilde{\vpi}'^2-\dfrac{c_s^2}{2}(\partial_i \tilde{\vpi})^2 -a^2(\eta)H^2\tilde{\vpi}^2\right]\\ \nonumber
    &+\dfrac{g^2}{2H^2}a^2(\eta)\tilde{\vpi}^2\,\dfrac{1}{-\eta^2\nabla^2+\mu^2+\frac{1}{4}}\sum_{n=0}^\infty\,\left[(\eta^2\partial_\eta^2+2\eta \partial_\eta)\dfrac{-1}{-\eta^2\nabla^2+\mu^2+\frac{1}{4}}\right]^n\,a^2(\eta)\tilde{\vpi}^2\,.
\end{align}
\end{tcolorbox}
\noindent By replacing the two-field action \eqref{S2-sigma}-\eqref{interpisigma} with this non-local single field EFT one can compute the corresponding seed four-point function $\hat{F}(u,v)$. However, as we will discuss shortly, the four-point function that results from summing over the infinite tower of non-local operators above, although very informative, cannot converge to the exact answer.  

\subsection{The four-point function from the non-Local EFT}
\label{4pt-from-EFT}

Assuming that it is legitimate to treat the time derivatives of $\Sigma$ perturbatively, one can exploit the effective non-local Lagrangian \eqref{EFTnonlocal} to compute $\tilde{\vpi}$'s four-point function. The equivalent of $\hat{F}(u,v)$ for this four-point function is given by
\begin{tcolorbox}[colframe=white,arc=0pt]
\begin{align}
\label{Fnl}
    \hat{F}_{\text{EFT}}(u,v)=-\dfrac{g^2}{2}\text{Im}\left\lbrace \int_{-\infty}^0 dx e^{i x/v} \dfrac{1}{x^2+\mu^2+\frac{1}{4}}\sum_{n=0}^{\infty}{\cal O}_n(x,\partial_x)e^{i x/u}+u\leftrightarrow v\right \rbrace\,,
\end{align}
\end{tcolorbox}
\noindent where
\begin{align}
\label{operatorO}
    {\cal O}_n(x,\partial_x)=\left[(x^2\partial_x^2+2x \partial_x)\left(\dfrac{-1}{x^2+\mu^2+\frac{1}{4}}\right)\right]^n\,,
\end{align}
$x= s \eta$ is a dimensionless time variable, and we have added the subscript ``EFT" to indicate that this four-point function is induced by the non-local EFT  \eqref{EFTnonlocal}.\\

\noindent The leading order four-point function ($n=0$ in \eqref{Fnl}) can be re-written as 
\begin{align}
\label{Fn0time}
    \hat{F}_{\text{EFT}}^{n=0}(u,v)=-g^2\left\lbrace \int_{-\infty}^0 dx \sin\left[\left(\frac{1}{u}+\frac{1}{v}\right) x \right]\, \dfrac{1}{x^2+\mu^2+\frac{1}{4}}\right\rbrace \,.
\end{align}
where the characteristic features of the non-local EFT appear: a single time integral  owing to the instantaneous force carried by $\sigma$ leading to a contact interaction, and the competition between the plane-wave oscillations of the cc field (the sin term) and the propagator of the heavy field.
Performing the integration results in
\begin{align}
\label{Fn0}
   \hat{F}_{\text{EFT}}^{n=0}= g^2\frac{e^{-c} \text{Ei}(c)-e^c \text{Ei}(-c)}{2 (\mu^2+1/4)^{1/2}}\,, \quad c\equiv \left(\mu^2+\frac{1}{4}\right)^{1/2}(1/u+1/v)\,,
\end{align}
where $\text{Ei}(z)$ is the exponential integral function. $\hat{F}_{\text{EFT}}^{n=0}$ is plotted (for a fixed $v\gg 1$) in Figure \ref{fig:FexactvsEFT}, and it is contrasted with the exact answer $\hat{F}(u,v)$ that Equations \eqref{correlatorfull} and \eqref{fullF} define. Consistently with the analysis of the previous section, we find that for $\mu=2$, as long as $u$ and $v$ are much greater than unity, $\hat{F}_{\text{EFT}}^{n=0}$ is in perfect agreement with the exact solution to the exchange diagram. In contrary, the mismatch between the two grows around $u\sim 1$ and the EFT result obviously does not contain the cosmological collider oscillations at small $u$. As for $\mu=1$, somewhat surprisingly, we find little difference between the two four-point functions, when $u\gg 1$. This might appear to contradict what we learned from the plot of $r(u,x)$ in Figure \ref{fig:r(u,x)}, which was that (for $\mu\lesssim 1$ and $u\gg 1$) the time derivatives are important around $x\sim m/H\sim {\cal O}(1)$. In retrospect, this overall agreement shows that integration around mass crossing of the intermediate momentum gives a tiny contribution to the whole four-point \eqref{psi4Sigma} in these kinematical configurations. Nevertheless, unlike the case with $\mu\gtrsim 1$, the corrections to the leading order four-point function induced by lighter fields cannot be captured by higher derivative operators in the EFT \eqref{EFTnonlocal}, as demonstrated in the left plot of Fig \ref{fig:FexactvsEFT}. Actually, doing so only makes the predictions worse as the would-be corrections become more and more important, signaling that the non-local EFT is simply not applicable there. Finally, for very heavy particle $\mu\gg 1$, $\hat{F}_{\text{EFT}}$ almost flawlessly reproduces the full answer, as demonstrated in Figure \ref{fig:heavymu}.  \\ 

\noindent It is instructive to draw a rough picture of the EFT four-point function behaviour by directly looking at the bulk time integral in Equation \eqref{Fn0time}. Since we are ultimately interested in the squeezed limit bispectrum, let us already switch to the bispectrum kinematics by setting $v=1/c_s\gg u=\frac{k_\L}{2c_s k_\S}$. Moreover, we take the intermediate field to be very heavy for the following discussion and refer to the integrand of \eqref{Fn0} by $I(x,k_\L/k_\S)$, where remember that $x=k_L \eta$ is the dimensionless conformal time.
We consider two limiting cases of interest, namely: 
\begin{itemize}
    \item  \textit{Short mode exits the sound horizon before the mass crossing of the long mode}
    ($\frac{k_\L}{2c_s k_\S}\gg m/H$): This regimes corresponds to the third timeline in Figure \ref{fig:overview-bispectrum}. It is useful to look at the qualitative behaviour of $I(x,k_\L/k_\S)$ for different values of $x$, i.e. 
    \begin{align}
    \label{qualitativeI}
        I(x,k_\L/k_\S)\sim g^2
        \begin{cases}
        &\dfrac{\sin(2c_s k_\S x/k_\L)}{x^2}\,,\qquad \frac{k_\L}{2c_sk_\S}\lesssim |x|<+\infty\,, \\
        &\\
        &\dfrac{2c_s k_\S}{k_\L}\dfrac{1}{x}\,,\qquad \qquad \frac{m}{H} \lesssim |x|\lesssim \frac{k_\L}{2c_sk_\S}\,,\\
        &\\
        &\dfrac{2c_s k_\S}{k_\L}\dfrac{H^2}{m^2}x\,,\qquad\qquad  0<|x|\lesssim \frac{m}{H}\,.
        \end{cases}
    \end{align}
    The first time interval corresponds to the time span across which the short mode is inside its sound horizon and the integrand is highly oscillatory. The second time stretch is between the sound horizon crossing of the short mode and the mass crossing of the long mode. During this time the massive field is \textit{relativistic}, therefore we expect that over this period the non-local interaction that it mediates (due to its supersonic character) is the largest. On the contrary, after the mass crossing of the long mode (the third line above), the massive particle slows down (becomes non-relativistic) and the associated interaction effectively turns into a local one, hence the $1/m^2$ suppression. \\

    \noindent The qualitative behaviour of $I(x,k_\L/k_\S)$ in \eqref{qualitativeI} indicates that $\hat{F}_{\text{EFT}}$ receives its dominant contribution from the second and the third intervals above, i.e. when the short mode is outside the sound horizon. As a result we find 
    \begin{align}
    \label{feftapprox}
        \hat{F}(k_\L/k_\S)\sim g^2\dfrac{2c_s k_\S}{k_\L}\left( -\log(\frac{2c_s m}{H}\frac{k_\S}{ k_\L})-1/2\right)\,.
    \end{align}
    Indeed this result  very well captures the general trend of the exact EFT four-point given by \eqref{Fn0}. This can be seen by taking the limit of $c\to 0$ (or equivalently $u,v\to \infty$), where we find 
    \begin{align}
\label{c0limit}
    \lim_{c\to 0} \hat{F}_{\text{EFT}}^{n=0}&=-g^2\left(\dfrac{1}{u}+\dfrac{1}{v}\right)\left(\log\left[\dfrac{u+v}{u v}(\mu^2+1/4)^{1/2}\right]-1+\gamma_E\right)\,,
\end{align}
which is close to the simplified result in \eqref{feftapprox} after the replacements $u\to k_\L/2c_sk_\S$ and $v\to 1/c_s$. 
Notice that, in harmony with the conclusions of Section \ref{modefunctionanal}, $\hat{F}_{\text{EFT}}^{n=0}$ agrees with the full four-point function in the same limit (Equation \eqref{Flimit}), at leading order in $1/\mu$. 
\item \textit{Short mode exits the sound horizon after the mass crossing of the long mode}
($\frac{k_\L}{2c_s k_\S}\ll \frac{m}{H}$): this regime corresponds to the first timeline in Figure \ref{fig:overview-bispectrum}. In this case we can approximate $I(x,k_\L/k_\S)$ by
    \begin{align}
        I(x,k_\L/k_\S)\sim g^2
        \begin{cases}
        &\dfrac{\sin(2c_s k_\S x/k_\L)}{x^2}\,,\qquad \frac{m}{H}\lesssim |x|< +\infty\,, \\
        &\\
        &\dfrac{m^2}{H^2}\sin(2c_s k_\S x/k_\L)\,,\qquad   \frac{k_\L}{2c_sk_\S}\lesssim |x|\lesssim \frac{m}{H} \,,\\
        &\\
        &\dfrac{H^2}{m^2}\dfrac{2c_s k_\S}{k_\L}x\,,\qquad\qquad  0<|x|\lesssim \frac{k_\L}{2c_sk_\S}\,.
        \end{cases}
    \end{align}
The first line corresponds to the time where the massive mode is highly relativistic. However, during this time frame the short mode is always deep inside the sound horizon, hence no important contribution to the four-point function arises. The same happens when the long mode become non-relativistic but the short mode is inside its sound horizon. Therefore the most important contribution comes after the sound horizon exit of the short mode (third line) where the massive particle is non-relativistic and it mediates a local interaction, leading to the following behaviour in the squeezed limit:  
    \begin{align}
    \label{feftapprox2}
        \hat{F}(k_\L/k_\S)\sim g^2\dfrac{H^2}{m^2}\dfrac{k_\L}{c_s k_\S}\,. 
    \end{align}
This is compatible with the asymptotic form of \eqref{Fn0} in the $c\to \infty$ limit, namely 
\begin{align}
    \lim_{c\to \infty}\hat{F}_{\text{EFT}}^{n=0}=g^2 \dfrac{u}{\mu^2+1/4}\,.
\end{align}
\item \textit{Short mode exits the sound horizon around the same time as the mass crossing of the long mode}($k_\L/k_\S\sim 2c_s\frac{m}{H}$): This is the resonance limit (second timeline in Figure \ref{fig:overview-bispectrum}). To see that  $\hat{F}_{\text{EFT}}$ should undergo a local maximum around this point, it is enough to observe that \eqref{feftapprox} grows by decreasing $k_\L/k_\S$ down to the minimum at which this formula is applicable, namely $k_\L/k_\S\sim 2c_s\frac{m}{H}$. For more squeezed configurations \eqref{feftapprox2} takes over and $\hat{F}(k_\L/k_\S)$ decreases. As a result, the four-point should reach a maximum somewhere in the resonance region, i.e. $k_\L/k_\S\sim 2c_s\frac{m}{H}$. 
\end{itemize}
\subsection{Limitations of the non-local EFT} 
\begin{figure}
    \centering
     \includegraphics[scale=0.72]{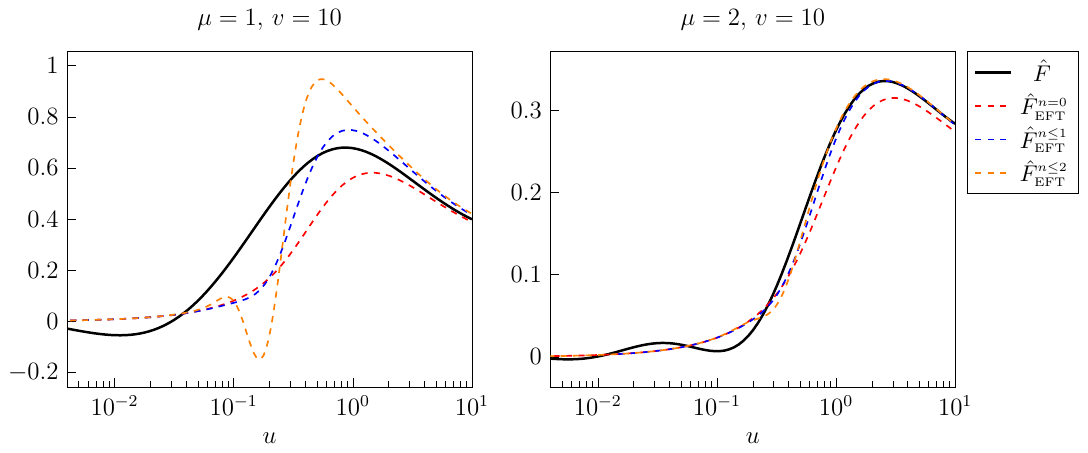}
         \caption{In this figure, we contrast the four-point function computed by means of the non-local EFT namely $\hat{F}_{\text{EFT}}$, with the full result $\hat{F}(u,v)$. Here, $\hat{F}^{n\leq m}_{\text{EFT}}$ indicates the four-point function that comes out of summing over the first $m+1$ operators in the action \eqref{EFTnonlocal}. \textit{Right}: for a typical heavy particle with $m/H \sim {\cal O}(1)$, the non-local EFT prediction for $u>1$ improves by including more higher (time) derivative terms in the action. In agreement with our analysis in Section \ref{modefunctionanal}, adding more terms does not help the precision for $u<1$. That is where the dynamics of $\sigma$ including the particle production effects and the associated oscillations become operative. \textit{Left}: for masses that are close to $3H/2$, including more terms in the action heightens the mismatch between the two results, signaling the divergence of the non-local EFT expansion with not so heavy intermediate particles.}
    \label{fig:FexactvsEFT}
\end{figure}

\begin{figure}
    \centering
    \includegraphics[scale=0.9]{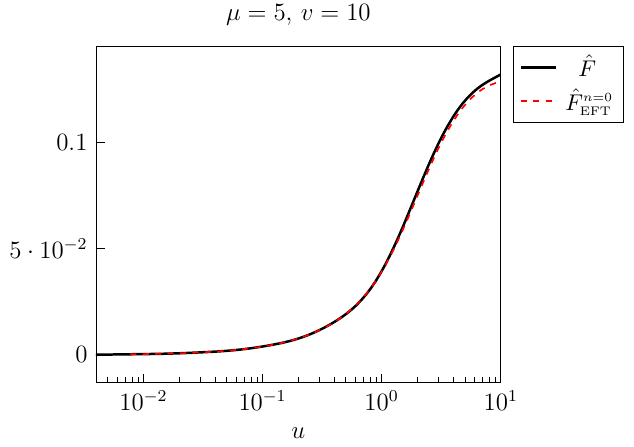}
    \caption{In this plot, the leading order four-point function $\hat{F}^{n=0}_{\text{EFT}}$ associated with a very heavy field ($\mu\gg 1$) is shown to be very close to the full result, irrespective of the size of $u$.}
    \label{fig:heavymu}
\end{figure}

The non-local EFT operators in Eq.~\eqref{EFTnonlocal} do not resum to the exact theory. This can be seen in a few different ways: 
\begin{itemize}
    \item Around $u=-1$ and $v=-1$, according to \eqref{partial}, the full four-point function $\hat{F}(u,v)$ has a logarithmic partial energy singularity. However, the four-point function $\hat{F}_{\text{EFT}}$ induced by the non-local EFT operators have no such singularity at any order in the (time) derivative expansion. This is simply because the latter is a sum over an infinite set of non-local contact terms, which can at best have total energy singularities but not partial energy ones. 
    \item Around $u=0$ (at fixed $v$), $\hat{F}_{\text{EFT}}$ is analytic, whereas the full correlator exhibits a non-analytic behaviour characterising the particle production effect in dS space (see e.g. Eq.~\eqref{squeezedF}). This discrepancy is very natural from the point of view of the non-local single-field EFT \eqref{EFTnonlocal}: the pair creation effect in dS comes from the dynamics of the massive field, which is ignored in this picture.  
   \item Related to the previous point, the dS pair creation affects the correlator not only in the squeezed limit but in any configuration. The mass dependence of such effects are non-perturbative in $1/\mu$ (like in the famous Boltzmann suppression factor $\exp(-\pi \mu)$). As a non-trivial example, consider the next to leading order correction to \eqref{Flimit} in the expansion around $u,v=\infty$, given by \eqref{NLOF}. 
   In the $\mu\to \infty$ limit, this correction is of order $\mu \exp(-\pi\mu)$, hence non-perturbative in $\mu^{-1}$. At the same time, the four-point function from the EFT, given by \eqref{Fnl}, cannot mimic this correction at any order in the time derivative expansion. It is so simply because 
   \begin{align}
       \hat{F}_{\text{EFT}}(-u,-v)=-\hat{F}_{\text{EFT}}(u,v)\,,
   \end{align}
   as can be easily seen from Eq.~\eqref{Fnl}, whereas the correction \eqref{NLOF} is quadratic in the energy ratios, hence invariant under $(u,v)\to (-u,-v)$. 
\end{itemize}
Despite the listed reasons above, in the approximation of $u\gtrsim 1$, $v\gtrsim 1$ and $\mu\gtrsim 1$, the numerical comparison in fig.~\ref{fig:FexactvsEFT} demonstrates that by including more terms in the derivative expansion \eqref{Fnl} the four-point function of the non-local EFT gets closer and closer to the full answer. It is beyond the scope of this work to provide a rigid proof for this statement (however, see Appendix \ref{appendix:F-asymptotic} for the analytical study of the NLO four-point, namely $\hat{F}_{\text{EFT}}^{n=1}$, obtained by keeping only the $n=1$ term in \eqref{Fnl}). We leave a dedicated study of this non-local EFT to future works.

\subsection{Non-local Lagrangian for \texorpdfstring{$\pi$}{e} and non-Gaussian templates for the low-speed collider}

Instead of resorting to the four-point function of the conformally coupled field in the limit where the non-local EFT \eqref{EFTnonlocal} applies, we could have directly derived a single field, non-local EFT for $\pi$. We note that this procedure has been first discussed in \cite{Gwyn:2012mw} in a slightly different but similar context, following the same logic, albeit without computing the resulting bispectra. Paralleling the same steps as before, up to cubic order in the field and leading order in time derivative, the Lagrangian \eqref{S2-sigma}-\eqref{interpisigma} turns into 
\begin{align}
\label{nonlocalEFTpi}
    S_{\pi,\textrm{induced}}=\int d\eta\, d^3\bfx\,a^2(\eta)&\left(\frac{\rho^2}{2} \pi'_c\,\dfrac{1}{m^2-2H^2-H^2\eta^2 \nabla^2}\pi'_c+\dfrac{\rho}{a(\eta)\Lambda_1}\pi'^2_c\,\dfrac{1}{m^2-2H^2-H^2\eta^2 \nabla^2}\pi'_c\right. \\ \nonumber
    &\left. +\dfrac{\rho c_s^2}{a(\eta)\Lambda_2}(\partial_i \pi_c)^2\,\dfrac{1}{m^2-2H^2-H^2\eta^2\nabla^2}\pi'_c\right)\,.
\end{align}
Obviously, computing the power spectrum and the bispectrum by means of this Lagrangian yields the same answer as acting with the proper weight-shifting operator on the four-point $\hat{F}^{n=0}_{\text{EFT}}$ \eqref{Fn0}. 
Explicitly, this gives
\begin{equation}
\left(\dfrac{\Delta P_{\zeta}}{P_{\zeta}}\right)_\textrm{EFT}=\frac{c_s^2 \rho^2}{H^2} \frac{e^{-2 \a} \text{Ei}(2 \a)-e^{2 \a} \text{Ei}(-2 \a)}{2 \a}\,,\end{equation}
in excellent agreement with the exact computation in the domain of validity of the EFT, and where we have defined $\a=c_s(\mu^2+1/4)^{1/2}$. As for the resulting shape for each diagram, they read 
\begin{tcolorbox}[colframe=white,arc=0pt]
\begin{subequations}
\label{nonlocalEFTs}
\begin{align}
    S_{\text{EFT}}^{\dot{\pi}^2 \sigma}(k_1,k_2,k_3)
    &=-\dfrac{1}{8} \frac{\Lambda_2}{\Lambda_1} \left(\frac{\rho}{H} \right)^2\dfrac{k_1 k_2}{k_3^2} \label{SEFT1}\\ \nonumber 
    & \left[\dfrac{2k_3}{k_T}+\a \exp(\frac{\a k_T}{k_3})\text{Ei}\left(-\frac{\a k_T}{k_3}\right)-\a \exp(-\frac{\a k_T}{k_3})\text{Ei}\left(\frac{\a k_T}{k_3}\right)\right]+ 2\,\text{perm}\,, \nonumber\\ 
     S_{\text{EFT}}^{(\partial_i \pi)^2\sigma}(k_1,k_2,k_3)
    &=\dfrac{1}{16} \left(\frac{\rho}{H} \right)^2 \dfrac{(k_3^2-k_1^2-k_2^2)}{k_1 k_2} \label{SEFT2} \\  
    & \left[-\dfrac{2k_1k_2}{k_3 k_T}+\dfrac{1}{\a}\left(1+\frac{\a k_1}{k_3}\right)\left(1+\frac{\a k_2}{k_3}\right)\exp(-\frac{\a k_T}{k_3})\text{Ei}\left(\frac{\a k_T}{k_3}\right)\right. \nonumber \\
    & \left.\quad-\dfrac{1}{\a}\left(1-\frac{\a k_1}{k_3}\right)\left(1-\frac{\a k_2}{k_3}\right)\exp(\frac{\a k_T}{k_3})\text{Ei}\left(-\frac{\a k_T}{k_3}\right)\right]+2\,\text{perm}\,, \nonumber
\end{align}
\end{subequations}
\end{tcolorbox}
\noindent Notice that at this order the normalized shapes of the bispectra depend on $c_s$ and $\mu$ only through the combination $\a \approx c_s m/H$, confirming the intuitive expectation developed in section \ref{sec:setup-overview} that this is the important ``order parameter'' for the low speed collider.\footnote{Actually, one can replace $\left(\mu^2+1/4\right)^{1/2}=\left(\frac{m^2}{H^2}-2\right)^{1/2}$ by $m/H$ without lack of rigor. Indeed, the appearance of $-2 H^2$ in the denominators in Eq.~\eqref{nonlocalEFTpi} comes from neglecting the time derivatives when integrating out the massive field at the level of the canonically normalized fields in conformal time \eqref{fieldredef}. Instead, these terms would be absent upon neglecting time derivatives when integrating out $\sigma$ in cosmic time. Using $m^2-2 H^2$ or $m^2$ simply results into a reorganization in the EFT between the leading order term and the higher-order corrections, but leads to the same all order result. The only advantage of using the first option is that it leads to a more rapidly convergent EFT, providing us with a better agreement with the exact result at leading-order.} Remarkably, these two one-parameter family of shapes generalise the two well known ones from the EFT of inflation --- generated by $\dot{\pi}^3$ and $\dot{\pi} (\partial_i \pi)^2$ interactions --- to which 
they boil down in the limit $\a \gg 1$, with subleading terms in a large $\alpha$ expansion systematically encoding the effects of successive higher-derivative corrections. When $\alpha$ drops below unity, these shapes become qualitatively different and accurately encode the physics of the low speed collider and the associated resonances described in section \ref{sec:resonances}.\\

\begin{figure}
    \centering
    \includegraphics[scale=0.8]{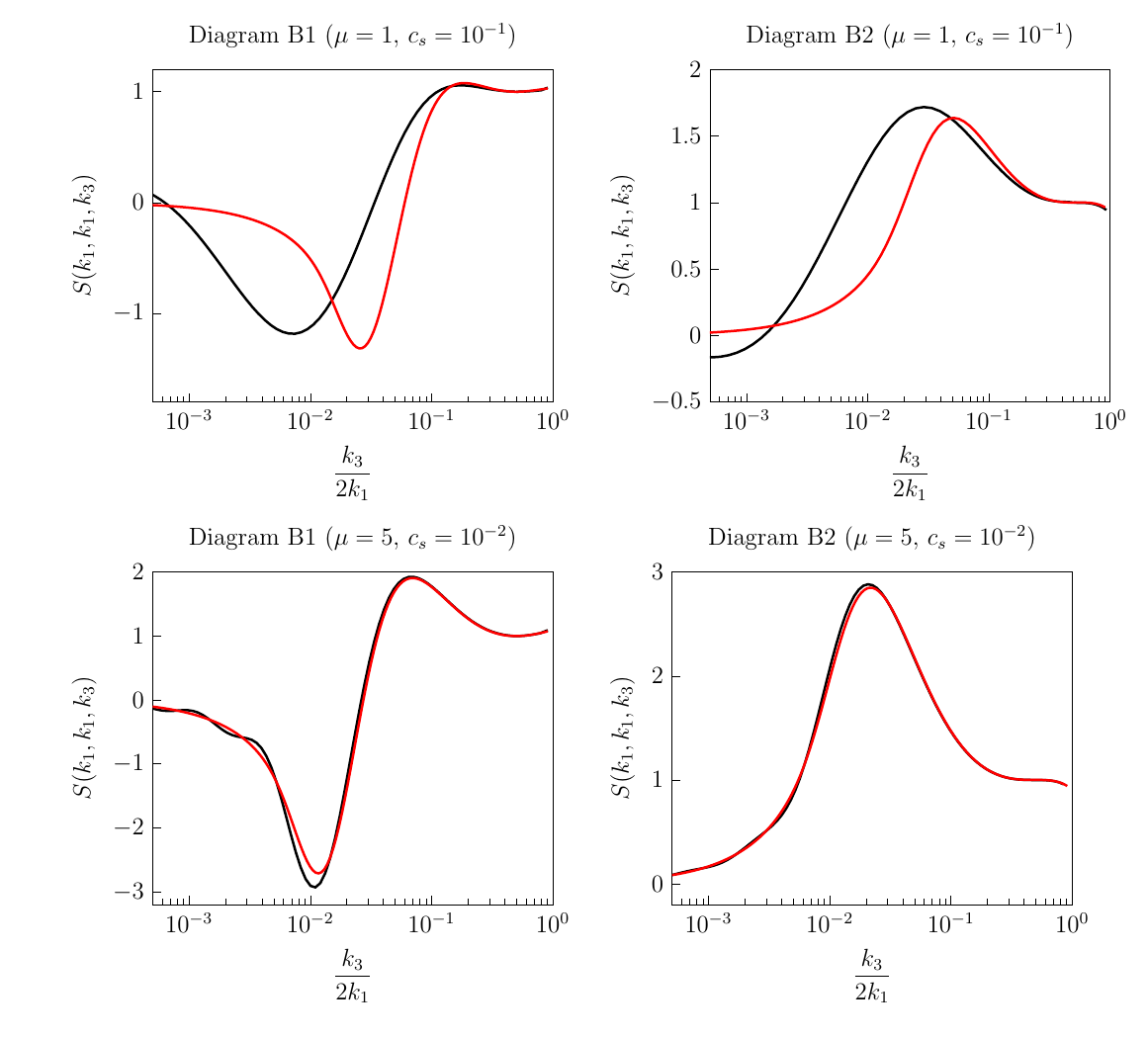}
    \caption{Shapes $S(k_1,k_1,k_3)$ of the bispectra for isosceles triangles, normalized to unity in equilateral configurations, for diagrams B1 (left) and B2 (right), and for ($\mu=1,c_s=0.1$) (top) and $(\mu=5,c_s=0.01)$ (bottom). The black curve is the exact bootstrap result and the red curve corresponds to the leading-order non-local EFT predictions \eqref{SEFT1}-\eqref{SEFT2}.}
    \label{fig:shapes-full-EFT-comparison}
\end{figure}

\noindent Naturally, when using the templates \eqref{SEFT1}-\eqref{SEFT2}, one should bear in mind their domains of validity. For instance, the two theories with $(\mu=1, c_s=0.1)$ and $(\mu=10,c_s=0.01)$ share the same parameter $\alpha$. Nonetheless, the latter theory is well within the realm of the EFT, whereas the former is not.
To further illustrate the (non) applicability and level of accuracy of the non-local EFT picture, in Figure \ref{fig:shapes-full-EFT-comparison} we have confronted the exact shape function of the bispectrum (for both diagrams) with the one \eqref{SEFT1}-\eqref{SEFT2} that the leading-order non-local EFT forecasts.
In sympathy with the analysis of Section \ref{modefunctionanal}, we see that for order one $\mu$'s the two curves start to diverge from each other once the triangle is squeezed to $k_3/k_1\sim 2c_s$. This corresponds to $u\sim 1$ for the $s-$ channel (for the other two channels we have $u\sim c_s^{-1}\gg 1$, indicating that their contribution is very well captured by $\hat{F}^{n=0}_{\text{EFT}}$). Even more, as we have seen in fig.~\ref{fig:FexactvsEFT} (left), the qualitative agreement between the exact and leading-order EFT results is somewhat an accident, as the EFT is simply not convergent for this set of parameters.
Instead, in the other depicted situation with a larger mass, $\mu=5$ and $c_s=0.01$, the leading order non-local EFT perfectly estimates the shape of the bispectrum.

\section{Conclusion}
In this work, we extended the reach of the cosmological bootstrap program to realistic and phenomenologically interesting situations where de Sitter boosts are explicitly broken by the subluminal speed of the curvature perturbation $\zeta$, i.e. $c_s\ll 1$. We showed that, using a set of bespoke weight-shifting operators, the boostless bispectra and trispectra of $\zeta$ induced by the exchange of 
massive particles can be linked to the de Sitter invariant exchange diagram of the conformally coupled field four-point correlator. In contrast with the ordinary case of $c_s=1$, the corresponding weight-shifting operators incorporate rescalings of the external energies of the conformally coupled field by the speed of sound,
while at the same time the momentum of the massive particle is held fixed. This implies that for the purpose of our computation the seed four-point function has to be analytically continued beyond the physical region 
delineated by momentum conservation $|\bfk_1+\bfk_2| \leq (k_1+k_2),(k_3+k_4)$. This continuation is non-trivial, and in fact the expressions provided for this seed solution in the literature are not globally applicable as they entail series expansions organized in powers of e.g. $|\bfk_1+\bfk_2|/(k_1+k_2)$, which do not converge outside the kinematical domain allowed for the four-point configuration. Therefore, one central task in our study was to bootstrap this four-point function from first principles in the appropriate domain before being able to derive useful formulae for the bispectra. This goal was achieved by employing some of the recently developed cosmological bootstrap techniques derived from locality, analyticity and unitarity in the form of a boundary equation that this four-point function satisfies alongside information about the structure of its singularities and finally a cutting rule that relates it to its three-point building blocks.\\ 

\noindent Following the concrete prescription outlined above, we  computed the bispectrum of $\zeta$ for any value of the mass of the heavy particle exchanged and any value of the (ratios between the) sound speed(s). We discovered that supersonic particles that are much lighter than the energy scale $H/c_s$ and are coupled to $\zeta$ leave a characteristic signature in the form of a resonance in the squeezed limit of the bispectrum. This resonance cannot be imitated by adding any number of local operators to the EFT of single field inflation, and it occurs around $k_\L/k_\S\sim c_s\frac{m}{H}$. 
We further showed that, unlike the case with $c_s=1$, the size of the signal has a logarithmic dependence on the mass of the new species. This logarithmic mass dependence originates from an IR divergence that accumulates over time after sound-horizon exit of the short mode $k_S$ until mass crossing of the long mode $k_L$. We also characterised the signal away from the resonance, be it the oscillations of the cosmological collider signal arising in ultra-squeezed configurations $k_\L/k_\S \ll c_s$, or the approximate local-shape behaviour for $k_\L/k_\S \gg c_s \frac{m}{H}$.\\

\noindent Beyond the non-perturbative effects of spontaneous particle production, clearly visible in the ultra-squeezed limit, we demonstrated that the features of the bispectrum described above can be alternatively explained with a simplified non-local single field picture. Indeed, because the interactions mediated by the heavy field propagate at a speed much faster than the one of $\zeta$, one can approximately consider that the former instantaneously responds to the dynamics of the latter. This leads to an effective single-field theory in terms of $\zeta$ only, which emerges after Taylor expanding the time derivatives of the massive field in its propagator; a manipulation that gives rise to an infinite set of non-local operators in the EFT, which are organized in positive powers of temporal derivatives. This non-local EFT provides one with simple templates for the bispectra: one-parameter families of shapes that depend on $\a \approx c_s m/H$, that generalise the ones from the EFT of inflation recovered in the large $\alpha$ limit, while describing the physics of the 
low speed collider and the associated resonances for small $\a$.
Nevertheless, we showed that the corresponding EFT breaks down for particles with masses of order the Hubble scale, for which only our exact bootstrap results are applicable.\\

\noindent Our work can be extended in a few directions: 
\begin{itemize}
   \item \textbf{Multiple exchange diagrams}. Incorporating other interactions between $\pi$ and $\sigma$ gives rise to more complicated double-and triple-exchange diagrams for the bispectrum. Such interactions can lead to a larger
   non-Gaussian signal and it would be tantalizing to relate those diagrams to a new set of higher order seed correlators (of the $\vpi$ field). The problem would then reduce to bootstrapping such seed solutions using bootstrap techniques similar  to the ones used in this work. 
  \item \textbf{Data analysis and prospects for detection}. It would be interesting to quantify the overlap of the new shapes described in this paper, notably the ones \eqref{SEFT1}-\eqref{SEFT2}, with the equilateral/orthogonal and local templates, to use Planck data to constrain the resonances associated with the low speed collider, including when treating the linear mixing $\dot{\pi}\sigma$ non-perturbatively, as well to assess to which extent these can be probed by future non-Gaussian searches.   
   \item \textbf{Non-local single field EFT}. In this work we showed that integrating out a heavy supersonic scalar that couples to the Goldstone boson $\pi$ through the specific interactions specified in \eqref{interpisigma} results in the non-local action \eqref{nonlocalEFTpi}. The set of non-local operators that appear in this action are not the most general ones that can in principle materialise. It would be interesting to systematically classify the consistent set of such operators, which may arise due to the effect of supersonic particles with generic couplings and spins.
  \end{itemize}

\section*{Acknowledgements}

We thank Daniel Baumann, Paolo Creminelli, Guillaume Faye, Jacopo Fumagalli, Sebastian Garcia Saenz, Harry Goodhew, Austin Joyce, Mang Hei Gordon Lee, Liam McAllister, Mehrdad Mirbabayi, Enrico Pajer, Lucas Pinol, Luca Santoni and Lukas Witkowski for stimulating discussions. We are grateful to Dong-Gang Wang and Guilherme Pimentel for discussions and for sharing the manuscript of a related work \cite{Pimentel:2022fsc}, to Denis Werth for very useful feedback throughout the course of this project, and David Stefanyszyn and Denis Werth for comments on the manuscript. SJ and S.RP are supported by the European Research Council under the European Union’s Horizon
2020 research and innovation programme (grant agreement No 758792, project GEODESI). We thank the HEP group at International Center for Theoretical Physics (ICTP) in Trieste for their hospitality when some parts of this work were under progress. 

\appendix

\section{Aspects of the seed four-point function}

\subsection{Derivation of \texorpdfstring{$a_{m,n}$}{e}'s and \texorpdfstring{$b_{m,n}$}{e}'s}
\label{detailsderivation}

Inserting the ansatz \eqref{ansatz} for the particular solution inside the boundary equation \eqref{boundaryeq1}, one finds the following recursive relations for $a_{m,n}$ and $b_{m,n}$: 
\begin{align}
 \left[\mu^2+1/4+(n-m)^2-(n-m)\right]b_{mn}&=(n-m-2)(n-m-1)b_{m+2,n}\,, \\ \nonumber
 \\\nonumber
 \left[\mu^2+1/4+(n-m)^2-(n-m)\right] a_{m,n}&=(n-m-2)(n-m-1)a_{m+2,n} \nonumber\\ \nonumber
  & +(2n-2m-3)b_{m+2,n}-(2n-2m-1)b_{m,n}\,,
\end{align}
alongside the following consistency conditions on elements with $m=0,1$: 
\begin{align}
&n(n-1)a_{1n}+(2n-1)b_{1n}= \dfrac{1}{2} g^2(-1)^{n+1}\,, \\ \nonumber
&n(n+1)a_{0n}+(2n+1)b_{0n}=0\,, \\ \nonumber
&b_{1n}=0\,\quad (n\geq 2)\,, \\ \nonumber
&b_{0n}=0\,\quad (n \geq 1) \,.
\end{align}
From the above equations we infer that 
\begin{align}
\label{infered}
    b_{0n} &=b_{2n}=0\,,\\ \nonumber
    b_{10} &=b_{11} =\dfrac{1}{2} g^2\,, \\ \nonumber
    a_{1n} &=\frac{g^2 (-1)^{n+1}}{2n(n-1)}\,\quad (n\geq 2)\,,\\ \nonumber
    a_{0n}&=0 \,\quad (n\geq 1)\,,\\ \nonumber
    a_{2n}&=0\, \quad (n\geq 3)\,.
\end{align}
It will come in handy to introduce an alternative set of variables:  
\begin{align}
  B_{k,n}\equiv b_{n-k,n}\,,\quad A_{k,n}\equiv a_{n-k,n}\,,\quad -\infty<k\leq n\,,
\end{align}
which satisfy 
\begin{align}
   & B_{k,n}(\mu^2+\frac{1}{4}+k^2-k)-B_{k-2,n}\,(k-2)(k-1)=0\,,\\ \nonumber
   & A_{k,n}(\mu^2+\frac{1}{4}+k^2-k)-A_{k-2,n}\,(k-2)(k-1)=(2k-3)B_{k-2,n}-(2k-1)B_{k,n}\,.
\end{align}
It follows from the first equation above that
\begin{align}
\label{bkn0}
    B_{k,n}=0\,\quad  (k\geq 1)\,,
\end{align}
and also that all the $B_{k,n}$ elements with $k\leq 0$ are fixed by $B_{0,n}$ and $B_{-1,n}$, and they are given by 
\begin{align}
\label{A7}
  &  B_{-(2l+1),n}=\dfrac{1}{6}(9/4+\mu^2)B_{-1,n}\,\dfrac{\left(\frac{7}{4}+\frac{i\mu}{2}\right)_{l-1}\,\left(\frac{7}{4}-\frac{i\mu}{2}\right)_{l-1}}{(2)_{l-1}\,(5/2)_{l-1}}\,\quad (l\geq 0)\,,\\ \nonumber
  & B_{-2l,n}=\dfrac{1}{2}(1/4+\mu^2)B_{0,n} \dfrac{\left(\frac{5}{4}+\frac{i\mu}{2}\right)_{l-1}\,\left(\frac{5}{4}-\frac{i\mu}{2}\right)_{l-1}}{(2)_{l-1}\,(3/2)_{l-1}}\,\quad (l\geq 0)\,.
\end{align}
In the expressions above, $q_l$ stands for the Pochhammer symbol $q_l\equiv \dfrac{\Gamma(q+l)}{\Gamma(q)}$. As for $A_k$'s, let us first look at the special cases $k=1$, $k=2$, for which we find
\begin{align}
    A_{1,n}=-\dfrac{B_{-1,n}}{(\mu^2+1/4)}\,,\quad A_{2,n}=\dfrac{B_{0,n}}{(\mu^2+9/4)}\,.
\end{align}
For $k>1$, \eqref{bkn0} implies 
\begin{align}
     & A_{2l,n}=\dfrac{B_{0,n}}{(\mu^2+9/4)}\dfrac{(1)_{l-1}\,(\frac{3}{2})_{l-1}}{(\frac{7}{4}-\frac{i\mu}{2})_{l-1}(\frac{7}{4}+\frac{i\mu}{2})_{l-1}}\,\quad (l\geq 1)\,,\\ \nonumber
     & A_{2l+1,n}=-\dfrac{B_{-1,n}}{1/4+\mu^2}\,\dfrac{(\frac{1}{2})_l\,(1)_l}{(\frac{5}{4}-\frac{i\mu}{2})_l\,(\frac{5}{4}+\frac{i\mu}{2})_l}\,\quad (l\geq 0).
\end{align}
The last two relations together with $a_{1,n\geq 2}$ given by \eqref{infered} fix the $B_{0,n}$ and $B_{-1,n}$ elements in terms of $g^2$, namely 
\begin{align}
    \quad B_{-1,2p}&=\dfrac{g^2}{2}\dfrac{ (\frac{1}{4}+\mu^2)}{2p(2p-1)}\dfrac{(\frac{5}{4}+\frac{i\mu}{2})_{p-1}(\frac{5}{4}-\frac{i\mu}{2})_{p-1}}{(\frac{1}{2})_{p-1}(1)_{p-1}}\,\quad (p\geq 1),\\ \nonumber
    B_{0,2p+1}&=\dfrac{g^2}{2} \frac{ (\frac{9}{4}+\mu^2)}{2p(2p+1)}\dfrac{(\frac{7}{4}+\frac{i\mu}{2})_{p-1}(\frac{7}{4}-\frac{i\mu}{2})_{p-1}}{(1)_{p-1}(\frac{3}{2})_{p-1}}\,\quad (p\geq 1)\,.
\end{align}
Furthermore, without loss of generality we can set 
\begin{align}
\label{bzero}
B_{-2l,2p}=B_{-2l-1,2p+1}=0\,\quad (p\geq 1\,,l\geq 0)\,.
\end{align}
Now we move onto the elements of $A_{k,n}$ with $k\leq 0$, the values of which are tied to the elements of $B_{k,n}$ through the following recursion relation: 
\begin{align}
\label{recur}
    A_{k,n}(\mu^2+\frac{1}{4}+k^2-k)-A_{k-2,n}(k-2)(k-1)=B_{k,n}\left[\dfrac{(2k-3)(\mu^2+\frac{1}{4}+k^2-k)}{(k-2)(k-1)}-(2k-1)\right]\,.
\end{align}
First of all, we have the freedom to set 
\begin{align}
    A_{0,2p+1}=A_{-1,2p}=0\,\quad (p\geq 0)\,,  
\end{align}
and it follows from \eqref{recur} and \eqref{bzero} that 
\begin{align}
    A_{2l,2p+1}=A_{-2l-1,2p}=0\,\quad (p\geq 0, l\geq 0)\,.
\end{align}
Then Equation \eqref{recur} in conjunction with \eqref{A7} leads us to the rest of the $A_{k,n}$ elements:
\begin{align}
      A_{-2l,2p+1}&=-\dfrac{g^2}{2}\dfrac{\cosh(\pi\mu)\,2^{2l+2p-2}}{\pi^2 \Gamma(2l+1)\Gamma(1+2p+1)}\\ \nonumber 
     &  \times \Gamma\left(\frac{1}{4}+l+\frac{i\mu}{2}\right) \Gamma\left(\frac{1}{4}+l-\frac{i\mu}{2}\right)  \Gamma\left(\frac{3}{4}+p+\frac{i\mu}{2}\right) \Gamma\left(\frac{3}{4}+p-\frac{i\mu}{2}\right) \\ \nonumber
     & \left(-H_{2l}-H_{-\frac{3}{4}+\frac{i\mu}{2}}+H_{-\frac{3}{4}+\frac{i\mu}{2}+l}+(\mu \to -\mu)\right)\,\quad (p\geq 1\,,\quad l\geq 0)\,, \\ \nonumber
      A_{-2l-1,2p}&=-\dfrac{g^2}{2}\dfrac{\cosh(\pi\mu)\,2^{2l+2p-2}}{\pi^2 \Gamma(2l+2)\Gamma(2p+1)}\\ \nonumber
      & \Gamma\left(\frac{3}{4}+l+\frac{i\mu}{2}\right)  \Gamma\left(\frac{3}{4}+l-\frac{i\mu}{2}\right)  \Gamma\left(\frac{1}{4}+p+\frac{i\mu}{2}\right)  \Gamma\left(\frac{1}{4}+p-\frac{i\mu}{2}\right) 
     \\ \nonumber
     & \left(1-H_{2l+1}-H_{-\frac{1}{4}+\frac{i\mu}{2}}+H_{-\frac{1}{4}+\frac{i\mu}{2}+l}+(\mu \to -\mu)\right)\,\quad (p\geq 1\,,\quad l\geq 0)\,.
\end{align}
It can be further verified that the above results for the $A_{k,n}$ and $B_{k,n}$ matrices are consistent with the earlier conditions in \eqref{infered}.  
Putting everything together and after doing some algebraic simplification we finally arrive at the results presented by Eqs.~\eqref{B-2ln}-\eqref{A-2l-1n}.
\subsection{Partial resummation of the series} 
It is possible to fully resum the logarithmic part of the series in \eqref{ansatz} by writing 
\begin{align}
\sum_{m\geq 0, n\geq 0}^{\infty}b_{m n}u^{-m} (u/v)^n &=\sum_{n \geq 0} \sum_{-\infty < k \leq 0} B_{k,n} \frac{1}{u^{-k}} \frac{1}{v^n} \nonumber \\
&=\sum_{l\geq 0,p\geq 0} B_{-2l,2p+1}\frac{1}{u^{2l}}\frac{1}{v^{2p+1}}+\sum_{l\geq 0,p\geq 0} B_{-(2l+1),2p}\frac{1}{u^{2l+1}}\frac{1}{v^{2p}}\,. 
\end{align}
Now a simplification arises due to the fact that the $B_{-2l,2p+1}$ and $B_{-(2l+1),2p}$ elements (given by Eqs.~\eqref{B-2ln} and \eqref{Boddeven}) have a factorised dependence on $l$ and $p$. Separating the $l$ and $p$ dependent blocks and exploiting the 
series expansion of the hypergeometric function ${}_2 F_1(a,b;c;z)=\sum_{n=0}^\infty \dfrac{(a)_n (b)_n}{(c)_n n!}z^n$ together with the simple fact that
$ \dfrac{(q)_n}{(q-1)_n}=\dfrac{n+q-1}{q-1}$ for $q\neq 1$, we arrive at Eq.~\eqref{Bpart}.\\

\noindent It is useful to partially resum the rest of the elements in the series, namely those that involve the $A_{k,n}$ components. Contrary to the previous case of \eqref{Bpart}, here the dependence on $u$ and $v$ will not factorise. To see this, we reorganize the series in the following manner
\begin{align}
    \label{eq:apart2}
    \sum a_{m,n}\dfrac{u^{n-m}}{v^n}&=\sum_{l=0}^{\infty}\sum_{p=0}^{\infty} A_{-2l,2p+1}\dfrac{1}{u^{2l}v^{2p+1}}+\sum_{l=0}^{\infty}\sum_{p=0}^{\infty}A_{-2l-1,2p}\dfrac{1}{u^{2l+1}v^{2p}}\\ \nonumber
    &+\sum_{p=0}^{\infty}\sum_{l=1}^{p}A_{2l,2p+1}\dfrac{u^{2l}}{v^{2p+1}}+\sum_{p=0}^{\infty}\sum_{l=0}^{p-1}A_{2l+1,2p}\dfrac{u^{2l+1}}{v^{2p}}\,. 
\end{align}
Due to the factorised dependence of $A_{k,n}$ on $k$ and $n$, the summation over $l$ and $p$ above can be separately performed within the first two terms. However, for the third and the fourth contributions, this is not possible simply because the upper limit of $l$ depends on $p$. After doing some algebra, the first two series simplify to the first two terms on the RHS of \eqref{simplifiedsum}, while the last two series above reduce to the third term, where the dependence on $u/v$ is fully resummed. 

\subsection{Singularity structure of \texorpdfstring{$\hat{F}_{++}$}{e} }
\label{singularity-structure-appendix}

\subsubsection*{Partial energy pole}

At we discussed before, the partial energy singularities of $\hat{F}_{++}$ emerge when either of $u$ or $v$ are sent to $-1+i\epsilon$. It can be directly inferred from the time integral that in this limit $\hat{F}_{++}$ behaves as \eqref{partial} (the $v\to -1$ limit follows from the symmetry under the exchange of $u$ and $v$). As a non-trivial cross-check, this behaviour can be checked at the level of the final answer \eqref{fullFpp}. We start with the particular solution: the part with $a_{m,n}$'s elements is singular in the $u\to -1$ limit because of the first two terms on the RHS of \eqref{eq:apart2}, which go as
\begin{align}
 &\sum_{p=0}^\infty\sum_{l=0}^\infty A_{-2l,2p+1}\dfrac{1}{u^{2l}v^{2p+1}}\to c_1(\mu)\log(u+1)f_-(v)\,,\\ \nonumber
    &\sum_{p=0}^\infty\sum_{l=0}^\infty A_{-2l-1,2p}\dfrac{1}{u^{2l+1}v^{2p}}\to -c_2(\mu)\log(u+1)f_+(v)\,,
\end{align}
while the second two terms remain finite in this limit. Here, the coefficients $c_i(\mu)$ are the same as those defined in the collinear limit \eqref{c1}-\eqref{c2}. Moreover,  using Eq.~\eqref{Bpart}, the logarithmic piece in the particular solution behaves as 
\begin{align}
\nonumber
       \lim_{u\to -1+i\epsilon} \log(u) & \sum_{n \geq 0}  \sum_{-\infty < k \leq 0} B_{k,n} \frac{1}{u^{-k}} \frac{1}{v^n}   =-\dfrac{i g^2 \pi^{3/2}}{4} \log(1+u)\\ 
    & \times    \left[\dfrac{1}{\Gamma(1/4+i\mu/2) \Gamma(1/4-i\mu/2)}f_-(v)-\dfrac{1}{\Gamma(3/4+i\mu/2) \Gamma(3/4-i\mu/2)}f_+(v)\right]    \,.
\end{align}
We have to add to all this the homogeneous solution of \eqref{fullF} which inherits the logarithmic divergences of $f_\pm(u)$ in the $u\to -1$ limit, namely 
\begin{align}
     \lim_{u\to -1+i\epsilon} f_+(u)&=-\dfrac{\sqrt{\pi}}{\Gamma(1/4+i\mu/2) \Gamma(1/4-i\mu/2)}\log(1+u)\,,\\ \nonumber
    \lim_{u\to -1+i\epsilon} f_-(u)&=\dfrac{\sqrt{\pi}}{\Gamma(3/4+i\mu/2) \Gamma(3/4-i\mu/2)}\log(1+u)\,.
\end{align}
Putting everything together and using the properties \eqref{fpmprop} to simplify $\hat{f}^*_3(-u^*)$, one finally arrives at Equation \eqref{partial}. 

\subsubsection*{Total energy singularity}
The second type of singularity emerges when the sum of the external energies in the diagram (i.e. $k_T$) vanishes. In this limit, from the knowledge of the time integral, we expect $\hat{F}_{++}$ to take the form in Equation \eqref{totalenergy}. Here we explicitly check this by looking at \eqref{correlatorfull}. 
We first note that the logarithmic part and the homogeneous piece in \eqref{correlatorfull} are both regular around $k_T=0$ inasmuch they are sums over factorised functions of $u$ and $v$, hence the analyticity of the $u\to -v$ limit. The same conclusion goes for the first two terms in \eqref{eq:apart2}. 
Consequently, the total energy singularity can arise only from the third and the fourth term in \eqref{eq:apart2}. Starting from the former, we find 
\begin{align}
   & \lim_{u\to -v} \sum_{p=0}^{\infty} \sum_{l=0}^p\,A_{2l,2p+1}\dfrac{u^{2l}}{v^{2p+1}} \nonumber\\ \nonumber
   &=\lim_{u\to -v}\sum_{l=0}^{\infty}\dfrac{g^2 u^{2l}}{2v^{2l+1}}\dfrac{\Gamma(2l)}{\Gamma(2l+2)}{}_3 F_2(1,3/4+l-i\mu/2,3/4+l+i\mu/2;1+l,3/2+l,v^{-2})\,\\ \nonumber
   &=\sum_{l=0}^{\infty} \dfrac{g^2}{8l^2}\dfrac{v}{v^2-1}(u/v)^{2l}\\ &=\dfrac{g^2}{4}\dfrac{v}{(v^2-1)}\log(1+u/v)(1+u/v)+\text{analytic in}\,\, u/v, 
\end{align}
where in the third line we have sent $l$ to infinity and used the following property of the hypergeometric functions
\footnote{We were unable to find this result in any standard text book of special functions. Nevertheless, using Mathematica, we numerically examined its validity to a high level of precision.}
\begin{align}
   \lim_{l\to \infty} {}_3 F_2(1,l+a,l+b;l+c,l+d,v^{-2})=v^2/(v^2-1)\,.
\end{align}
The non-analytic part of the fourth term in \eqref{eq:apart2} can be extracted in a similar way: 
\begin{align}
\nonumber
    & \lim_{u\to -v} \sum_{p=0}^{\infty} \sum_{l=0}^p\,A_{2l+1,2p}\dfrac{u^{2l+1}}{v^{2p}}\\ \nonumber
    &=-\lim_{u\to -v}\sum_{l=0}^{\infty} \dfrac{g^2}{4 (2l^2+3l+1)} \dfrac{u^{2l+1}}{v^{2l+2}}\,{}_3 F_2(1,5/4+l+i\mu/2,5/4+l-i\mu/2;3/2+l,2+l,1/v^2)\,,\\ 
    &=g^2 \dfrac{v}{v^2-1}\sum_l \dfrac{1}{8l^2}(u/v)^{2l}=\dfrac{g^2}{4}\dfrac{v}{(v^2-1)}\log(1+u/v)(1+u/v)+\text{analytic in}\,\, u/v \,.
\end{align}
As a result we arrive at Eq.~\eqref{totalenergy}. 

\section{The asymptotic limit of \texorpdfstring{$\hat{F}(u,v)$}{e} at \texorpdfstring{$u,v\to \infty$}{e}}
\label{appendix:F-asymptotic}
In this appendix, we first compute the NLO and NNLO corrections to the asymptotic limit of $\hat{F}(u,v)$ in \eqref{Flimit}. Then we compare the result with the predictions of the non-local EFT (defined with \eqref{EFTnonlocal}).\\

\noindent Having \eqref{Ycoef}, it is straightforward to read off the NLO term in the $u,v\to \infty$ limit of \eqref{correlatorfull} (while $u/v$ is held fixed). It turns out that, up to cubic order in the inverse of the energy ratios $u$ and $v$, $\hat{F}(u,v)$ only entails elementary functions of the ratio $u/v$, and it is given by
\begin{align}
\label{deltaF3}
    \dfrac{1}{g^2}\hat{F}(u,v) &\approx -\left(\dfrac{1}{u}+\dfrac{1}{v}\right)\left(\log(C(\mu)\dfrac{u+v}{uv})+\gamma_E-1\right)\\ \nonumber
    & -\frac{(u+v) \left(\left(\mu ^2+\frac{9}{4}\right) \left(u^2+v^2\right)+\left(2 \mu
   ^2-\frac{3}{2}\right) u v\right) \left(\log \left(C(\mu)\frac{ (u+v)}{u v}\right)+\gamma_E\right)}{6 u^3 v^3}\\ \nonumber
   & +\dfrac{11}{36}\frac{(u+v) \left(\left(\mu
   ^2+\frac{27}{44}\right) \left(u^2+v^2\right)-\left(\frac{9}{22}-2 \mu ^2\right) u v\right)}{
   u^3 v^3}+\frac{\Gamma \left(\frac{3}{4}-\frac{i \mu }{2}\right)^2 \Gamma \left(\frac{i \mu
   }{2}+\frac{3}{4}\right)^2}{\pi  u v}\,.
\end{align}
This result shows in particular that \eqref{Flimit} is a viable estimation of $\hat{F}(u,v)$ as long as $\mu\ll \text{max}(u,v)$.\\ 

\noindent It is instructive to compare \eqref{deltaF3} with the prediction of the non-local EFT \eqref{EFTnonlocal}. In the (time) derivative expansion, the leading order four-point \eqref{Fn0} is found to be
\begin{align}
\label{fnl0app}
    \hat{F}_{\text{EFT}}^{n=0}&=-g^2\left(\dfrac{1}{u}+\dfrac{1}{v}\right)\left(-1+\gamma_E+\log\left[(\mu^2+1/4)^{1/2}\left(\dfrac{1}{u}+\dfrac{1}{v}\right)\right]\right) \\ \nonumber
    &-\dfrac{g^2}{36}(\mu^2+1/4)\left(\dfrac{1}{u}+\dfrac{1}{v}\right)^3 (6 \log \Big[(\mu^2+1/4)^{1/2}\left(\dfrac{1}{u}+\dfrac{1}{v}\right)\Big]+6 \gamma_E -11)+{\cal O}(c_s^4)\,.
\end{align}
Up to this order in the speed of sound (namely ${\cal O}(c_s^3)$) and at leading order in the large mass regime $(\mu^2+1/4)\gg 1$, this result agrees with \eqref{deltaF3}
\footnote{The agreement to order ${\cal O}(c_s^3)$ in the large mass limit is somewhat an accident as it is evident from the plot of Figure \ref{fig:FexactvsEFT} that around $u\sim 1$ (namely for larger $c_s$'s in a fixed kinematics) the LO Lagrangian fails to reproduce the four-point function $\hat{F}$. The deviations between the two arise at higher orders in the expansion around $c_s=0$.}
. The NLO Lagrangian in \eqref{EFTnonlocal} corrects \eqref{fnl0app} by the following amount
\begin{align}
     &\hat{F}_{\text{EFT}}^{n=1}=\dfrac{g^2}{6}\frac{1}{\mu^2+1/4}\left(\dfrac{1}{u}+\dfrac{1}{v}\right)-\dfrac{g^2}{3}\left(\dfrac{1}{u^3}+\dfrac{1}{v^3}\right) \log\left[(\mu^2+1/4)^{1/2}\left(\dfrac{1}{u}+\dfrac{1}{v}\right)\right]\\ \nonumber
        & +\dfrac{g^2}{36}\left(\dfrac{1}{u}+\dfrac{1}{v}\right)\left((5-12
   \gamma )
   \left(\frac{1}{u^2}+\frac{1}{v^2}\right)+\frac{4
   (3 \gamma -2)}{u v}\right)+{\cal O}(c_s^4)\,.
\end{align}
By explicit comparison with the full four-point function \eqref{deltaF3}, it can be seen that $\hat{F}_{\text{NL}}^{n=1}$ corrects the analytical mass dependence of \eqref{fnl0app} up to order ${\cal O}\left((\mu^2+1/4)^{-1}\right)$. In conclusion, except for the last term in \eqref{deltaF3} which is non-perturbative in $1/\mu$ (hence invisible at any order in the derivative expansion) the remaining terms are captured by the LO and NLO operators in the non-local EFT \eqref{EFTnonlocal}, up to order ${\cal O}\left((\mu^2+1/4)^{-1}\right)$ in the large mass regime.
\bibliographystyle{JHEP}
\bibliography{References}

\end{document}